\documentclass[12pt]{aastex63}

\usepackage{float}
\usepackage{graphicx}           
\usepackage{latexsym}           
\usepackage{amsmath}            
\usepackage{nccmath}            
\usepackage{bm}                 
\usepackage[english]{babel}
\usepackage{color}              
\usepackage{booktabs}           
\usepackage{multirow}           
\usepackage{natbib}

\newcommand{\DELETED}[1]{\relax}%
{\relax}%

\definecolor{violet}  {rgb}{1.0,0.0,1.0}

\definecolor{dviolet} {rgb}{0.75,0.0,1.0}

\definecolor{blue}    {rgb}{0.0,0.7,1.0}

\definecolor{lblue}   {rgb}{0.5,1,1}

\definecolor{dblue}   {rgb}{0.0,0.0,1.0}

\definecolor{blgr}    {rgb}{0.70,0.80,1.00}

\definecolor{navy}    {rgb}{0.00,0.00,0.48}

\definecolor{green}   {rgb}{0.7,1.0,0.0}

\definecolor{dgreen}  {rgb}{0.0,0.6,0.0}

\definecolor{lgreen}  {rgb}{0.0,0.8,0.0}

\definecolor{dg}      {rgb}{0.0,0.6,0.0}

\definecolor{orange}  {rgb}{1.0,0.5,0.0}

\definecolor{dorange} {rgb}{1.0,0.6,0.0}

\definecolor{brown}   {rgb}{0.1,0.1,0.0}

\definecolor{lbrown}  {rgb}{0.7,0.5,0.0}

\definecolor{red}     {rgb}{1,0.0,0.0}

\definecolor{dred}    {rgb}{0.6,0.0,0.0}

\definecolor{grey}    {rgb}{0.1,0.1,0.1}

\definecolor{lgrey}   {rgb}{0.5,0.5,0.5}

\definecolor{black}   {rgb}{0.0,0.0,0.0}

\newcommand\n            {\noindent}

\newcommand\m            {\medskip}
\newcommand\s            {\smallskip}

\newcommand\si           {\smallskip\indent}
\newcommand\bn           {\bigskip\noindent}
\newcommand\mn           {\medskip\noindent}
\newcommand\sn           {\smallskip\noindent}
\newcommand\cl           {\centerline}
\newcommand\ve           {\vfill\eject}

\newcommand\arcspt       {{$\buildrel{\prime\prime}\over .$}}
\newcommand\degree       {{\ifmmode^\circ\else$^\circ$\fi}} 
\newcommand\arcm         {{\ifmmode {'\ }\else$'     $\fi}} 
\newcommand\arcs         {{\ifmmode{''\ }\else$''    $\fi}} 

\newcommand\bEcl         {{$b^{\rm Ecl}$}}
\newcommand\lEcl         {{$l^{\rm Ecl}$}}

\newcommand\bII          {{$b^{\rm II}$} }
\newcommand\lII          {{$l^{\rm II}$}}

\newcommand{\bul}        {$\bullet$\ }
\newcommand\cge          {{$\gtrsim$}}
\newcommand\cle          {{$\lesssim$}}
\newcommand\degsq        {{deg$^{2}$}}

\newcommand\eg           {{\it e.g.},}
\newcommand\ie           {{\it i.e.},}
\newcommand\emin         {{$e^{-}$}}
\newcommand\eminpixsec   {{$e^{-}$/pixel/sec}}

\newcommand\etal         {{et\thinspace al.}}

\newcommand\Inu          {{$I_{\nu}$}}
\newcommand\Ho           {{$H_{0}$} }

\newcommand\HAB          {{$H_{AB}$} }
\newcommand\JAB          {{$J_{AB}$} }
\newcommand{\jit}        {\emph{jit}}

\newcommand\kmsMpc       {{km\ s$^{-1}$\ Mpc$^{-1}$} }

\newcommand\Lya          {{Ly$\alpha$} }

\newcommand\magarc       {{mag\ arcsec$^{-2}$}}

\newcommand\mum          {{\micron}}

\newcommand\nWsqmsr      {{nW\ m$^{-2}$\ sr$^{-1}$}}

\newcommand\WsqmHz       {{$W\ m^{-2}\ Hz^{-1}$}}

\newcommand\re           {{$r_{\rm e}$} }

\newcommand\alphaSun     {{$\alpha_{\odot}$}}

\newcommand\ProFit       {\texttt{ProFit}}
\newcommand\ProFound     {\texttt{ProFound}}
\newcommand\SExtractor   {\texttt{SourceExtractor}}

\newcommand\texp         {{$t_{\rm exp}$}}

\newcommand\Teff         {{$T_{\rm eff}$}}

\defcitealias{Driver2016b}{D16}
\defcitealias{Windhorst2011}{W11}
\defcitealias{Carleton2022}{SKYSURF-2}
\defcitealias{Windhorst2022}{SKYSURF-1}

\def\ltsima{$\; \buildrel < \over \sim \;$}
\def\lsim{\lower.5ex\hbox{\ltsima}}
\def\gtsima{$\; \buildrel > \over \sim \;$}
\def\gsim{\lower.5ex\hbox{\gtsima}}

\newlength{\txw}
\newlength{\txh}

\begin{document}

\vspace*{-0.50cm}
\title{SKYSURF: Constraints on Zodiacal Light and Extragalactic Background Light
through Panchromatic HST All-Sky Surface-Brightness Measurements: I. Survey
Overview and Methods} 

\author{Rogier A. Windhorst}
\affiliation{School of Earth and Space Exploration, Arizona State University,
Tempe, AZ 85287-1404}

\author{Timothy Carleton} 
\affiliation{School of Earth and Space Exploration, Arizona State University,
Tempe, AZ 85287-1404}

\author{Rosalia O'Brien} 
\affiliation{School of Earth and Space Exploration, Arizona State University,
Tempe, AZ 85287-1404}

\author{Seth H. Cohen} 
\affiliation{School of Earth and Space Exploration, Arizona State University,
Tempe, AZ 85287-1404}

\author{Delondrae Carter} 
\affiliation{School of Earth and Space Exploration, Arizona State University,
Tempe, AZ 85287-1404}

\author{Rolf Jansen} 
\affiliation{School of Earth and Space Exploration, Arizona State University,
Tempe, AZ 85287-1404}

\author{Scott Tompkins} 
\affiliation{School of Earth and Space Exploration, Arizona State University,
Tempe, AZ 85287-1404}

\author{Richard G. Arendt} 
\affiliation{NASA Goddard Space Flight Center, Greenbelt, MD 21771}

\author{Sarah Caddy} 
\affiliation{Macquarie University, Sydney, NSW 2109, Australia}

\author{Norman Grogin} 
\affiliation{Space Telescope Science Institute, 3700 San Martin Drive,
Baltimore, MD 21210}

\author{Anton Koekemoer} 
\affiliation{Space Telescope Science Institute, 3700 San Martin Drive,
Baltimore, MD 21210}

\author{John MacKenty} 
\affiliation{Space Telescope Science Institute, 3700 San Martin Drive,
Baltimore, MD 21210}

\author{Stefano Casertano} 
\affiliation{Space Telescope Science Institute, 3700 San Martin Drive,
Baltimore, MD 21210}

\author{Luke J. M. Davies} 
\affiliation{The University of Western Australia, M468, 35 Stirling Highway, 
Crawley, WA 6009, Australia}

\author{Simon P. Driver} 
\affiliation{International Centre for Radio Astronomy Research (ICRAR) and the
International Space Centre (ISC), The University of Western Australia, M468,
35 Stirling Highway, Crawley, WA 6009, Australia}

\author{Eli Dwek} 
\affiliation{NASA Goddard Space Flight Center, Greenbelt, MD 21771}

\author{Alexander Kashlinsky} 
\affiliation{NASA Goddard Space Flight Center, Greenbelt, MD 21771}

\author{Scott J. Kenyon} 
\affiliation{Smithsonian Astrophysical Observatory, 60 Garden Street,
Cambridge, MA 02138} 

\author{Nathan Miles} 
\affiliation{Space Telescope Science Institute, 3700 San Martin Drive,
Baltimore, MD 21210}

\author{Nor Pirzkal} 
\affiliation{Space Telescope Science Institute, 3700 San Martin Drive,
Baltimore, MD 21210}

\author{Aaron Robotham} 
\affiliation{International Centre for Radio Astronomy Research (ICRAR) and the
International Space Centre (ISC), The University of Western Australia, M468,
35 Stirling Highway, Crawley, WA 6009, Australia}

\author{Russell Ryan} 
\affiliation{Space Telescope Science Institute, 3700 San Martin Drive,
Baltimore, MD 21210}


\author{Haley Abate} 
\affiliation{School of Earth and Space Exploration, Arizona State University,
Tempe, AZ 85287-1404}

\author{Hanga Andras-Letanovszky} 
\affiliation{Steward Observatory, University of Arizona, Tucson, AZ 85721-0065}

\author{Jessica Berkheimer} 
\affiliation{School of Earth and Space Exploration, Arizona State University,
Tempe, AZ 85287-1404}

\author{John Chambers} 
\affiliation{School of Earth and Space Exploration, Arizona State University,
Tempe, AZ 85287-1404}

\author{Connor Gelb} 
\affiliation{School of Earth and Space Exploration, Arizona State University,
Tempe, AZ 85287-1404}

\author{Zak Goisman} 
\affiliation{School of Earth and Space Exploration, Arizona State University,
Tempe, AZ 85287-1404}

\author{Daniel Henningsen} 
\affiliation{School of Earth and Space Exploration, Arizona State University,
Tempe, AZ 85287-1404}

\author{Isabela Huckabee} 
\affiliation{School of Earth and Space Exploration, Arizona State University,
Tempe, AZ 85287-1404}

\author{Darby Kramer} 
\affiliation{School of Earth and Space Exploration, Arizona State University,
Tempe, AZ 85287-1404}

\author{Teerthal Patel} 
\affiliation{School of Earth and Space Exploration, Arizona State University,
Tempe, AZ 85287-1404}

\author{Rushabh Pawnikar} 
\affiliation{School of Earth and Space Exploration, Arizona State University,
Tempe, AZ 85287-1404}


\author{Ewan Pringle} 
\affiliation{School of Earth and Space Exploration, Arizona State University,
Tempe, AZ 85287-1404}

\author{Ci'mone Rogers} 
\affiliation{School of Earth and Space Exploration, Arizona State University,
Tempe, AZ 85287-1404}

\author{Steven Sherman} 
\affiliation{School of Earth and Space Exploration, Arizona State University,
Tempe, AZ 85287-1404}

\author{Andi Swirbul} 
\affiliation{School of Earth and Space Exploration, Arizona State University,
Tempe, AZ 85287-1404}

\author{Kaitlin Webber} 
\affiliation{School of Earth and Space Exploration, Arizona State University,
Tempe, AZ 85287-1404}

\email{Rogier.Windhorst@asu.edu}

\begin{abstract}
We give an overview and describe the rationale, methods, and testing of the
Hubble Space Telescope (HST) Archival Legacy project ``SKYSURF.'' SKYSURF uses
HST's unique capability as an {\it absolute photometer} to measure the
$\sim$0.2--1.7 \mum\ sky surface brightness (SB) from 249,861 WFPC2, ACS, and
WFC3 exposures in $\sim$1400 independent HST fields. SKYSURF's panchromatic
dataset is designed to constrain the discrete and diffuse UV to near-IR sky
components: Zodiacal Light (ZL), Kuiper Belt Objects (KBOs), Diffuse Galactic
Light (DGL), and the discrete plus diffuse Extragalactic Background Light
(EBL). We outline SKYSURF's methods to: (1) measure sky-SB levels between
detected objects; (2) measure the discrete EBL, most of which comes from
AB$\simeq$17--22 mag galaxies; and (3) estimate how much truly {\it diffuse
light} may exist. Simulations of HST WFC3/IR images with known sky-values and
gradients, realistic cosmic ray (CR) distributions, and star plus galaxy counts
were processed with nine different algorithms to measure the {\it ``Lowest
Estimated Sky-SB'' (LES)} in each image between the discrete objects. The best
algorithms recover the LES values within 0.2\% when there are no image
gradients, and within 0.2--0.4\% when there are 5--10\% gradients. We provide a
proof of concept of our methods from the WFC3/IR F125W images, where any
residual diffuse light that HST sees in excess of Zodiacal model predictions
does {\it not} depend on the total {\it object flux} that each image contains.
This enables us to present our first SKYSURF results on diffuse light in
\citet{Carleton2022}. 
\end{abstract}

\bn 


\n \section{Introduction} \label{sec1}

\sn The Hubble Space Telescope (HST) was designed in the 1960s and 1970s to 
observe very faint objects at UV to near-IR wavelengths above the Earth's
atmosphere \citep[\eg][]{Smith1993}. HST's ability to observe outside the
Earth's atmosphere has resulted in very significant gains over ground-based
telescopes in four main areas, namely the ability to: (1) observe in the vacuum
ultraviolet; (2) observe with very stable, repeatable, and narrow Point-Spread
Functions (PSFs); (3) observe against very dark foregrounds and backgrounds;
and (4) perform precision point-source photometry at (very) high
time-resolution. As a consequence, HST also has the unique ability to
accurately measure the surface brightness of foregrounds or backgrounds on
timescales of decades. It is precisely this rather unused capability of HST
that project ``SKYSURF'' will focus on in this paper: measuring the {\it
sky-surf}ace brightness (sky-SB) in all eligible HST Archival images and
analyzing the results to constrain astronomical foregrounds or backgrounds. 

As of April 24, 2022, HST has been in orbit for over 32 years. After successful
correction of the spherical aberration in its primary mirror in December 1993,
HST has produced an unprecedented wealth of high-quality data that has
fundamentally changed our understanding of the Universe. The HST
Archive~\footnote{\url{https://archive.stsci.edu}} presently contains more than
1.5 million exposures from both its imagers and spectrographs. By design, HST
studies frequently targeted faint stars or faint galaxies, but HST has also
produced very dramatic results on, \eg\ planetary and Solar System objects,
exoplanets around nearby stars, Galactic star-formation regions, nearby
galaxies, massive black holes in galaxies, and distant quasars. Of particular
relevance for project SKYSURF are HST's most-used wide field-of-view (FOV)
cameras: the Advanced Camera for Surveys/Wide Field Channel (ACS/WFC), Wide
Field Planetary Camera 2 (WFPC2), and Wide Field Camera 3 (WFC3/UVIS \&
WFC3/IR). 

During the early days of HST before and just after the first Space Shuttle
Servicing Mission (SM1), and before the Hubble Deep Field (HDF) project
\citep{Williams1996}, HST images were not dithered at the sub-pixel level
\citep[\eg][]{Windhorst1992, Windhorst1994b}, because at that time it was not
clear that deliberate image offsets could be done with the required sub-pixel
accuracy. With the introduction of the deep HDF imaging data set
\citep{Williams1996}, it was shown that sub-pixel accuracy dithering could, in
fact, be done, and indeed resulted in much better-sampled image PSFs and
correspondingly increased image depth over the Zodiacal foreground compared to
non-dithered images \citep[see, \eg][]{Driver1995, Odewahn1996, Windhorst1998}.
Since 1995, a properly dithered HST imaging dataset in a given filter has been
traditionally processed using ``drizzling'' techniques, described by, \eg\
\citet{Fruchter2002}, \citet{Lauer1999}, \citet{Koekemoer2011}, 
\citet{Grogin2011}, and \citet{Koekemoer2013}. 

Since 1995, the standard HST drizzling process traditionally removed the 
sky-foreground levels by subtracting a surface fit to the image with the
discrete objects masked out, hence setting the image sky-SB values to zero.
While the original and subtracted sky-SB value may have been preserved in the
reduced image FITS headers, the image sky-values are often not kept in the
subsequent data products, nor is the information about sky-SB gradients that
were removed from the images during the drizzling process. Most HST users have
thus subtracted their image sky-SB values since 1995. This mode of operation
is, in general, not an issue and, in fact, the desired way of proceeding, since
the very large majority of HST targets have been point sources or nearly
point-like sources, and the users' intended interest has usually been the
(almost point-like) faint object flux at certain wavelengths over the local
sky-foreground. Hence, removing the sky-SB and its gradient during the
drizzling process has been, for almost all purposes, a necessary step. However,
for SKYSURF, we need to {\it precisely preserve and measure} the sky-SB in all
eligible HST images on timescales of decades, which we describe below. This
paper will therefore summarize the diffuse astronomical foregrounds and
backgrounds that one may expect in the HST images (\S\ \ref{sec2}), as well as
the instrumental foregrounds that need to be identified, subtracted or
discarded \citep[\S\ \ref{sec3}--\ref{sec4} and][referred to as
\citetalias{Carleton2022} throughout]{Carleton2022}, before these astronomical
foregrounds and backgrounds can be assessed. 

Many of the procedures and methods in this paper are by necessity {\it
non-conventional}, even after 32 years of Hubble Space Telescope use, as
explained above. SKYSURF will reprocess most of the HST images acquired since
1994 on servers provided by Amazon Web Services (AWS). As a result, we simply
cannot plan to repeat this process many times. Hence, the focus of this first
SKYSURF overview paper is to publish our survey rationale and methods early.
This will allow the community to comment on our methods as early as possible
and give our SKYSURF team the opportunity to improve upon those methods before
they are all executed on AWS. 

Our paper is organized as shown in the Table of Contents, where the
(sub-)section headings list all the steps needed to justify (\S\ \ref{sec2}),
define and organize (\S\ \ref{sec3}), calibrate and re-process (\S\
\ref{sec41}) the SKYSURF database with close attention to systematics that may
affect the sky-SB levels in HST images (\S\ \ref{sec411}--\ref{sec48}). This
includes methods that are anchored in simulations to measure the object-free
sky-SB, a sky-SB preserving implementation of the drizzle algorithm, the
flagging of images with orbital straylight, and our methods to do star-galaxy
separation and make panchromatic discrete object catalogs. We discuss our
findings in \S\ \ref{sec5} and summarize our conclusions in \S\ \ref{sec6}.
Appendices give details on the HST orbital parameters and straylight
(\ref{secAppA}), the specific requirements for SKYSURF's image drizzling and
removal of images with artifacts or large extended objects (\ref{secAppB}), and
SKYSURF's procedures to make object catalogs, do star-galaxy separation and
Galactic extinction corrections (\ref{secAppC}). In \citetalias{Carleton2022},
we estimate the sky-SB in all individual WFC3/IR exposures in the F125W, F140W,
and F160W filters, make corrections for the WFC3/IR Thermal Dark signal,
present our first constraints on diffuse light at 1.25--1.6 \mum, and summarize 
our main results thus far. 

The various astronomical foregrounds and backgrounds that exist in the SKYSURF
images are discussed in more detail in \S\ \ref{sec2}. They form the core
reason for carrying out the SKYSURF project. In summary, they are the
following. The Zodiacal Light (ZL) is the main foreground in most HST images,
and SKYSURF will measure it in \S\ \ref{sec42}, and model it in
\citetalias{Carleton2022} as well as possible with available the tools. All
stars in our galaxy (except the Sun) and all other galaxies are beyond the
InterPlanetary Dust Cloud, so the ZL is thus always referred to as a
``foreground''. The Diffuse Galactic Light is caused by scattered star-light in
our Galaxy and can be a background (to nearby stars), or a foreground (to more
distant stars and all external galaxies; see Appendix \ref{secAppC2}). Most
objects in an average moderately deep (AB\cle 25--26 mag) HST image are faint
galaxies close to the peak in the cosmic star-formation history at z\cle 2
\citep[\eg][]{Madau2014}. Most of the Extragalactic Background Light (EBL)
therefore comes from distant galaxies and AGN \citepalias[\S\ \ref{sec23},
\ref{sec47} and][]{Carleton2022}, and is thus referred to as a ``background''.
Before SKYSURF can quantify and model these astronomical foregrounds and
backgrounds, it needs to address the main contaminants, which are residual
detector systematics (\S\ \ref{sec41}), orbital phase-dependent straylight from
the Earth, Sun, and/or Moon (\S\ \ref{sec43}), and the WFC3/IR Thermal Dark
signal \citepalias{Carleton2022}.

Throughout we use Planck cosmology \citep{Ade2016}: \Ho~=~66.9 $\pm$ 0.9\ 
\kmsMpc, matter density parameter $\Omega_{m}$=0.32$\pm$0.03 and vacuum energy
density $\Omega_{\Lambda}$=0.68$\pm$0.03, resulting in a Hubble time of 13.8
Gyr. When quoting magnitudes, our fluxes are all in AB-magnitudes (hereafter
AB-mag), and our SB-values are in AB-\magarc\ \citep{OkeGunn1983} or MJy/sr, 
using flux densities $F_{\nu}$ = 10$^{-0.40({\rm AB} - 8.90\ {\rm mag})}$ in
Jy. Sky-SB values can be converted to units of \nWsqmsr\ by multiplying the
MJy/sr units by 10$^{-11}$(c/$\lambda_{c}$), where $\lambda_{c}$ is the filter
central wavelength. Further details on the flux density scales used are given
in Fig.~\ref{fig:fig1} and \S\ \ref{sec415}.


\n \section{SKYSURF Goals in the Context of Astronomical Foregrounds and 
Backgrounds} \label{sec2} 

\sn For the sake of clarity, we will make a distinction between diffuse {\it
foregrounds} and diffuse {\it backgrounds}. In the following subsections and
\citetalias{Carleton2022}, we define and summarize the physical phenomena from
which these diffuse foregrounds and backgrounds arise, as they form the core of
the SKYSURF project. SKYSURF has two main science goals:

\mn {\it (1)\ SKYSURF-SB:}\ Measure the panchromatic HST ACS, WFPC2, and WFC3
sky-SB --- free of discrete object flux --- across the celestial sphere, and
derive the best possible constraints to the Zodiacal Light (ZL), Diffuse
Galactic Light (DGL), and the Extragalactic Background Light (EBL); and

\mn {\it (2)\ SKYSURF-Sources:}\ Measure the panchromatic integrated 
background from {\it discrete object catalogs} (Galactic stars, galaxies)
across the sky, and derive independent measurements over \cge 1400
representative HST fields far enough apart in the sky to average over the
effects of cosmic variance more accurately than existing HST surveys alone can
do.

\n \subsection{The UV--near-IR Zodiacal Foreground} \label{sec21} 

\begin{figure*}[!hptb]
\n\cl{
\includegraphics[width=0.950\txw]{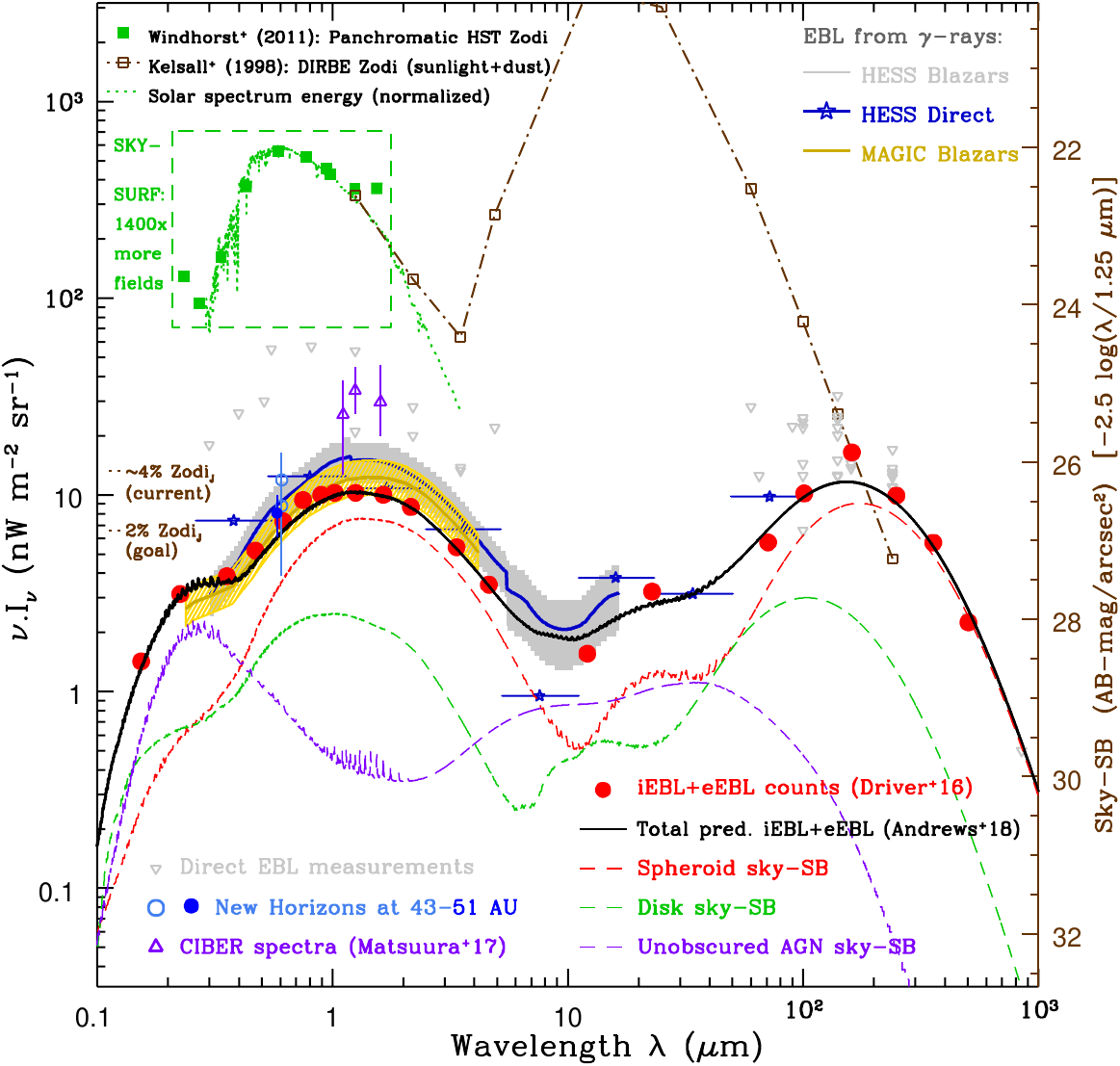}
}

\vspace*{+0.10cm}
\n \caption{
Summary of astrophysical foreground and background energy relevant to SKYSURF.
The left scale indicates the total energy $\nu$.\Inu\ in \nWsqmsr, and the right
scale the corresponding sky-SB in AB-\magarc\ at 1.25 \mum\ (which can be scaled
to other wavelengths as indicated). The discrete measurements of 
\citetalias{Driver2016b} from {\it integrated and extrapolated galaxy counts
(iEBL+eEBL)} (red-filled circles) and other published data are shown (\S\
\ref{sec2}). Grey triangles indicate {\it total EBL measurements} that require
accurate modeling of DGL and ZL, and are $\sim$3--5$\times$ higher than the
discrete iEBL+eEBL, leaving room for a substantial {\it diffuse light
component}. Green squares are panchromatic HST sky-SB measurements compared to
the Solar spectrum. (The 0.23 and 1.6 \mum\ filters shown did not get
\texttt{LOW-SKY}; see \S\ \ref{sec2} \& \ref{sec43}). Brown squares indicate
the COBE/DIRBE data and the Zodiacal dust model of \citet{Kelsall1998}. The
grey and orange wedges and blue stars are $\gamma$-ray Blazar EBL constraints
from the HESS and MAGIC TeV experiments. The black line is the sum of the
\citet{Andrews2018} EBL models for star-formation in spheroids (red), disks
(green), and unobscured AGN (purple). {The purple triangles are the 
\citet{Matsuura2017} CIBER spectral 1.1 \& 1.6 \mum\ estimates of diffuse 
light in excess of the \citet{Kelsall1998} model prediction, as well as the
\citet{Sano2020} 1.25 \mum\ estimate, where the iEBL+eEBL has also been
subtracted.} The light and dark blue circles with error bars at 0.608 \mum\ are
the \citet{Lauer2021, Lauer2022} diffuse light estimates with New Horizons at
43--51 AU. SKYSURF's currently achieved calibration+zeropoint accuracy in the
1.25--1.6 \mum\ HST sky-SB estimates is \cle 4\% of the Zodiacal sky-SB
(Table~\ref{tab:tab5}), as indicated. Our goal is a $\sim$2\%-accurate sky-SB
model across the sky at 0.2--1.7 \mum\ wavelengths to address the discrepancy
between the total EBL and the {\it discrete} iEBL+eEBL. In
\citetalias{Carleton2022} we discuss SKYSURF's resulting upper limits to the
{\it diffuse light} in detail.}
\label{fig:fig1}
\end{figure*}

\sn Much of the area surveyed with HST cameras consists of relatively empty
sky surrounding targets of interest for which the observations were made. {\it
At 0.6--1.25 \mum\ wavelength, over 95\% of the photons in the HST Archive come
from the Zodiacal Light in the InterPlanetary Dust (IPD) cloud, \ie\ from
distances less than 5 AU.} This fraction is illustrated by the ratio of a 
typical ZL spectrum (green squares and green dotted line in
Fig.~\ref{fig:fig1}) to the {\it discrete} EBL integral (red circles plus
black model in Fig.~\ref{fig:fig1}, see also \S\ \ref{sec23}). These photons are
precisely the sky-SB photons present in nearly all HST images between the
discrete objects that are of interest to our SKYSURF study. These sky photons
come primarily from the ZL foreground, which is caused by sunlight scattered by
dust and small particles in the IPD at distances r\cle 3--5 AU, or from even
closer light sources such as Earthshine or Geocoronal emission, scattered light
in the Optical Telescope Assembly (OTA) or thermal foregrounds in the camera
detectors, as discussed in \S\ \ref{sec4} and \citetalias{Carleton2022}.
Constraints on ZL are obtainable from the HST Archive, yet no precise all-sky
panchromatic measurements of the HST sky-SB exist. Ground-based telescopes are
{\it unable} to make absolute measurements of the ZL due to atmospheric
absorption, OH-lines, air glow, and light pollution unless very special
measures are taken 

\ve 


\vspace*{-1.00cm}
\hspace*{-0.00cm}
\n



\n \citep[\eg][]{Bernstein2002, Leinert1998}. Located above the Earth's
atmosphere, HST thus has the unique and rather unexplored capability to serve
as an {\it absolute photometer} for low surface brightness foreground and
background measurements in the 0.2--1.7 \mum\ range on timescales of decades. 

\n \subsection{Discrete HST Objects: Stars and Galaxies} \label{sec22} 

\sn Other than planetary and other moving targets, the main science interest in
HST images has, in general, been stellar objects and galaxies from the brightest
observable stars and star-forming (SF) regions in our own Galaxy and nearby
galaxies to the faintest galaxies visible in the deepest HST images, such as
the Hubble UltraDeep Field \citep[HUDF, \eg][]{Beckwith2006}. Stellar objects
here will include Quasi-Stellar Objects (QSOs) or (weak) Active Galactic Nuclei
(AGN). By selection, the large majority of objects observed in the HST Archive
are nearly point-like objects. This is, of course, because HST was designed to
observe faint objects at UV to red or near-IR wavelengths outside the Earth's
atmosphere (\S\ \ref{sec1}), and faint objects tend to be compact \citep[the
effects of SB-selection on the HST catalog completeness are discussed in \S\
\ref{sec47}; see also][]{Windhorst2008, Windhorst2021}. 

To date, the Hubble Legacy Archive (HLA)~\footnote{\url{http://hla.stsci.edu}}
contains over 1.5 million HST observations, and the Hubble Source Catalog
(HSC)~\footnote{\url{http://archive.stsci.edu/hst/hsc/}} contains at least 3.7
million objects. Following the detailed description of \citet{Budavari2012} and
\citet{Whitmore2016}, the HLC Version 1 object catalogs are derived from
subsets of the WFPC2, ACS/WFC, WFC3/UVIS, and WFC3/IR \SExtractor\ source
lists from the HLA data release version 10 (DR10). This incorporated
cross-matching and relative astrometry of overlapping images to minimize
offsets between closely aligned sources in different images. After correction
for such offsets, the astrometric residuals of cross-matched sources are
significantly reduced, with median errors less than 8 m.a.s. The absolute
astrometry of the HLA is anchored into Gaia DR1, Pan-STARRS, the Sloan Digital
Sky Survey (SDSS), and 2MASS. 

The HLA and HLC are an outstanding permanent legacy of HST's 30$^+$ year record.
SKYSURF's main goal is not to replicate the extensive work that the HLA and HLC
have done to create its object catalogs. Instead, SKYSURF focuses on the
249,861 ACS/WFC, WFPC2, WFC3/UVIS, and IR images in principle suitable for
SKYSURF's main sky-SB science goals, as discussed in \S\
\ref{sec25}--\ref{sec4}. Of these images, 220,657 have exposure times
\texp$\ge$200 sec, and are also eligible for drizzling, panchromatic object
catalogs and object counts, as discussed in \S\ \ref{sec45}--\ref{sec46} \&
Appendix \ref{secAppB}--\ref{secAppC}. Using the WFC3/IR F125W filter as the
fiducial wavelength in this paper, two aspects are essential for SKYSURF:

\si (1) The Galactic star-counts have very flat slopes, while the galaxy counts
have much steeper count slopes, and they cross over with about equal surface
densities at average Galactic latitudes around AB$\simeq$18 mag at 1.25 \mum\ 
\citep[\eg][see also \S\ \ref{sec47} and its Figures here]{Windhorst2011}.

\si (2) The galaxy counts change from non-converging to converging slopes in
the range 17\cle AB\cle 22 mag with only a mild dependence on wavelength
\citep{Windhorst2011, Driver2016b}. Therefore, while the vast majority of
objects detected in HST images of average (\cle 1--2 orbits) depth are
moderately faint (AB\cle 26 mag) galaxies, most of the \emph{total energy}
emitted by discrete objects at UV--optical--near-IR wavelengths is produced by
those galaxies already detected in \emph{single}-exposure HST images
\citep[][\S\ \ref{sec23} \& \ref{sec47} here]{Driver2016b}.

\si The consequences of these two facts for SKYSURF are rather profound: to
accurately measure both the integrated discrete galaxy counts {\it and} the
sky-SB from all SKYSURF images, we must have: (a) very accurate star-galaxy
separation procedures, especially at brighter fluxes (AB\cle 18 mag) where
stars dominate the object counts; and (b) very accurate procedures to grow the
light profiles of all detectable stars and galaxies, especially those with
17\cle AB\cle 22 mag, where most of the EBL is produced, {\it and} remove their
discrete object light from the images before the best estimates of the ZL and
EBL can be made. Hence, SKYSURF must measure and account for the light from
all discrete objects from 220,657 HST images in a manner that differs from that
adopted for the HLA/HSC, as described below. For this, we will use the
star-galaxy separation methods of \citet{Windhorst2011}, which on shallow HST
images are generally robust to AB\cle 25--26 mag (\S\ \ref{sec47}).


\n \subsection{Integrated and Extrapolated Extragalactic Background Light from
Discrete Objects (iEBL+eEBL)} \label{sec23} 

\sn The Extragalactic Background Light is defined as the flux received from all
sources of photon production since recombination at far-UV ($\lambda$\cge 0.1
\mum) to far-IR ($\lambda$\cle 1000 \mum) wavelengths
\citep[\eg][]{McVittie1959, Partridge1967a, Partridge1967b, Hauser2001,
Lagache2005, Kashlinsky2005a, Finke2010, Dominguez2011, Dwek2013, Khaire2015,
Driver2016b, Koushan2021, SaldanaLopez2021}. That is, the EBL reflects the
energy production of the Universe from z$\simeq$1090 until today and consists
mainly of light from stars, AGN, and reprocessed light from dust, with some
contribution from material heated by accretion \citep[\eg][]{Alexander2005,
Jauzac2011, Andrews2018}. The EBL observed today thus results from the cosmic
star-formation history, AGN activity (\ie\ accretion onto super-massive black
holes), and the evolution of cosmic dust over the past $\sim$13.5 billion
years. The EBL can be divided into two roughly equal components: one covering
the UV--near-IR (0.1--8\mum; the Cosmic Optical Background, COB) and one
covering the mid--far-IR (8--1000\mum; the Cosmic Infrared Background, CIB;
\citealp{Dwek1998b, Kashlinsky2000, Andrews2018}; Fig.~\ref{fig:fig1} here). 

With the advent of space-based and ground-based facilities, deep fields have
been obtained across the entire far-UV to far-IR wavelength range. For
instance, \citet{Driver2016b} and \citet{Koushan2021} combined recent wide and
deep panchromatic galaxy counts from the Galaxy And Mass Assembly survey (GAMA;
\citealp{Driver2011, Driver2016a, Hopkins2013, Liske2015}), COSMOS/G10
\citep{Davies2015, Andrews2017}, the HST Early Release Science field (ERS;
\citealp{Windhorst2011}), and Ultra-Violet Ultra-Deep Fields (UVUDF;
\citealp{Teplitz2013, Rafelski2015}), plus near-, mid- and far-IR datasets from
ESO, {\it Spitzer} and {\it Herschel}. To estimate the EBL from discrete
objects, great care was taken in each dataset to produce object catalogs, total
fluxes and object counts across a broad wavelength range. 

\begin{figure*}[!hptb]
\n \cl{
\hspace*{+0.000cm}
\includegraphics[width=0.4900\txw]{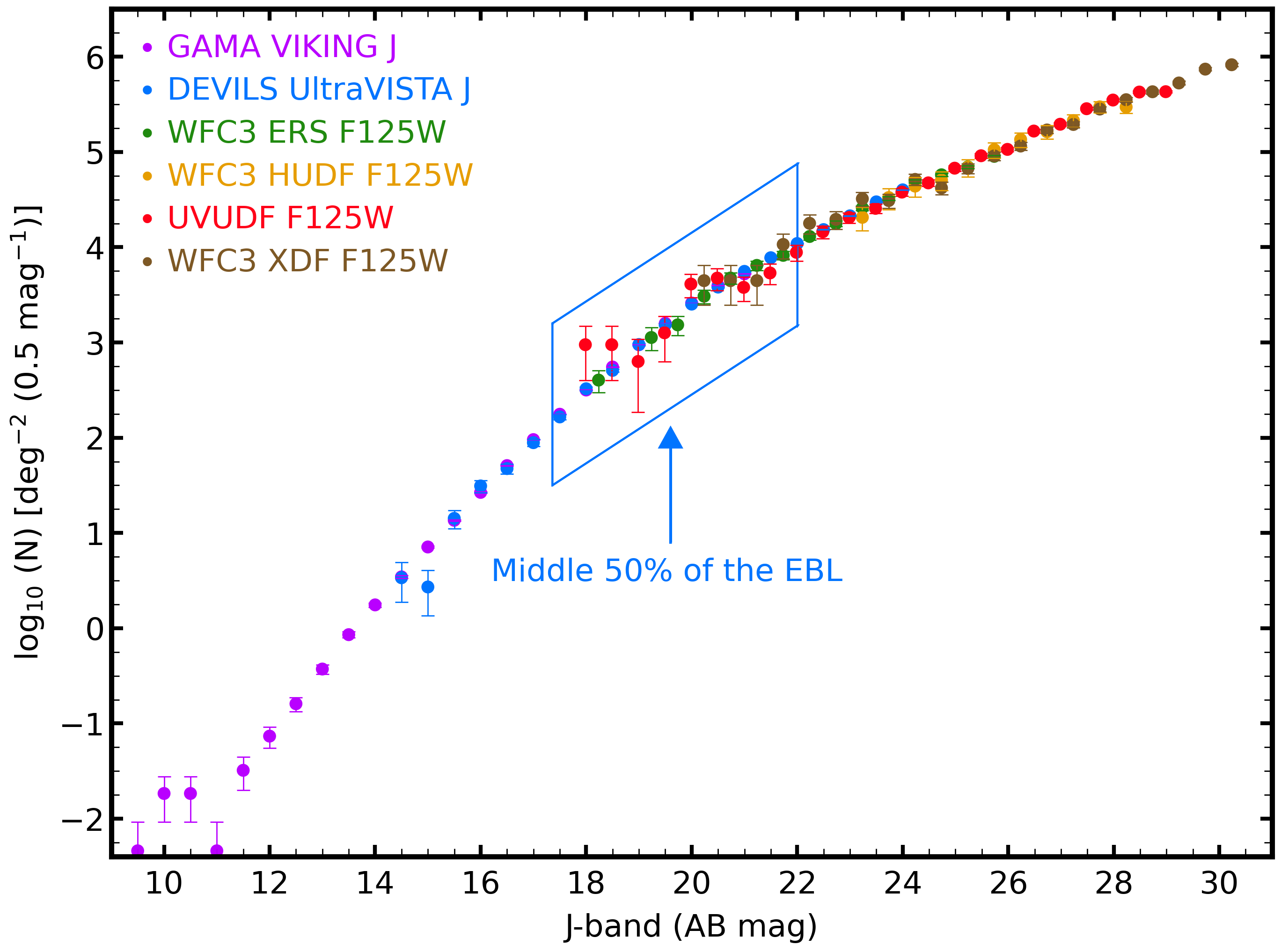}
\hspace*{+0.000cm}
\includegraphics[width=0.5100\txw]{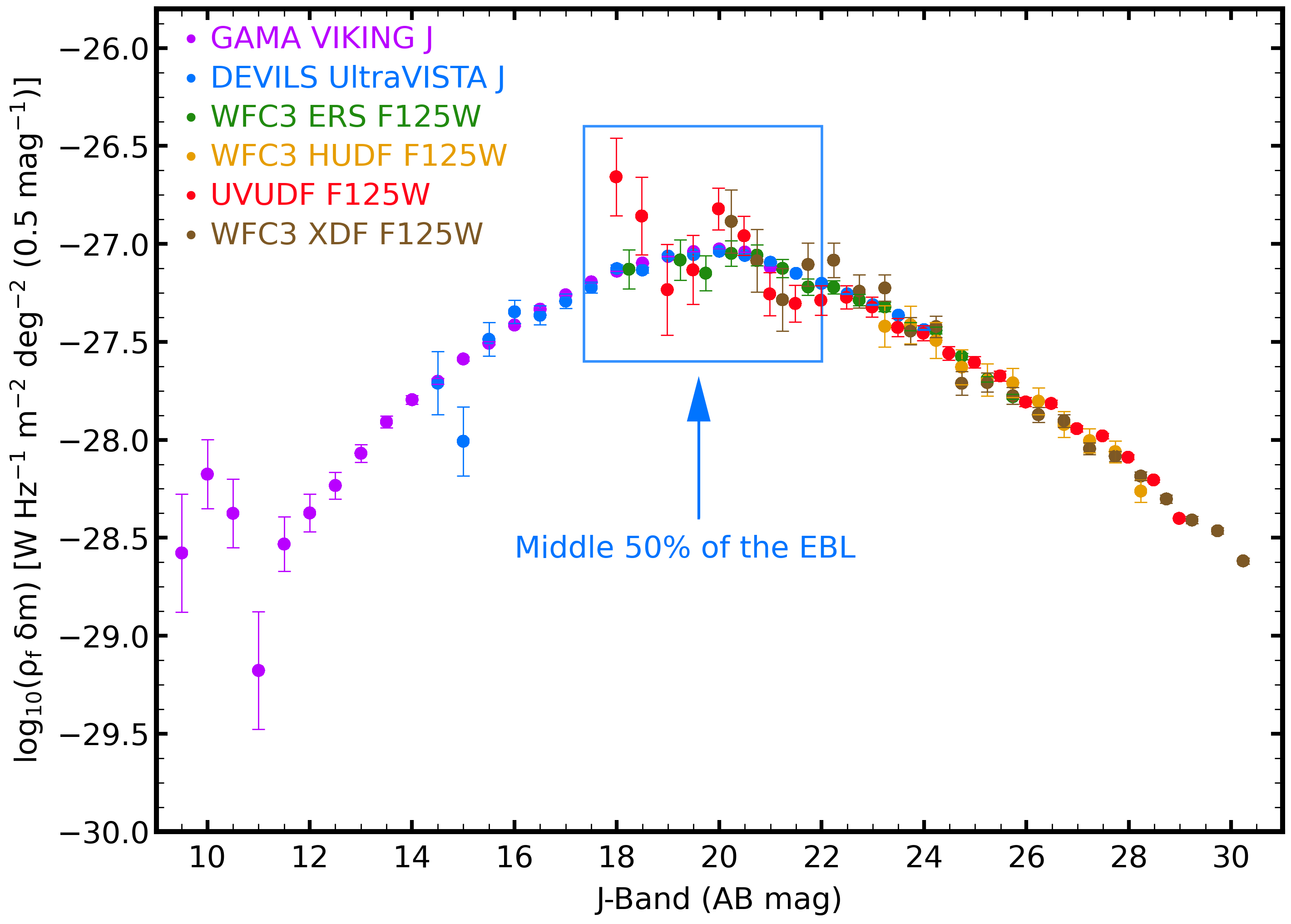}
}

\n \cl{
\hspace*{+0.000cm}
\includegraphics[width=0.4900\txw]{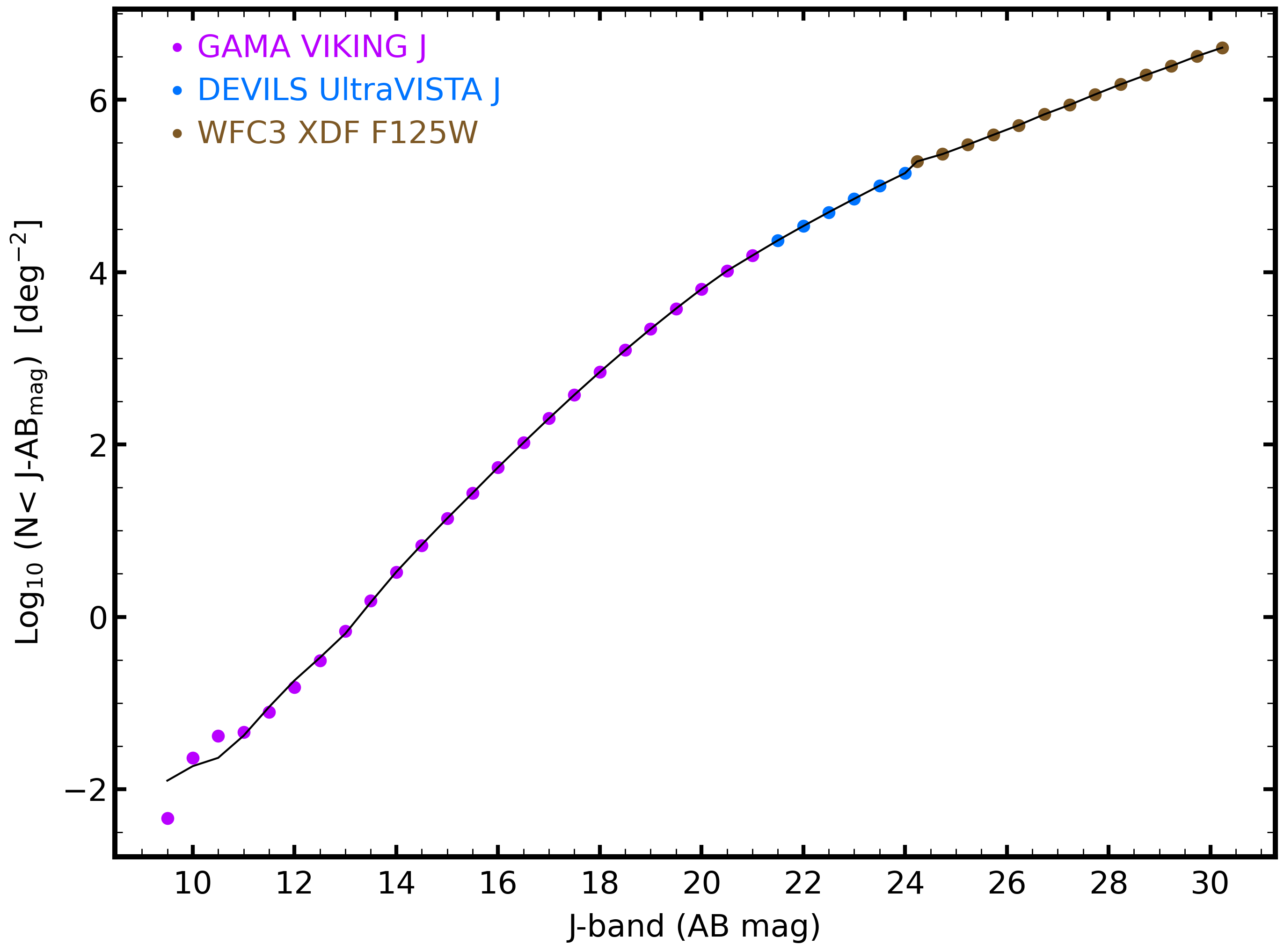}
\hspace*{+0.100cm}
\includegraphics[width=0.5100\txw]{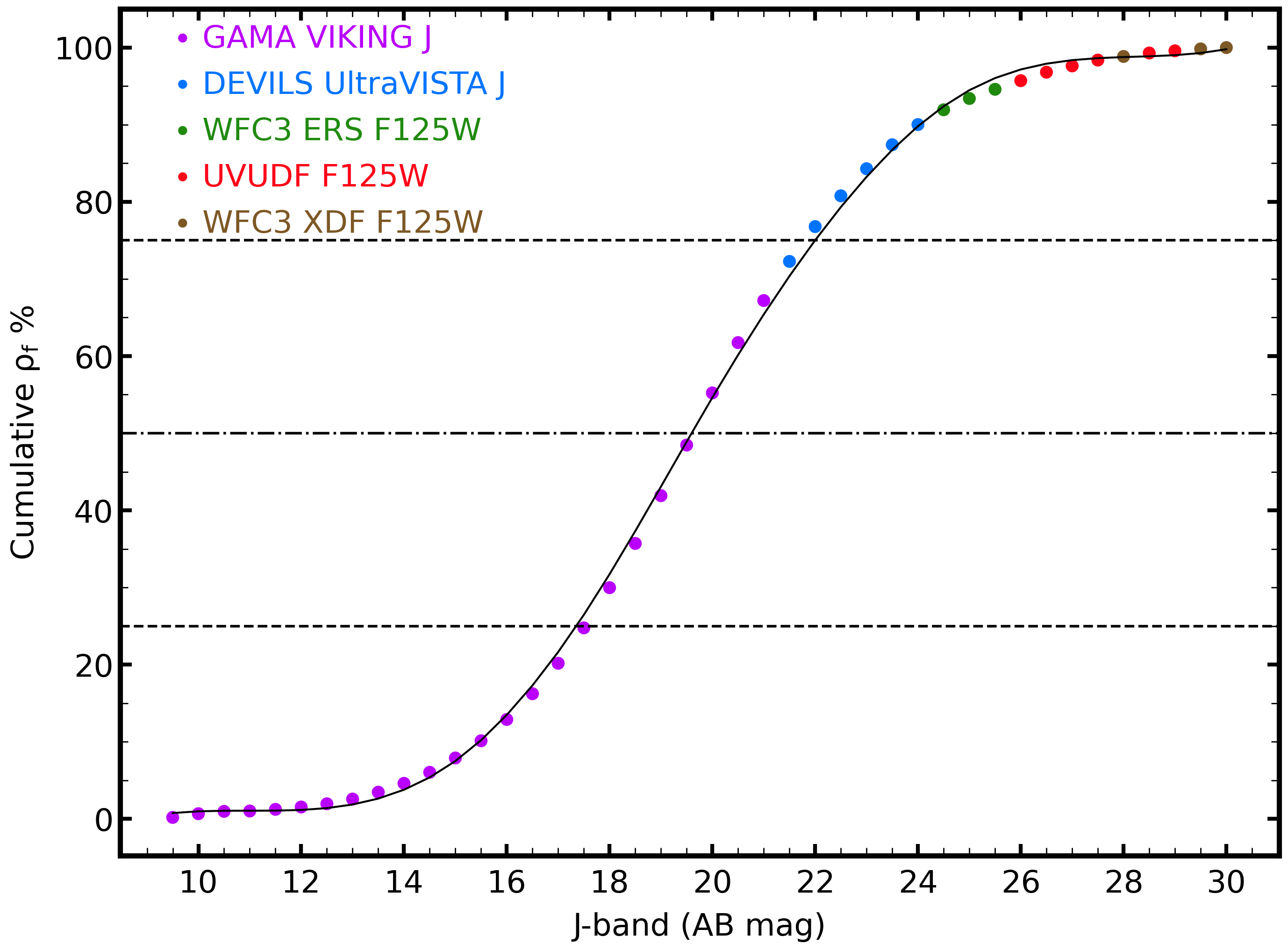}
}

\vspace*{-0.00cm}
\n \caption{
{\bf (a) [Top Left]:}\ Differential galaxy counts in the J-band or WFC3/IR F125W
filter. Combined ground-based + HST-surveys cover 10\cle AB\cle 30 mag
\citep{Driver2016b}. The galaxy counts start to reach a converging slope
($\gamma$ $<$ 0.40) over the general flux range of AB\cge 17--22 mag, and over
a similar magnitude interval for the other filters between 0.2--1.7 \mum\ 
\citep{Windhorst2011}.
{\bf (b) [Top Right]:}\ Differential J-band counts of panel (a) normalized to
the $\gamma$=0.40 slope, yielding the EBL energy contribution
$\rho_f$.$\delta$m from each 0.5 mag-wide flux interval in units of
W/Hz/m$^2$/deg$^2$. The converging sky-integral constitutes the integrated
iEBL. Cosmic Variance (CV) inside a single HST WFC3/IR FOV can be \cge 30\%.
SKYSURF will average the panchromatic galaxy counts over $\sim$1400 usable
independent HST fields, reducing Cosmic Variance-errors in the iEBL to $\sim$3\%
(\S\ \ref{sec252}). 
{\bf (c) [Bottom Left]:}\ Integral galaxy counts in the J-band filter, obtained
by integrating panel (a). The integral galaxy counts are used to calculate the
natural confusion limit that partially limits catalog completeness in
Fig.~\ref{fig:fig10}, as discussed in \S\ \ref{sec47}. 
{\bf (d) [Bottom Right]:}\ Integrated 1.25 \mum\ EBL contribution from panel
(b). The black curves in panels (c) and (d) indicate cubic polynomial fits. The
{\it brightest 25\%} of the discrete EBL level is reached at \JAB\cle 17.36
mag, while the {\it middle 50\%} of the EBL level peaks at \JAB=19.60 mag, and
the {\it faintest 25\%} of the EBL level occurs at \JAB\cge 22.01 mag. The
interquartile range or middle 50\% of the discrete J-band EBL thus comes from
galaxies as bright as 17.36 \cle \JAB\cle 22.01 mag, which have a median
redshift z\cle 1. The interquartile range is indicated by the blue boxes in
panel (a)--(b), and by the corresponding blue wedges in 
Fig.~\ref{fig:fig10}--\ref{fig:fig11}.}
\label{fig:fig2}
\end{figure*}

Fig.~\ref{fig:fig2} gives an example of this process as relevant for the 
current SKYSURF analysis. Fig.~\ref{fig:fig2}a shows the galaxy counts in the
J-band or F125W filter from the above datasets. Fig.~\ref{fig:fig2}b shows
these galaxy counts {\it normalized} to the converging magnitude-slope of
$\gamma$=0.40 \citep[][]{Driver2016b}, which yields the EBL energy
contribution $\rho_f$.$\delta$m from each 0.5 mag-wide flux interval. Earlier
examples of the integrated galaxy counts and the resulting EBL are given by,
\eg\ \citet{Madau2000}, \citet{Hopwood2010}, \citet{Xu2005},
\citet{Totani2001}, \citet{Dole2006}, \citet{Keenan2010}, \citet{Berta2011},
and \citet{Bethermin2012}, as summarized in \citet{Driver2016b} and
\citet{Koushan2021}. The galaxy contribution to the integrated light is bounded
since the faint galaxy count slope falls well below the critical value for
convergence (\ie\ $\gamma$ = $\delta$log N/$\delta$m$<$0.4). 

This {\it integral} over the discrete galaxy counts converging down to the
detection limit is referred to as the ``iEBL'', and the {\it extrapolated}
converging integral of the discrete galaxy counts beyond the detection limit
as the ``eEBL'' (Fig.~\ref{fig:fig2}bd). The {\it discrete} EBL is defined as
the sum of the iEBL and eEBL, which is indicated by the red-filled circles in
Fig.~\ref{fig:fig1}. The {\it discrete} EBL is distinct from the {\it diffuse}
EBL which is defined in \S\ \ref{sec24}. 

\citet{Driver2016b} and \citet{Koushan2021} used Monte Carlo spline fits to
extrapolate the observed discrete galaxy counts to beyond the detection limits
of the deepest available images, which provided a range in allowed
extrapolated slopes and corresponding uncertainties in the resulting eEBL.
These simulations are consistent with the range in faint-end power-law slopes
of the galaxy luminosity function over the relevant redshift range
\citep[\eg][]{Ilbert2005, Hathi2010, Windhorst2021}, and result in eEBL
integrals that, in general, converge very quickly for AB\cge 26 mag 
(Fig.~\ref{fig:fig2}b). The integrated discrete iEBL as extrapolated with the
eEBL in Fig.~\ref{fig:fig2}d is thus an estimate of the {\it discrete} EBL
that comes from galaxies. In \S\ \ref{sec47}, we will correct the discrete eEBL
for the fraction of fainter objects {\it known} to exist in deeper HST images
that are missing due to SB-incompleteness effects in the shallower SKYSURF
images. Fig.~\ref{fig:fig1} also shows the 3-component EBL model prediction of
\citet{Andrews2018} that links spheroid formation dominating at high-redshift to
later disk formation and (unobscured) AGN, as well as reprocessing of UV
photons by dust. The model predictions of \citet[\eg][]{Cowley2018} match
these iEBL+eEBL measurements. 

Fig.~\ref{fig:fig2}d shows that the {\it brightest 25\%} of the discrete iEBL
comes from galaxies brighter than \JAB$\simeq$17.36 mag, while the {\it faintest
25\%} is due to galaxies fainter than \JAB$\simeq$22.01 mag. The interquartile
range of 17.36\cle \JAB\cle 22.01 mag --- indicated by the blue boxes in 
Fig.~\ref{fig:fig2}ab, and by the corresponding blue wedges in the Figures of
\S\ \ref{sec47} --- accounts for the {\it middle 50\%} of the discrete J-band
iEBL and is due to galaxies with a median redshift z \cle 1. Thus, {\it most of
the discrete iEBL flux comes from moderately faint galaxies already detected 
in short SKYSURF exposures at AB\cle 26 mag,} where the change in count-slope
occurs at all UV--optical--near-IR wavelengths.

\ve 


\vspace*{-1.50cm}
\n



\n \subsection{Diffuse Extragalactic Background Light (dEBL)} \label{sec24} 

\sn The total EBL is defined as the sum of the integrated and extrapolated
discrete EBL of \S\ \ref{sec23} and any {\it diffuse} EBL component that may
exist: 

\begin{equation}
Total\ EBL\ =\ (iEBL\ +\ eEBL)\ +\ dEBL\ \ \equiv\ \ eIGL\ +\ dEBL
\label{eq:eq1}
\end{equation}

\n Fig.~\ref{fig:fig1} compares the {\it discrete} EBL (iEBL+eEBL) of
\citet[][which they define as ``eIGL'']{Driver2016b} from the far-UV to the
far-IR to various {\it total} EBL estimates or upper limits as reviewed by
\citet{Dwek2013} and \citet{Ashby2015}. Many of these methods estimate the
total EBL directly, which are plotted as grey triangles
\citep[\eg][]{Puget1996, Fixsen1998, Dwek1998a, Hauser1998, Lagache1999,
Dole2006, Bernstein2002, Bernstein2007, Cambresy2001, Matsumoto2005,
Matsumoto2011}. {More recent work that constrained the absolute EBL level
can be found in \citet[\eg][]{Matsuura2011} for the far-IR CIB through AKARI
measurements, in \citet{Tsumura2013b}, \citet{Matsuura2017} and
\citet{Sano2020} for NIR EBL constraints, and in \cite{Kawara2017} and
\citet{Mattila2017} for optical EBL constraints.} {Fig.~\ref{fig:fig1} also
shows the New Horizons constraints on diffuse light observed at $\sim$4.7--51
AU from the Sun \citep{Zemcov2017, Lauer2021, Lauer2022}, where the ZL
contribution is much smaller.} 

In the far-IR, the discrete EBL agrees fairly well with the directly measured
CIB \citep{Bethermin2012, Magnelli2013}, but Fig.~\ref{fig:fig1} shows a
significant optical--near-IR discrepancy between the iEBL+eEBL data (red-filled
circles) and the total EBL estimates (grey triangles). This difference amounts
to as much as a factor of $\sim$3--5, and is often attributed to a possible
component of {\it diffuse} Extragalactic Background Light (dEBL). {We note
that earlier groups plotted the {\it total} EBL signal (i.e., before the
iEBL+eEBL was subtracted) in figures like Fig.~\ref{fig:fig1}, while more
recent work {\it did} subtract the iEBL+eEBL from their data, either by
modeling and subtracting it directly \citep[\eg][]{Lauer2021, Lauer2022}, or by
using CIBER spectra including the Ca-triplet to estimate and subtracting the
Zodiacal foreground \citep[\eg][]{Matsuura2017, Korngut2022}. Hence, their
Zodi+iEBL+eEBL subtracted diffuse light values have been plotted in
Fig.~\ref{fig:fig1}. Our HST SKYSURF analysis in \S\ \ref{sec3}--\ref{sec4}
below already automatically subtracts from the diffuse light signal: a) almost
all the starlight, b) \cge 95\% of the {\it discrete} EBL integral from objects
detected in the HST images with AB\cle 26.5 mag; and c) estimates and subtracts
the undetected eEBL integral for AB\cge 26.5 mag, which is \cle 5\% of the
total {\it discrete} EBL in \citet{Carleton2022}. Hence, our SKYSURF results
will be directly comparable to these most recent results. We return to this 
point in \S\ \ref{sec5}.}

HESS/MAGIC $\gamma$-ray Blazar studies (\eg\ \citealp{Biteau2012, Dwek2013,
HESS2013, Lorentz2015, FermiLAT2018}; grey and orange wedges in
Fig.~\ref{fig:fig1}) provide independent constraints to the total EBL from
deviations of the Blazar TeV spectra from a power-law, which is explained by
pair-production involving $\gamma$-ray and EBL photons. \citet{Desai2019} and
\citet{HAWC2022} similarly find low numbers based on GeV--TeV from Fermi-LAT
and HAWC, respectively. Hence, $\gamma$-ray Blazar studies would imply a lower
level of dEBL than these direct studies that constrain the total EBL.

Direct estimates of the true level of dEBL rely on a robust subtraction of
three other sources of light: ZL, DGL, and the iEBL+eEBL \citep{Hauser2001,
Mattila2006}. SKYSURF is designed to investigate this apparent discrepancy
between the total EBL signal and the discrete iEBL+eEBL. If real, this rather
large discrepancy could be caused by a number of systematic errors that may
result in larger foregrounds. In order of increasing distance from the HST
instrument A/D converters, these are:

\n (1) Uncorrected systematics in the HST sky-SB measurements, \eg\ detector
systematics (\S\ \ref{sec41}) or Thermal Dark signal \citepalias{Carleton2022};

\n (2) Close sources of straylight (\eg\ Earthshine or scattered Sunlight; \S\
\ref{sec43}); 

\n (3) Systematic deviations from, or missing components in the ZL model 
\citepalias{Carleton2022};

\n (4) Systematic deviations from and uncertainties in the DGL model 
\citepalias[see references in][]{Carleton2022};

\n (5) Contributions by Intra-galaxy Halo Light (IHL) or (undetected) low
SB galaxies \citepalias{Carleton2022}; and

\n (6) Diffuse light from Reionization (\citet[\eg][]{Windhorst2018}). 

\sn Since we do not know the true cause of this discrepancy in
Fig.~\ref{fig:fig1}, we will hereafter refer to light sources not accounted
for by HST systematics, identifiable straylight, the ZL and DGL models or the
discrete EBL more generally as ``{\it diffuse light}'' and not as ``dEBL''. 
Further details on possible sources of diffuse light are given in
\citetalias{Carleton2022}. 

\sn In summary, most of the {\it discrete} EBL comes from moderately faint
galaxies at 17\cle AB\cle 22 mag in the redshift range 0.5\cle z\cle 2. The
true level and source of any diffuse light is as yet unclear. SKYSURF is
designed to help reconcile the total EBL measurements with the integrated and
extrapolated EBL (Fig.~\ref{fig:fig1}--\ref{fig:fig2}), and investigate how
much room may be left for a truly diffuse light component, whatever its nature. 

\n \subsection{SKYSURF's High-Level Project Outline} \label{sec25} 

\sn SKYSURF has two main science parts, and two essential supporting parts, as
illustrated in the colored columns and rows in Fig.~\ref{fig:fig3}. We
highlight both science parts briefly here, with details discussed in \S\
\ref{sec3}--\ref{sec4}.

\ve 

\n \subsubsection{SKYSURF-SB: All-Sky Constraints to ZL and DGL} \label{sec251}

\sn As indicated by the orange columns in Fig.~\ref{fig:fig3}, SKYSURF will
estimate the \emph{absolute} sky-SB at 0.2--1.7\mum\ using the methods of \S\ 
\ref{sec4}. From 249,861 ACS+WFC3 images in the Archive, we select those with
the lowest contamination due to Earthshine, Sun and Moon. The measured
SB-values sample the entire sky and can be modeled as:

\begin{equation}
SB(\lambda, l^{Ecl}, b^{Ecl}, l^{II}, b^{II}, t, SA, T)\ =\ TD(\lambda, T)\ 
+\ SL(\lambda, t) +\ ZL(\lambda, l^{Ecl}, b^{Ecl}, t, SA)\ +\ DGL(\lambda,
l^{II}, b^{II})\ +\ EBL(\lambda),
\label{eq:eq2}
\end{equation}

\n Here, ZL(t) and DGL can be fit simultaneously on scales of degrees as a
function of wavelength, Ecliptic coordinates ($l^{Ecl}$, $b^{Ecl}$), Galactic
coordinates ($l^{II}$, $b^{II}$), time of the year ($t$) or Modified Julian
Date (MJD), and Sun Angle (SA), to match SKYSURF's very large number of
panchromatic sky-SB measurements. The time- or SA-dependence is the key factor
that distinguishes the ZL from other SB components. The HST data do not sample
the temporal and spatial parameter space as deeply and uniformly as the
COBE/DIRBE data \citep[\eg][]{Kelsall1998}, but the HST sky-SB data do sample a
wider range of solar elongations and cover a full calendar year (multiple
times). The TD parameter on the right-hand side is the WFC3/IR Thermal Dark
signal that depends on wavelength and HST's ambient temperature $T$. This
near-IR thermal component needs to be modeled and subtracted from any diffuse
light signal that we observe \citepalias[\S\ \ref{sec414} and][]{Carleton2022}.
The SL parameter indicates the straylight that the HST telescope + instruments
receive from the Earth, the Sun and the Moon, which we attempt to minimize
using the methods in \S\ \ref{sec43} and \citetalias{Carleton2022} when
assessing our constraints on the ZL, DGL and EBL. The SL depends on wavelength
and time or orbital phase, which determines the angles to the Earth's Limb, Sun
and Moon (\S\ \ref{sec43}). 

In \citetalias{Carleton2022}, we will identify any large differences between
the HST sky-SB measurements and existing ZL models, which is most
straightforwardly done at wavelengths $\lambda$$\simeq$1.25--1.6 \mum\ as a
function of Ecliptic Latitude \bEcl. A major goal of SKYSURF is to update the
ZL models to cover the full 0.2--1.7 \mum\ wavelength range observed by
SKYSURF, and the range of (\lEcl, $b^{Ecl}$) and SA values sampled by HST. 

\n \subsubsection{SKYSURF-Sources: Panchromatic Counts and iEBL/eEBL Averaged
over Cosmic Variance} \label{sec252}

\begin{figure*}[!hptb]
\n \cl{
 \includegraphics[width=0.850\txw]{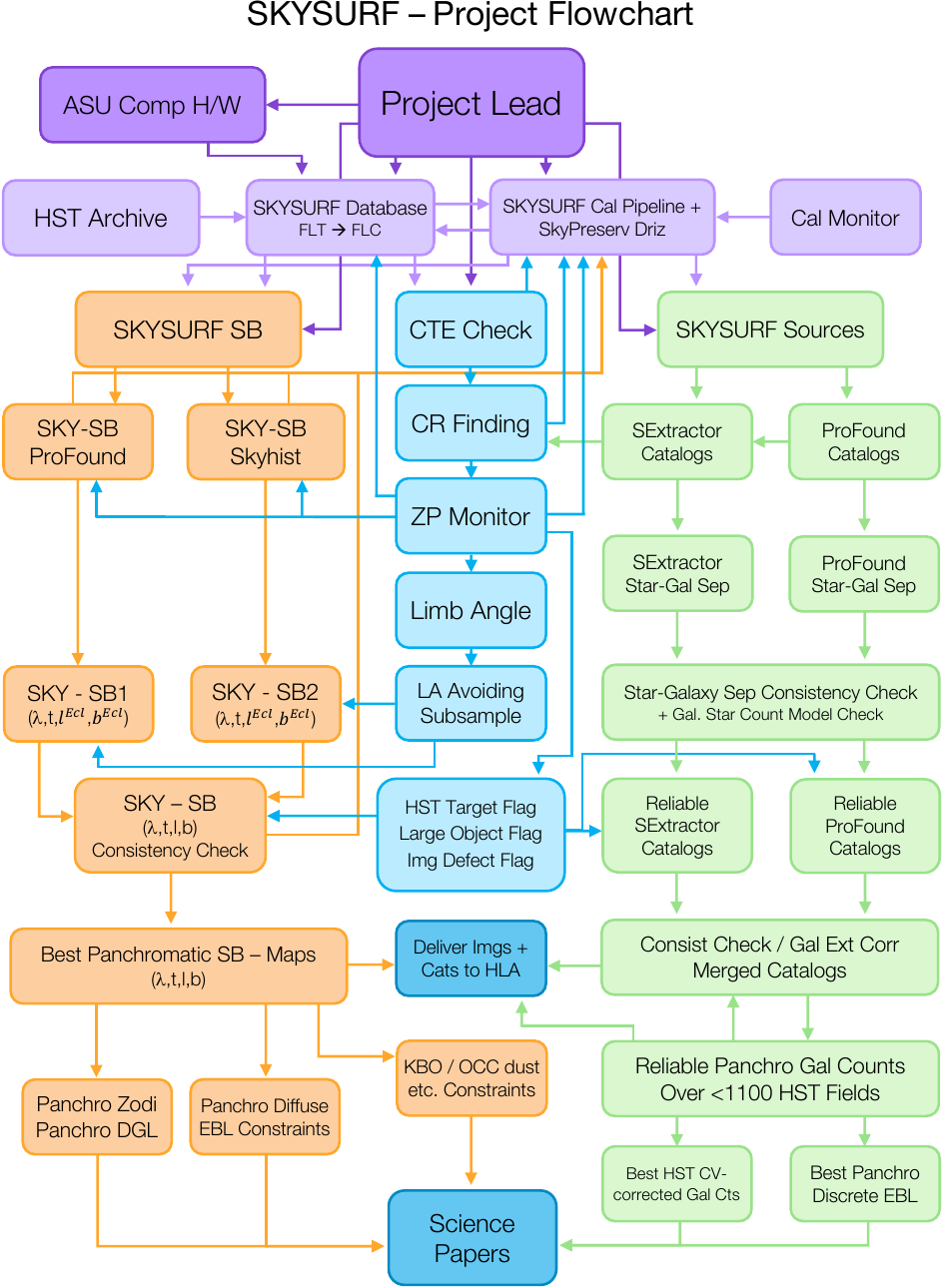}
}

\vspace*{+0.20cm}
\n \caption{
Flowchart for project SKYSURF. The orange columns indicate the two independent
methods to do SKYSURF's sky-SB studies, while the green columns indicate the
two independent methods to perform SKYSURF's panchromatic discrete object
counts over $\sim$1400 independent HST fields. The light purple row summarizes
SKYSURF's database building and standard processing steps. The light blue
column summarizes SKYSURF's non-standard processing steps. ``Skyhist'' indicates
our best percentile-clip method to estimate sky-SB values. Details are given in
\S\ \ref{sec25} \& \ref{sec42}.} 
\label{fig:fig3}
\end{figure*}

\begin{figure*}[!hptb]
\n \cl{
 \includegraphics[width=1.000\txw,angle=+0]{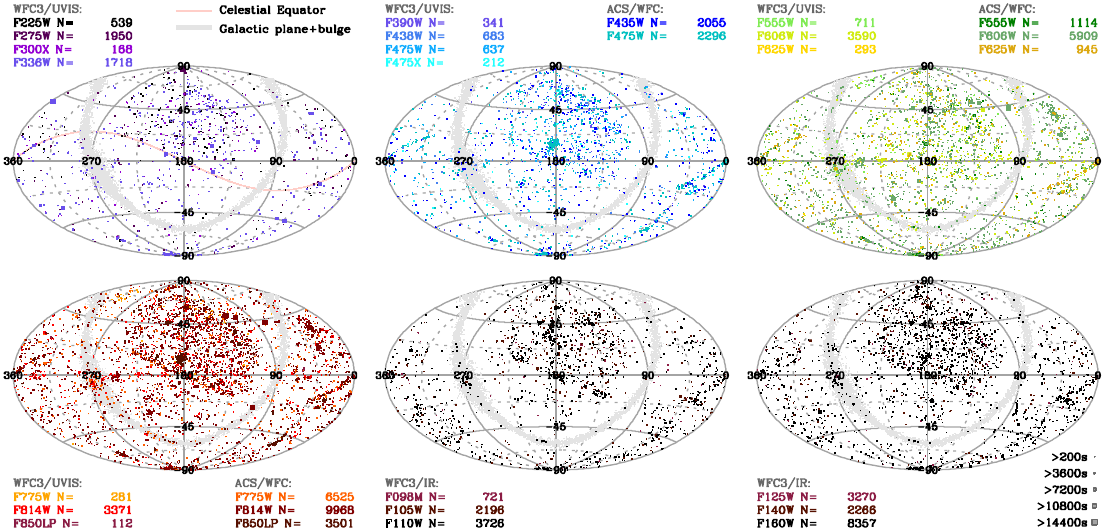}
}

\vspace*{+0.30cm}
\n \caption{
Aitoff equal-area projections in Ecliptic coordinates of {\it all} ACS/WFC and
WFC3/UVIS+IR images with \texp$\ge$200 sec that are publicly available as of
Sept. 23, 2019. The Galactic plane and bulge are represented by the light grey
band, and the Celestial Equator is indicated by the pink curve [{\it upper
left}]. SKYSURF measures the absolute all-sky surface brightness S($\lambda$,
t, $l^{Ecl}$, $b^{Ecl}$) in 12 main broad-band filters at $\sim$0.2--1.7 \mum\
in wavelength from 249,861 HST Archival images in $\sim$1400 independent HST 
fields (Eq.~\ref{eq:eq2} in \S\ \ref{sec2}).}
\label{fig:fig4}
\end{figure*}

\n The discrete panchromatic object counting part of SKYSURF is indicated by 
the green columns in Fig.~\ref{fig:fig3}, which provides discrete object
catalogs, star-galaxy separation, and object counts in the main HST broad-band
filters across the sky. Because the normalized differential galaxy counts
flatten with a converging slope for AB\cge 22 mag (Fig.~\ref{fig:fig2}b), most
of the EBL-energy (and its uncertainty) comes from galaxies with
AB$\simeq$17--22 mag at a median redshift z\cle 1. Their Cosmic Variance (CV)
over a {\it single} HST FOV is $\sim$30--40\% at these redshifts
\citep[\eg][]{Somerville2004, Trenti2008, Moster2011, Driver2016b}. SKYSURF's
goal is therefore to average the panchromatic galaxy counts over 1400 {\it
well-separated} representative HST fields to reduce the iEBL-errors from cosmic
variance to \cle 2\%, or \cle 4\% when accounting for the systematic and 
zeropoint errors in \S\ \ref{sec415}. Even a {\it contiguous} HST survey region
of 1400 fields (\eg\ COSMOS) would still give $\sim$8\% CV, and such fields are
not available in the 12 main broad-band HST filters. Hence, SKYSURF's {\it
all-sky} distribution of the $\sim$1400 independent HST fields in
Fig.~\ref{fig:fig4} is {\it essential} to reduce CV in the resulting galaxy
counts \citep{Driver2010}. Further details are given in \S\ \ref{sec32},
\ref{sec45} and Tompkins \etal\ (2022, in preparation). 

In what follows, we will define {\it independent HST fields} as those
instrument FOVs that are far enough apart in the sky (\cge 1\degree) to
provide faint galaxy counts that are sufficiently independent to average over
CV. Fig.~\ref{fig:fig4} shows that there are 4,858 independent HST FOVs using
this definition, not all of which are usable for objects counts (\S\
\ref{sec32} \& Appendix \ref{secAppB3}). As discussed in \S\ \ref{sec3}, the
total number of instrument FOVs that SKYSURF has covered is $\sim$3.5$\times$
larger, as many HST users have covered their targets on average by a number of 
adjacent instrument FOVs. 

To the typical 5$\sigma$ completeness limits of AB\cle 26--28 mag of most
drizzled SKYSURF images we expect an integrated galaxy surface density of
(3--5)$\times$10$^5$ deg$^{-2}$ (\eg\ Fig.~\ref{fig:fig2}ac). For the estimated
total usable SKYSURF area of \cge 10 \degsq\ (\S\ \ref{sec32}), this implies 
a total of $\sim$(3--5)$\times$10$^6$ galaxies. Hence, SKYSURF will provide
galaxy counts for a sample as large as the SDSS \citep{York2000}, but
distributed over the whole sky and reaching $\sim$5 mag deeper. Unlike SDSS,
the SKYSURF discrete object sample does {\it not} cover a contiguous area. But
because it sparsely samples the whole sky, SKYSURF's discrete object catalogs
will be well suited to minimize Cosmic Variance in the galaxy counts. A key
element of this SKYSURF goal is HST's photometric stability over 11--18 years
of data (\S\ \ref{sec415}).



\vspace*{-0.00cm}
\hspace*{-1.00cm}
\n




\hspace*{+0.00cm}
\n


\ve 

\vspace*{-0.50cm}
\n \subsubsection{SKYSURF Plan} \label{sec253}

\n Because the SKYSURF database contains 249,861 HST Archival images, it does, 
in general, {\it not} lack sufficient statistics, not even after conservative 
removal of large foreground targets and image defects (Appendix 
\ref{secAppB3}). Instead, SKYSURF is limited by systematic errors, and for this
reason, we need to carefully account for possible systematics summarized in \S\
\ref{sec4}. Hence, SKYSURF carries out each of its two main science
goals --- accurate panchromatic sky-SB measurements and object counts --- 
{\it along two main independent paths each}, indicated by the two orange and
two green columns in Fig.~\ref{fig:fig3} with significant cross-checks. The
purple rows in Fig.~\ref{fig:fig3} show SKYSURF's database building and its
data processing (\S\ \ref{sec3}--\ref{sec4}), while the blue column shows its
specific data flagging steps (\S\ \ref{sec4} and Appendix \ref{secAppB3}).


\n \section{Project SKYSURF --- Database Overview} \label{sec3} 

\sn In this section, we summarize the selection of the SKYSURF instruments,
filters, and exposures (\S\ \ref{sec31}), and the resulting SKYSURF database
and total usable survey area (\S\ \ref{sec32}). The database considered for
SKYSURF ranges from each instrument's launch date until January 2020, when we
started building its database. Summaries of each HST instrument used in
SKYSURF and their data reduction pipelines can be found in the Instrument
Handbooks (IHBs), Data Handbooks (DHBs), and Instrument Science Reports (ISRs)
listed on the STScI
website~\footnote{\url{https://www.stsci.edu/hst/instrumentation}}. Specific
HST instrument details relevant for SKYSURF are discussed in \S\ \ref{sec41}.

\n \subsection{SKYSURF Instruments, Filters, and Exposures} \label{sec31} 

\sn {\bf HST Instruments Used:}\ \ SKYSURF's Archival data come from HST's
primary wide field imaging instruments: ACS/WFC, WFPC2, WFC3/UVIS and WFC3/IR.
These data span more than 17 years for ACS (2002--2020), $\sim$16 years for
WFPC2 (1994-2009), and $\sim$11 years for WFC3 (2009--2020). Despite its much
older detectors, broad-band WFPC2 images were included in SKYSURF because they
provide an earlier time baseline (1994--2009). ACS/WFC and WFC3/UVIS have
higher throughput than WFPC2, but due to its much larger pixels WFPC2's
sensitivity to SB is comparable to that of ACS/WFC and WFC3/UVIS. For many
targets WFPC2 provides broad-band exposures in the common ``BVI'' filters
(F555W, F606W, and F814W) that were later replicated in the same filters with
ACS/WFC or WFC3/UVIS. Hence, we will compare the older WFPC2 sky-SB estimates
for the same targets observed at the same time of the year (\ie\ at similar Sun
Angles) to those observed later in the {\it same} filters with ACS/WFC or
WFC3/UVIS. This provides SKYSURF with an independent assessment of subtle
instrument systematics in the sky-SB measurements and zeropoint drifts over the
decades. Details are given in O'Brien \etal\ (2022, in preparation).

\sn {\bf WFPC2, ACS/WFC, and WFC3/UVIS+IR Images not used in SKYSURF:}\ \ We
did {\it not} retrieve from the HST Archive all of the following ACS/WFC, WFPC2
and WFC3/UVIS and WFC3/IR image types for SKYSURF: (1) grism, narrow-band and
most medium-band images; (2) images taken with Quad or Linear Ramp filters; (3)
images that use subarrays or time-series; (4) images of (fast) moving targets;
(5) WFC3/UVIS or WFC3/IR images that were produced by spatial scans; and (6)
ACS polarization images where a Polarizer is crossed with a broad-band filter.
This is because these images are harder to calibrate and because their sky-SB 
would be much harder to measure, since it may not be uniform across these
specialized images by their very design.

\sn {\bf Other HST Cameras not used in SKYSURF:}\ \ The following other HST
cameras have been onboard the HST spacecraft part of the last 32 years, but are
not used for SKYSURF: WF/PC-1, FOC, ACS/HRC \& SBC, NICMOS NIC1, NIC2, NIC3,
STIS/CCD, and STIS/MAMA. For WF/PC-1, this is because the instrument was in
HST's spherically aberrated beam, which affects both point source detection and
accurate SB-measurements \citep[\eg][]{Windhorst1992, Windhorst1994a}. The
ACS/HRC \& SBC, FOC, NICMOS NIC1, NIC2, NIC3, STIS/CCD, and STIS/MAMA images
are not used, because they cover very small FOVs, and/or have rather unusual or
very broad-band filter sets that would be hard to compare to measurements in the
standard modern filter sets present in ACS/WFC, WFPC2, or WFC3. NICMOS also has
significant time-dependent dark-current levels (the ``pedestal'' effect) that
would make dark-current subtraction and absolute sky-SB measurements rather
uncertain, despite the advantage of significantly lower foregrounds in the
near-IR over the other early HST instruments. 

\sn {\bf SKYSURF Pointings:}\ \ The HST pointings used for SKYSURF are not
completely randomly distributed across the sky (Fig.~\ref{fig:fig4}). They are
sparser in the South than in the North, with a tendency to avoid the Galactic
plane ($\vert$$b_{II}$$\vert$\cle 20\degree) and favoring the celestial
equator (Decl.$\simeq$0\degree). These biases can be due to, \eg\ HST observers
selecting targets from ground-based surveys in equatorial regions that can be
accessed by ground-based telescopes in both hemispheres, and from the SDSS
whose footprint is clearly visible through its higher density of HST targets in
Fig.~\ref{fig:fig4} (see \S\ \ref{sec32}). 

\sn {\bf SKYSURF Filters:}\ \ We use the 12 main broad-band filters between
0.2--1.7 \mum\ (UV [F225W]--{\it H} [F160W]). Fig.~\ref{fig:fig4} shows that
SKYSURF has 28 broad-band ACS and WFC3 filters in total. Many of these filters
are very similar in wavelength and may be grouped together \citep[after small
zeropoint corrections and differential K-corrections; see ][]{Windhorst2011}
when combining them into the panchromatic galaxy counts. All 28 filters will be
used for panchromatic sky-SB determination at their appropriate effective
wavelengths ($\lambda_{eff}$), but the galaxy counts may be combined in very
similar filters. Filter red leaks and blue leaks are discussed in \S\
\ref{sec41}.

\sn {\it Wide Field Planetary Camera 2 data since 1994:}\ \ The main SKYSURF
filters used for WFPC2 are the F300W, F336W, F439W, F450W, F555W, F606W, F675W,
F702W, F814W, and F850LP filters, plus several other less-used broad-band
filters summarized in Table~\ref{tab:tab1}. 

\sn {\it Advanced Camera for Surveys/Wide Field Channel data since 2002:}\ \ The
main SKYSURF filters used for the ACS/WFC are the F435W, F475W, F555W, F606W,
F775W, F814W, and F850LP filters, also broadly referred to as the ACS ``$BViz$''
filters, plus several other less-used broad-band filters summarized in
Table~\ref{tab:tab2}a. 

\sn {\it Wide Field Camera 3 UVIS data since 2009:}\ \ The main SKYSURF filters
used for WFC3/UVIS are the vacuum UV filters F225W, F275W, F336W, and the
F438W, F555W, F606W, and F814W, filters, also broadly referred to as the WFC3
``$BVi$'' filters, plus several other less-used broad-band WFC3/UVIS filters, 
including F775W and F850LP, summarized in Table~\ref{tab:tab2}b. Where possible,
these WFC3/UVIS filters are used as external checks on the ACS/WFC sky-SB values
measured in the same filters on the same targets observed at similar times of
the year. 

\sn {\it Wide Field Camera 3 IR data since 2009:}\ \ The main SKYSURF filters
used for WFC3/IR are the F098W, F105W, F110W, F125W, F140W, and F160W filters,
as summarized in Table~\ref{tab:tab3}, plus several other less-used medium-band 
or narrow-band filters used for the WFC3/IR Thermal Dark signal calibration in 
\citetalias{Carleton2022}. 

\begin{figure*}[!hptb]
\n \cl{
 \includegraphics[width=0.500\txw]{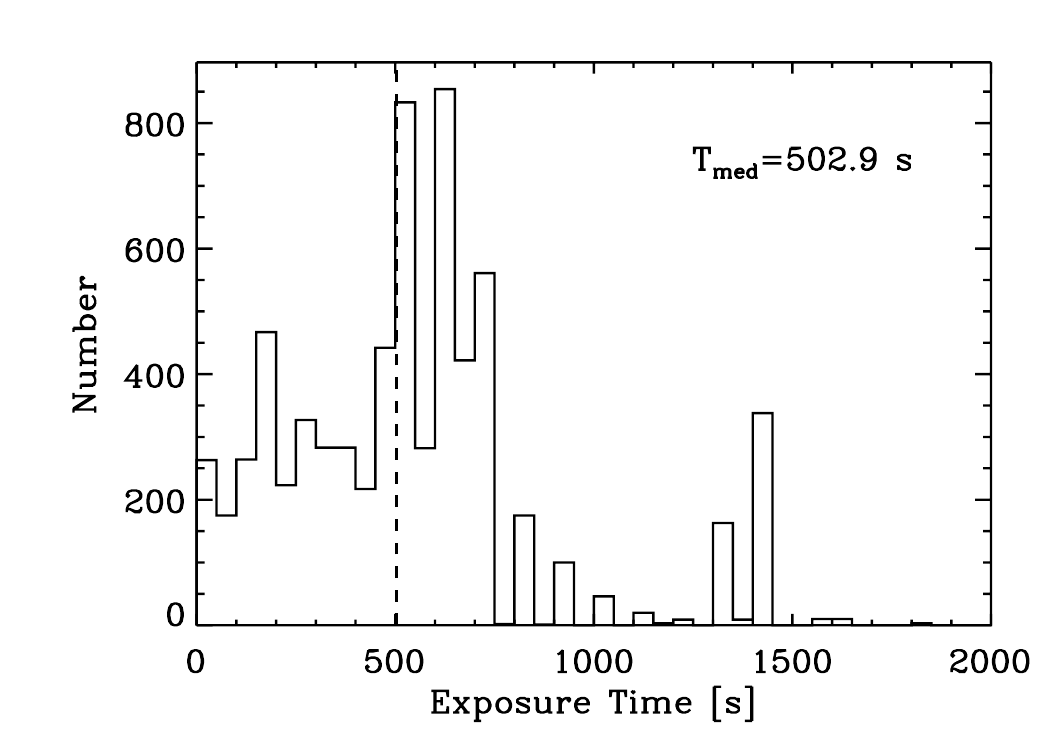}
}

\vspace*{-0.20cm}
\n \caption{
Exposure time distribution for all WFC3/IR F125W images available to SKYSURF.
The median exposure time of the individual F125W images is \texp$\simeq$500
sec. The exposure time distributions for the other WFC3/IR filters are similar.
According to the WFC3 IHB \citep{Dressel2021} and Fig.~\ref{fig:fig10} here, at
this median 500 sec depth, the individual WFC3 F125W images reach AB\cle 26.0
mag for point source detection, so that most faint compact objects with AB\cge
26.5 mag would have been missed in these short exposures. This detection limit
somewhat affects the sky-SB estimates from these images after repeated
mode-fitting or n$\sigma$ clipping to remove the detected objects, which masks
the detected objects with AB\cle 26.0 mag, but does not remove the flux from
non-detected objects at AB\cge 26.5 mag.}
\label{fig:fig5}
\end{figure*}

\mn {\bf SKYSURF Exposures and Exposure times:}\ \ We initially considered
{\it all} ACS/WFC, WFPC3, WFC3/UVIS and WFC3/IR exposures for SKYSURF
processing, regardless of whether or not the \texttt{LOW-SKY} or
\texttt{SHADOW} Special Requirements were specified by the HST observer in the
Astronomers Proposal Tool (``APT''~\footnote{\url{
https://www.stsci.edu/scientific-community/software/astronomers-proposal-tool-apt}}). 
For sky-SB measurements we sub select exposures of sufficient duration to allow
robust estimates of the background level. For drizzling and discrete object
catalog generation, we sub select images with exposure times \texp$\ge$ 200
sec, which constitute the vast majority of images and ensures sufficient depth
for our purposes. These are generally the exposures where the sky-noise
exceeds the read-noise (except in the UV due to significant Charge Transfer
Inefficiency trails; see \S\ \ref{sec41} \& Appendix \ref{secAppB2}), and so
they are potentially the most useful for galaxy counts over as large of a
random area as possible. As an example, the distribution of exposure times for
all 6796 WFC3/IR F125W images available to SKYSURF is shown in
Fig.~\ref{fig:fig5}. The median exposure time is \texp$\simeq$500 sec, where a
typical F125W image reaches AB \cle 26 mag (5$\sigma$) for compact objects. In
general, comparable median exposure times are found for SKYSURF's other
broad-band filters listed in Tables~\ref{tab:tab1}--\ref{tab:tab3}. These
Tables also list the corresponding median image depths derived from the
instrument Exposure Time Calculators. 


\vspace*{-0.00cm}
\begin{deluxetable*}{| c | cc | cccc | ccc | ccc |}
\tablecaption{SKYSURF Data Summary --- WFPC2}\label{tab:tab1}
\setlength{\tabcolsep}{4pt}
\tablewidth{0pt}
\tablehead{
\colhead{{\bf Instr}/}                                              $\vert$ & 
\colhead{N$_{Exp}$}                                                         & 
\colhead{Disk}                                                      $\vert$ & 
\colhead{N$_{Exp}$}                                                         & 
\colhead{$t_{\rm exp}^{50\%}$}                                              & 
\colhead{$t_{\rm exp}^{25\%}$}                                              & 
\colhead{$t_{\rm exp}^{75\%}$}                                      $\vert$ & 
\colhead{$t_{\rm exp}^{50\%}$}                                              & 
\colhead{$t_{\rm exp}^{25\%}$}                                              & 
\colhead{$t_{\rm exp}^{75\%}$}                                      $\vert$ & 
\colhead{$50\%$-limit}                                                      & 
\colhead{$25\%$-limit}                                                      & 
\colhead{$75\%$-limit}                                              $\vert$ \\[-4pt]
\colhead{Filter}                                                    $\vert$ &
\colhead{ }                                                                 &
\colhead{Space}                                                     $\vert$ &
\multicolumn{4}{c}{--- only $t_{\rm exp}\ge 200$ sec ---}           $\vert$ &
\multicolumn{3}{c}{--- for all $t_{\rm exp}$ ---}                   $\vert$ &
\multicolumn{3}{c}{--- for images with all $t_{\rm exp}^{\ a}$ ---} $\vert$ \\[-4pt]
\colhead{ }                                                         $\vert$ &
\colhead{ }                                                                 &
\colhead{(GB)}                                                      $\vert$ &
\colhead{ }                                                                 &
\colhead{(s) }                                                              &
\colhead{(s) }                                                              &
\colhead{(s) }                                                      $\vert$ &
\colhead{(s) }                                                              &
\colhead{(s) }                                                              &
\colhead{(s) }                                                      $\vert$ &
\colhead{(AB-mag)}                                                          &
\colhead{(AB-mag)}                                                          &
\colhead{(AB-mag)}                                                  $\vert$         
}
\startdata
{\bf WFPC2}     &        &       &       &     &     &      &     &     &      &       &       &       \\
F255W           &    796 &  97.0 &   796 & 500 & 300 &  800 & 500 & 300 &  800 & 22.54 & 22.00 & 23.03 \\
F300W           &  11019 &  97.0 & 10744 & 800 & 400 & 1000 & 800 & 400 & 1000 & 25.17 & 24.44 & 25.40 \\
F336W           &   2514 &  22.0 &  2134 & 600 & 400 &  900 & 600 & 400 & 1000 & 24.71 & 24.28 & 25.24 \\
F380W           &     89 &   0.8 &    89 & 600 & 500 & 1000 & 600 & 500 & 1000 & 25.16 & 24.98 & 25.67 \\
F439W           &   1298 &  12.0 &  1298 & 500 & 313 &  700 & 500 & 313 &  700 & 24.65 & 24.16 & 24.99 \\
F450W           &   5991 &  84.0 &  5988 & 600 & 400 & 1000 & 600 & 400 & 1000 & 25.90 & 25.51 & 26.36 \\
F547M           &    611 &   5.3 &   611 & 400 & 300 &  600 & 400 & 300 &  600 & 25.25 & 24.97 & 25.63 \\
F555W           &   6829 &  59.0 &  6457 & 500 & 350 & 1100 & 600 & 350 & 1200 & 26.34 & 25.88 & 26.88 \\
F569W           &     44 &  0.37 &    44 & 800 & 500 & 1100 & 800 & 500 & 1100 & 26.37 & 25.97 & 26.62 \\
F606W           &  24265 & 205.0 & 24168 & 600 & 500 & 1000 & 600 & 500 & 1000 & 26.63 & 26.49 & 27.00 \\
F622W           &    186 &   1.6 &   186 & 900 & 600 & 1000 & 900 & 600 & 1000 & 26.57 & 26.25 & 26.65 \\
F675W           &   1926 &  17.0 &  1822 & 500 & 400 &  700 & 500 & 400 &  700 & 25.90 & 25.71 & 26.17 \\
F702W           &   2006 &  17.0 &  2000 & 700 & 400 & 1200 & 700 & 400 & 1200 & 26.47 & 26.03 & 26.86 \\
F785LP          &    274 &   2.4 &   274 & 500 & 500 &  500 & 500 & 500 &  500 & 25.01 & 25.01 & 25.01 \\
F791W           &    478 &   4.1 &   471 & 350 & 260 &  600 & 375 & 260 &  600 & 25.33 & 25.00 & 25.73 \\
F814W           &  18759 & 160.0 & 18659 & 600 & 400 & 1100 & 600 & 400 & 1100 & 25.86 & 25.52 & 26.33 \\
F850LP          &   1002 &   8.8 &  1002 & 400 & 400 &  600 & 400 & 400 &  600 & 24.17 & 24.17 & 24.57 \\
\hline
Subtot          &  78087 &   793 & 76743 &     &     &      &     &     &      &       &       &       \\ 
\hline
\enddata
\mn
\tablenotetext{a}{Detection limit is the AB-magnitude for 5$\sigma$ point
sources at the median (50\%) exposure time $t_{\rm exp}^{50\%}$ listed. The
25\% and 75\% columns indicate the exposure times and corresponding 5$\sigma$
point source detection limits for the shallowest 25\% and 75\% of the images,
respectively.}
\end{deluxetable*}



\vspace*{-0.00cm}
\begin{deluxetable*}{| c | cc | cccc | ccc | ccc |}
\tablecaption{SKYSURF Data Summary --- ACS/WFC and WFC3/UVIS}\label{tab:tab2}
\setlength{\tabcolsep}{4pt}
\tablewidth{0pt}
\tablehead{
\colhead{{\bf Instr}/}                                              $\vert$ & 
\colhead{N$_{Exp}$}                                                         & 
\colhead{Disk}                                                      $\vert$ & 
\colhead{N$_{Exp}$}                                                         & 
\colhead{$t_{\rm exp}^{50\%}$}                                              & 
\colhead{$t_{\rm exp}^{25\%}$}                                              & 
\colhead{$t_{\rm exp}^{75\%}$}                                      $\vert$ & 
\colhead{$t_{\rm exp}^{50\%}$}                                              & 
\colhead{$t_{\rm exp}^{25\%}$}                                              & 
\colhead{$t_{\rm exp}^{75\%}$}                                      $\vert$ & 
\colhead{$50\%$-limit}                                                      & 
\colhead{$25\%$-limit}                                                      & 
\colhead{$75\%$-limit}                                              $\vert$ \\[-4pt]
\colhead{Filter}                                                    $\vert$ &
\colhead{ }                                                                 &
\colhead{Space}                                                     $\vert$ &
\multicolumn{4}{c}{--- only $t_{\rm exp}\ge 200$ sec ---}           $\vert$ &
\multicolumn{3}{c}{--- for all $t_{\rm exp}$ ---}                   $\vert$ &
\multicolumn{3}{c}{--- for images with all $t_{\rm exp}^{\ a}$ ---} $\vert$ \\[-4pt]
\colhead{ }                                                         $\vert$ &
\colhead{ }                                                                 &
\colhead{(GB)}                                                      $\vert$ &
\colhead{ }                                                                 &
\colhead{(s) }                                                              &
\colhead{(s) }                                                              &
\colhead{(s) }                                                      $\vert$ &
\colhead{(s) }                                                              &
\colhead{(s) }                                                              &
\colhead{(s) }                                                      $\vert$ &
\colhead{(AB-mag)}                                                          &
\colhead{(AB-mag)}                                                          &
\colhead{(AB-mag)}                                                  $\vert$         
}
\startdata
{\bf ACS/WFC}   &        &       &       &     &     &      &     &     &      &       &       &       \\
F435W           &   5898 &  1250 &  5461 & 661 & 500 & 1200 & 650 & 440 & 1200 & 26.09 & 25.71 & 26.66 \\
F475W           &   6280 &  1380 &  5417 & 522 & 370 &  700 & 470 & 365 &  674 & 26.12 & 25.89 & 26.46 \\
F555W           &   2555 &   560 &  2317 & 540 & 385 &  700 & 520 & 370 &  697 & 25.88 & 25.55 & 26.15 \\
F606W           &  16930 &  3730 & 15990 & 530 & 400 &  784 & 515 & 382 &  767 & 26.50 & 26.25 & 26.82 \\
F625W           &   1839 &   380 &  1479 & 532 & 382 &  600 & 467 & 340 &  577 & 25.89 & 25.60 & 26.08 \\
F775W           &   8953 &  2000 &  8675 & 510 & 404 &  716 & 503 & 400 &  608 & 25.70 & 25.48 & 25.87 \\
F814W           &  30278 &  6710 & 27536 & 525 & 450 &  800 & 509 & 400 &  752 & 25.90 & 25.68 & 26.22 \\
F850LP          &   8884 &  2000 &  8586 & 507 & 400 &  675 & 500 & 400 &  669 & 24.65 & 24.43 & 24.92 \\
\hline
Subtot          &  81617 & 18010 & 75461 &     &     &      &     &     &      &       &       &       \\ 
\hline
{\bf WFC3/UVIS} &        &       &       &     &     &      &     &     &      &       &       &       \\
F225W           &   1600 &   280 &  1126 & 560 & 400 &  700 & 516 & 368 &  699 & 25.23 & 24.89 & 25.54 \\
F275W           &   5622 &   920 &  3975 & 660 & 484 & 1212 & 528 & 190 &  800 & 25.24 & 24.20 & 25.65 \\
F300X           &    366 &    61 &   141 & 609 & 351 &  869 & 450 & 100 &  600 & 25.87 & 24.37 & 26.14 \\
F336W           &   4616 &   970 &  3999 & 645 & 470 &  820 & 600 & 408 &  800 & 25.91 & 25.52 & 26.19 \\
F390W           &   1038 &   230 &   912 & 597 & 558 &  850 & 596 & 482 &  790 & 26.43 & 26.22 & 26.65 \\
F438W           &   1851 &   260 &  1009 & 430 & 350 &  783 & 350 & 205 &  511 & 25.41 & 25.26 & 25.73 \\
F475W           &   1977 &   240 &   905 & 800 & 400 & 1308 & 325 & 150 &  720 & 26.05 & 25.37 & 26.67 \\
F475X           &    525 &    80 &   309 & 524 & 360 &  798 & 300 & 175 &  580 & 26.45 & 25.87 & 26.76 \\
F555W           &   2271 &   350 &  1334 & 477 & 378 &  600 & 356 & 140 &  531 & 26.04 & 25.24 & 26.36 \\
F606W           &   7794 &  1350 &  5484 & 599 & 400 &  843 & 425 & 300 &  700 & 26.37 & 26.11 & 26.71 \\
F625W           &    804 &   100 &   425 & 515 & 400 &  700 & 370 & 180 &  621 & 25.85 & 25.27 & 26.26 \\
F775W           &    688 &   170 &   279 & 606 & 400 &  699 & 320 & 200 &  507 & 25.23 & 24.90 & 25.56 \\
F814W           &  10602 &  1880 &  6467 & 595 & 400 &  867 & 400 & 242 &  653 & 25.56 & 25.15 & 25.91 \\
F850LP          &    330 &    50 &   192 & 374 & 364 &  473 & 349 & 200 &  379 & 24.42 & 23.87 & 24.49 \\
\hline
Subtot          &  40084 &  6941 & 26557 &     &     &      &     &     &      &       &       &       \\ 
\hline
\enddata
\mn
\tablenotetext{a}{Detection limit is the AB-magnitude for 5$\sigma$ point
sources at the median (50\%) exposure time $t_{\rm exp}^{50\%}$ listed. The
25\% and 75\% columns indicate the exposure times and corresponding 5$\sigma$
point source detection limits for the shallowest 25\% and 75\% of the images,
respectively.}
\end{deluxetable*}



\vspace*{-0.00cm}
\begin{deluxetable*}{| c | cc | cccc | ccc | ccc |}
\tablecaption{SKYSURF Data Summary --- WFC3/IR}\label{tab:tab3}
\setlength{\tabcolsep}{4pt}
\tablewidth{0pt}+
\tablehead{
\colhead{{\bf Instr}/}                                              $\vert$ & 
\colhead{N$_{Exp}$}                                                         & 
\colhead{Disk}                                                      $\vert$ & 
\colhead{N$_{Exp}$}                                                         & 
\colhead{$t_{\rm exp}^{50\%}$}                                              & 
\colhead{$t_{\rm exp}^{25\%}$}                                              & 
\colhead{$t_{\rm exp}^{75\%}$}                                      $\vert$ & 
\colhead{$t_{\rm exp}^{50\%}$}                                              & 
\colhead{$t_{\rm exp}^{25\%}$}                                              & 
\colhead{$t_{\rm exp}^{75\%}$}                                      $\vert$ & 
\colhead{$50\%$-limit}                                                      & 
\colhead{$25\%$-limit}                                                      & 
\colhead{$75\%$-limit}                                              $\vert$ \\[-4pt]
\colhead{Filter}                                                    $\vert$ &
\colhead{ }                                                                 &
\colhead{Space}                                                     $\vert$ &
\multicolumn{4}{c}{--- only $t_{\rm exp}\ge 200$ sec ---}           $\vert$ &
\multicolumn{3}{c}{--- for all $t_{\rm exp}$ ---}                   $\vert$ &
\multicolumn{3}{c}{--- for images with all $t_{\rm exp}^{\ a}$ ---} $\vert$ \\[-4pt]
\colhead{ }                                                         $\vert$ &
\colhead{ }                                                                 &
\colhead{(GB)}                                                      $\vert$ &
\colhead{ }                                                                 &
\colhead{(s) }                                                              &
\colhead{(s) }                                                              &
\colhead{(s) }                                                      $\vert$ &
\colhead{(s) }                                                              &
\colhead{(s) }                                                              &
\colhead{(s) }                                                      $\vert$ &
\colhead{(AB-mag)}                                                          &
\colhead{(AB-mag)}                                                          &
\colhead{(AB-mag)}                                                  $\vert$         
}
\startdata
{\bf WFC3/IR}   &        &       &       &     &     &      &     &     &      &       &       &       \\
F098M           &   1158 &     7 &  1103 & 703 & 603 & 1003 & 703 & 553 & 1003 & 25.98 & 25.80 & 26.23 \\
F105W           &   5412 &    33 &  4792 & 603 & 299 &  903 & 403 & 228 &  803 & 25.98 & 25.53 & 26.46 \\
F110W           &   8847 &    54 &  6473 & 353 & 253 &  603 & 288 & 203 &  503 & 26.08 & 25.82 & 26.47 \\
F125W           &   6810 &    39 &  5554 & 553 & 453 &  703 & 503 & 299 &  653 & 26.06 & 25.68 & 26.24 \\
F140W           &   5647 &    35 &  4691 & 349 & 228 &  603 & 299 & 203 &  553 & 25.80 & 25.49 & 26.24 \\
F160W           &  22199 &   140 & 19283 & 503 & 399 &  653 & 453 & 303 &  603 & 25.69 & 25.38 & 25.89 \\
\hline
Subtot          &  50073 &   308 & 41896 &     &     &      &     &     &      &       &       &       \\ 
\hline
Total           & 249861 & 26052 &220657 &     &     &      &     &     &      &       &       &       \\ 
\enddata
\mn
\tablenotetext{a}{Detection limit is the AB-magnitude for 5$\sigma$ point
sources at the median (50\%) exposure time $t_{\rm exp}^{50\%}$ listed. The
25\% and 75\% columns indicate the exposure times and corresponding 5$\sigma$
point source detection limits for the shallowest 25\% and 75\% of the images,
respectively. The last row gives the grand total over
Tables~\ref{tab:tab1}--\ref{tab:tab3}.}
\end{deluxetable*}




\vspace*{-1.00cm}
\hspace*{-0.00cm}
\n



\n \subsection{The Panchromatic SKYSURF HST Database and Total Usable Area}
\label{sec32} 

\sn {\bf Number of Exposures and Retrieval:}\ \ We retrieved from the HST
Archive all 249,861 available images (81,617 ACS/WFC + 78,087 WFPC2 + 40,084
WFC3/UVIS + 50,073 WFC3/IR exposures), or \cge 26 TB in total
(Fig.~\ref{fig:fig4} and Tables~\ref{tab:tab1}--\ref{tab:tab4}). These images
are all public as of 2020, and have exposure times up to one full orbit. 
Since processing and retrieval of such a vast amount of data posed some
demands on the HST Archive, we spread ingestion over the Spring of 2020 with a
typical transfer rate of $\sim$175 GB per day. Complete disk copies of the
SKYSURF database are kept at ASU in Arizona and at ICRAR at the University of
Western Australia. 

\sn {\bf All-Sky Maps of Available Panchromatic SKYSURF Images:}\ \ All-sky
maps of all images eligible for SKYSURF analysis are shown in 
Fig.~\ref{fig:fig4}a--\ref{fig:fig4}f. The SDSS footprint appears as the
better-sampled tilted rectangle in Fig.~\ref{fig:fig4}, since the SDSS has
provided many targets for HST survey and SNAPshot programs, and many of those
images are suitable for SKYSURF. In our all-sky sky-SB analysis, SKYSURF will
appropriately weigh the uneven sampling of panchromatic sky-SB values due to
this higher HST field-density inside the SDSS footprint (Eq.~\ref{eq:eq2}), as
well as the resulting all-sky discrete object counts over $\sim$1400
independent HST fields (\S\ \ref{sec45} \& Appendix \ref{secAppC}), as needed. 

\sn {\bf Estimated Total Usable SKYSURF Area:}\ \ Table~\ref{tab:tab4}
summarizes the total number of exposures per SKYSURF instrument, to estimate
the maximum usable area that HST has covered with these data since 1994. Each
instrument uses between 1--3 detectors per camera, and Col. 5 lists the total
number of SKYSURF {\it single exposures} of the full cameras (except for WFPC2
where the PC1 data were discarded). Col. 3 lists the FOV (in arcsec) for each
of the {\it full} camera exposures, and Col. 4 the total area per full exposure
in each camera. Col. 6 lists for each camera the approximate average number of
exposures per filter and the approximate average number of filters used on each
HST pointing, as well as their product. Since 1994, the average HST user of
WFPC2, ACS, or WFC3 has used an average of $\sim$8 exposures per filter and
$\sim$1.8 filters per pointing. The total number of filters per FOV ranges from
1 for single-exposure SNAPshot targets to 13 for the HUDF. Col. 7 lists the
estimated number of independent HST pointings or FOVs in each full camera,
which are considered those that are more than 1.0 FOV (or \cge 6\arcm) apart in
their pointing centers, given the single detector FOV values in Col. 3. In \S\
\ref{sec45} we discuss how the drizzle footprints were defined that determined
these associations. Col. 8 lists the maximum SKYSURF area covered by each
camera, which is not yet corrected for repeat visits of a given pointing with a
different camera in the same filter. This will be done when the footprints and
drizzling of all SKYSURF data are finished on AWS (\S\
\ref{sec45}--\ref{sec46}). Hence, only an upper limit to the total {\it
unique} SKYSURF area is listed that may be usable for independent object counts
across the sky. 

Of the 249,861 individual exposures in the SKYSURF database, 220,657 images 
have \texp$\ge$200 sec and are spread out over 16,822 HST pointings or FOVs
across the sky (Fig.~\ref{fig:fig4}). The 249,861 SKYSURF exposures from
Tables~\ref{tab:tab1}--\ref{tab:tab3} contain \cge 878,000 individual detector
read-outs, including the 50,073 WFC3/IR exposures which we split into their
individual ramp-readouts to better monitor sky-SB vs. orbital phase (\S\
\ref{sec43}). All 249,861 SKYSURF exposures are processed through the initial
SB-measurement steps of \S\ \ref{sec42}, as it cannot be determined a priori
whether they are useful for SKYSURF's sky-SB goals or not. We estimate that
about one-third of all these images have LOW-SKY or SHADOW flags or equivalent
low background levels, such that they can constrain the ZL, DGL, or any diffuse
light. 

The subset of 220,657 images with \texp$\ge$200 sec is used for drizzling,
object catalogs and counts, and has covered \cle 32 \degsq\ across the sky
since 1994 (Table~\ref{tab:tab4}). Of this total area, not all images are
usable for SKYSURF background object counts, \eg\ due to large targets that
overfill the FOV, Galactic plane targets, or large artifacts (Appendix
\ref{secAppB3}). We estimate that about 30--50\% of these 16,822 HST FOVs or
$\sim$10 \degsq\ are in principle usable for object counts. In total, 4,858 of
the 16,822 HST FOVs are \cge 1\degree\ away from the nearest-neighbor HST
field. Here, we assume that angular distances \cge 1\degree\ at z$\simeq$1--2
--- which corresponds to \cge 30 Mpc in Planck Cosmology \citep{Ade2016} ---
make the galaxy counts in such fields sufficiently independent to average over
CV \citep{Driver2010}. Of these 4,858 independent FOVs, we also expect 
$\sim$30--50\% to survive the large target or large defect filtering above, so
we expect that $\sim$1400 of these HST targets can be meaningfully used to
reduce CV in the galaxy counts. Henceforth, we refer to these as our
``$\sim$1400 independent HST fields'' suitable for galaxy counts.



\begin{deluxetable*}{| l | ccc | ccc | c |}
\tablecolumns{10}
\tablewidth{1.0\linewidth}
\tablecaption{Total Number of Exposures per SKYSURF Instrument to be Drizzled,
and Maximum Area Covered
\label{tab:tab4}}
\tablehead{
\colhead{$\vert$\ SKYSURF \ \ \ \ \ \ }                 $\vert$ & 
\colhead{N$_{chip}$$^b$}                                        & 
\colhead{FOV}                                                   &
\colhead{Area/Exp}                                      $\vert$ &
\colhead{N$_{Exp}$}                                             &
\colhead{N$_{Exp}$/Filt $\times$}                               &
\colhead{N$_{FOV}$$^d$}                                 $\vert$ &
\colhead{Max. Total\ \ \ \ \ \ }                        $\vert$ \\[-4pt]
\colhead{$\vert$\ \ Instrument\ \ \ \ \ \ }             $\vert$ & 
\colhead{}                                                      &
\colhead{/chip}                                                 &
\colhead{(arcm$^2$)}                                    $\vert$ &
\colhead{\ }                                                    &
\colhead{N$_{Filt}$/Point$^c$}                                  &
\colhead{\ }                                            $\vert$ &
\colhead{Area\ (deg$^2$)$^e$\ \ }                       $\vert$ \\[-4pt]
\colhead{$\vert$\ \ \ \ \ \ \ \ (1)}\ \ \ \ \ \ \ \ \ \ $\vert$ &
\colhead{(2)}                                                   &
\colhead{(3)}                                                   &
\colhead{(4)}                                           $\vert$ &
\colhead{(5)}                                                   &
\colhead{(6)}                                                   &
\colhead{(7)}                                           $\vert$ &
\colhead{(8)}                                           $\vert$         
}
%
\startdata
WFPC2$^a$     &3  & 75\arcs$\times$75\arcs  & 4.69      & 76,743   &$\sim$6.01$\times$1.77$\simeq$10.61   & 7,230 &    9.42 \\
ACS/WFC       &2  &202\arcs$\times$101\arcs &11.33      & 75,461   &$\sim$9.39$\times$1.70$\simeq$16.00   & 4,717 &   14.85 \\
WFC3/UVIS     &2  &162\arcs$\times$81\arcs  & 7.29      & 26,557   &$\sim$6.24$\times$1.92$\simeq$11.97   & 2,219 &    4.49 \\
WFC3/IR       &1  &136\arcs$\times$123\arcs & 4.65      & 41,896   &$\sim$8.86$\times$1.78$\simeq$15.77   & 2,656 &    3.43 \\
\hline
Total SKYSURF$^f$ &    &                  &$<$7.27$>$ &220,657 &$<$7.74$\times$1.77$>$$\simeq$$<$13.65$>$ &16,822 &$\sim$32 \\
\hline
\enddata
\mn
\tablenotetext{a}{The WFPC2 FOV consists of three 800$^2$ pixel CCDs (WF2--WF4),
each with 0\arcspt 10 pixels and a usable area of 750$^2$ pixels. Each WF CCD
thus covers 1.56 arcmin$^2$ for a total WF area of 4.69 arcmin$^2$. In
addition, WFPC2 has one 800$^2$ pixel CCD (PC1) with 0\arcspt 046 pixels that
covers 0.33 arcmin$^2$. PC1 is not used in SKYSURF because of its higher noise
per arcsec$^2$ and its larger calibration systematics.}

\tablenotetext{b}{Col. 2 lists the number of detectors typically used per
camera, and therefore the number of detector readouts that form one exposure in
a given FITS file ({\footnotesize \eg\ 
{\url{https://hst-docs.stsci.edu/acsdhb/chapter-2-acs-data-structure/2-2-acs-file-structure}}})
and its extensions. Col. 3 lists the FOV (in arcsec) for each of the {\it
single} detectors, and Col. 4 the total area covered by the {\it full} camera
in each single exposure (in arcmin$^2$).}

\tablenotetext{c}{Col. 5 lists the total number of SKYSURF FITS files that
contain the single exposure data from each {\it full} camera. Col. 6 lists for
each camera the approximate average number of exposures per filter and the
approximate average number of filters used on each HST pointing, as well as
their product.} 

\tablenotetext{d}{Col. 7 lists the estimated number of independent HST
pointings or FOVs in each camera, which are considered those that are more than
1.0 FOV diameter (or \cge 6\arcm) apart in their pointing centers. This is
simply Col. 5 divided by the result of Col. 6. \S\ \ref{sec45} gives details
of how the independent FOVs and subsequent drizzle footprints were defined.} 

\tablenotetext{e}{Col. 8 lists the {\it maximum} SKYSURF area covered by each
camera, which is not yet corrected for repeat visits of a given pointing with
a different camera in the same filter. This will be done when the footprints
and drizzling of all SKYSURF data are finished on AWS (\S\ \ref{sec32} \&
\ref{sec45}--\ref{sec46}).} 

\tablenotetext{f}{The bottom row lists the full camera area weighted over all
exposure files, and the average number of exposures per filter times the
average number of filters used per pointing ($<$N$_{Exp}$/Filt$>$ $\times$
$<$N$_{Filt}$/Point$>$) --- each weighted with the total number of exposures
in each camera (Col. 5) --- as well as the {\it maximum total} area that 
SKYSURF may cover.} 
\end{deluxetable*}



\vspace*{-0.00cm}
\n \section{High-level SKYSURF Methods} \label{sec4} 

\sn In this section, we discuss our methods to produce both sky-SB measurements
and object catalogs from SKYSURF's images, with the details needed to assess
their accuracy, reliability and completeness across the sky. This includes the 
calibration methods applied, the image zeropoints (ZP) and ZP monitors as a
function of time, our algorithms to make object-free estimates of the sky-SB,
the orbital sky-SB dependence and sources of straylight, and our treatment of
sky-SB gradients. Because SKYSURF's object catalogs affect our estimates of the
object-free sky-SB, we also summarize SKYSURF image drizzling strategy and
drizzle footprints, as well as our star-galaxy separation method and catalog
reliability and completeness.


\n \subsection{Calibration with Best Available Calibration Files, and Other
General Calibration Aspects} \label{sec41} 

\sn In this section, we summarize the standard calibration of all SKYSURF
images with the best available calibration files and other calibration
considerations for SKYSURF's specific purposes. This includes any sources that
{\it systematically add or remove} electron (\emin) signal from the image sky-SB
levels, as well as the zeropoints and ZP monitoring over time of each HST
instrument from which data is used here. This first sub-section discusses the
effects that all instruments have in common, while the following sub-sections
discuss specific aspects of each individual HST instrument as they may affect
SKYSURF's sky-SB measurements. 

The relative sky-SB errors induced by each of the main aspects of the
calibration process below are summarized in Table~\ref{tab:tab5} as a
percentage of the average sky-SB levels measured, with references to the
sections below where details are given. {All errors are 1-$\sigma$
compared to the mean trends in the calibration parameters discussed in or
estimated from the ISRs or IHBs cited below.} In some cases, a range is given
for the relative errors which may depend on wavelength or the presence of image
gradients. The bottom row of Table~\ref{tab:tab5} lists the total relative
error in each of the instrument sky-SB estimates, which assumes that the
individual error components are independent. When an error range is listed, the
largest of the percentage errors are propagated into the total error. Hence, we
consider the total relative sky-SB errors to be conservative estimates. 

\mn {\bf Standard calibration:}\ \ SKYSURF calibrates each image using the
latest on-orbit reference files and flux scale, which includes the standard
bias-subtracted, dark-frame subtracted, flat-fielded images (the $\_flt$
files), which have also been CTE-corrected (the $\_flc$ files). The total of
249,861 images from Tables~\ref{tab:tab1}--\ref{tab:tab3} were retrieved from
the Mikulski Archive for Space Telescopes
(MAST~\footnote{\url{https://archive.stsci.edu}}) in Jan-May 2020 using the
pipelines in effect as of that time period. For ACS these are the
\texttt{calacs} pipeline version 10.2.1, and for WFC3 the \texttt{calwf3}
pipeline version 3.5.0. The WFC3/UVIS images were downloaded again in early
2022 calibrated with \texttt{calwf3} pipeline version 3.6.2 to implement the
2021 CTE corrections (Appendix \ref{secAppB2}) and to automatically correct for
the slowly time-varying filter zeropoints as a function of wavelength (\S\
\ref{sec415}). O'Brien \etal\ (2022, in preparation) summarize the differences
in the ACS/WFC and WFC3/UVIS detector design and the resulting subtle
differences in their calibration pipelines as relevant for SKYSURF. 

All these calibrated images have {\it not} been sky-subtracted, and their
calibration quality and flatness (in the absence of large bright objects, see
\S\ \ref{sec42}) is critical for SKYSURF. The errors due to bias+dark-frame
subtraction and flat-fielding are retrieved from the Instrument Handbooks and
the Instrument Science Reports (ISRs). All these standard calibration errors
are expressed as relative errors of the low--average sky-SB levels measured,
and are summarized in the error budgets of Table~\ref{tab:tab5} 
\citepalias[\S\ \ref{sec416}, \ref{sec48} and][]{Carleton2022}. We note the
following pipeline calibration details that are relevant for all of SKYSURF's
instruments below: 

\sn {\bf Geometrical Distortion Corrections (GDC):}\ \ The calibrated SKYSURF
images can be directly used to measure extended emission or sky-SB values
before the images are drizzled. The flat-fielding process corrects each
pixel's SB for high-frequency (pixel-to-pixel) variations, and to first order
for low-frequency, large-scale structures due to camera-, chip- or illumination
properties across the FOV. The flat-field process is thus designed to produce
$\_flt$ files that would have constant values in all pixels if the original
source had a perfectly uniform SB. However, due to the significant Instrument
Distortion Corrections in each of WFPC2, ACS, and WFC3 cameras, a Pixel Area
Map (PAM) would need to be applied if one were to use the undrizzled images for
{\it point-source photometry}, since the flat-fielding process is not
explicitly designed to make point-source photometry uniform across the images.
This is because instrument distortion causes some pixels cover more area on the
sky than others, so point-source photometry is location-dependent on the
detectors. Once the overall sky-SB is measured on each SKYSURF image, the
drizzling process (\S\ \ref{sec45}--\ref{sec46}) explicitly performs the full
GDCs, so that photometry on compact and extended sources will now {\it both} be
accurate on the drizzled images. Hence, drizzling replaces the need for
applying a PAM for point-source photometry. 

\sn {\bf Drizzling Pixel Scale:}\ \ Drizzled images (\S\
\ref{sec45}--\ref{sec46}) have the proper GDC applied, and therefore give the
correct photometry for {\it both extended and point sources using the same
images}. The WFC3 IHB \citep{Dressel2021} states specifically that ``In 
drizzled images ($\_drz$ files), photometry is correct for both point and
extended
sources.''~\footnote{\url{https://hst-docs.stsci.edu/wfc3dhb/chapter-7-wfc3-ir-sources-of-error/7-8-ir-flat-fields}}\ 
In \S\ \ref{sec45}--\ref{sec46} we will drizzle {\it all} SKYSURF images to the
same pixel scale of 0\arcspt 060/pixel, including {\it all single} exposures,
so they may be used for discrete object finding and photometry. This will lead
to some PSF undersampling of the cameras with the finest pixel scales (ACS/WFC
with $\sim$0\arcspt 05/pixel and WFC3/UVIS with $\sim$0\arcspt 039/pixel), but
that is acceptable for SKYSURF's first goal of all-sky panchromatic sky-SB
measurements. It will also lead to some minor loss in point-source sensitivity
for the ACS/WFC and WFC3/UVIS images, but SKYSURF has such a large dynamic
range in flux and area that this will not be a limitation to its second goal of
accurate all-sky panchromatic object counts from $\sim$1400 independent HST
fields (\S\ \ref{sec252}). This choice of drizzled pixel size also significantly
reduces the storage requirements of SKYSURF's final output images, and the AWS
processing costs, as compared to smaller pixels. 

\sn {\bf Corrections for CCD Preflash or Postflash Levels:}\ \ Charge Transfer
Efficiency (CTE) degradation occurs in CCDs due to the heavy CR bombardment
over time and is especially noticeable at low sky-SB levels, hence, in all
WFC3/UVIS vacuum-UV filters, and also in all WFPC2, ACS/WFC, and WFC3/UVIS
broad-band filters well after each instruments' Shuttle launch. When CTE
effects are severe, then CTE-corrections as applied in the pipeline
\citep[\eg][]{Anderson2010} may not be sufficient. Most observers will have
anticipated this by adding a ``preflash'' level to their WFPC2 exposures, or a
``postflash'' to their ACS/WFC or WFC3/UVIS CCD exposures, to bring the sky-SB
up to a level where the CTE traps are largely filled. Therefore, SKYSURF needs
to verify if the WFPC2 preflash and ACS/WFC and WFC3/UVIS postflash levels in
the broad-band filters were properly subtracted in the pipelines before
reliable sky-SB measurements can be made. All preflash or postflash levels are
prescribed by the observer, and the best-estimate preflash or postflash frames
are subtracted in the instrument pipelines. 

The WFC3/UVIS postflash frames have low-level gradients of \cle 20\%, with
overall amplitudes that depend somewhat non-linearly on the duration of the
postflash level selected by the user \citep{Biretta2013}. These authors state
that ``examination of the long-term stability of the postflash LEDs shows no
evidence of systematic fading over 9 months''. \citet{Biretta2013} find
quasi-random LED brightness fluctuations with rms amplitude of $\sim$0.6--1.2\%
(\eg\ their Fig. 14--16). Since CTE degradation has steadily increased over
the years, the recommended postflash levels to fill in the traps have increased
from 0 \emin/pix in 2009 to \cge 20 \emin/pix in 2020 and beyond. 

Taking the F606W filter as an example, Fig.~\ref{fig:fig1} shows that a
typical Zodiacal sky-SB is $\sim$562 \nWsqmsr\ \ or V$\sim$22.86 AB-\magarc.
With the WFC3/UVIS F606W zeropoint of 26.08 AB-mag (for 1.00 \emin/sec) and
0\arcspt 0397 pixel, this corresponds to a Zodiacal sky-SB of $\sim$0.031
\emin/pix/sec. In an average $\sim$500 sec F606W exposure, the F606W sky level
then amounts to $\sim$15.3 \emin/pix. Hence, when an average LED postflash of
10 \emin/pix gets added and subsequently subtracted in the pipeline, the above
$\sim$1.2\% postflash subtraction error corresponds to a $\sim$0.5\% error (\ie\
$\sim$0.12/(10+15)) in the inferred sky-SB, with some variance around this
number depending on the actual postflash level used. In the bluer WFC3/UVIS
filters, the relative error due to the postflash subtraction will be larger
than in F606W, but for ACS it will be somewhat smaller because of its larger
0\arcspt 05 pixels and its $\sim$0.4 mag higher throughput in the optical
compared to WFC3/UVIS. We adopt $\sim$1\% of the average Zodiacal sky-SB as the
CCD postflash subtraction error in Table~\ref{tab:tab5}. A discussion of CTE
effects on low-SB fluxes in the WFC3/UVIS UV filters --- after the required
postflash application and removal --- is given by, \eg\ \citet{Smith2018,
Smith2020}. Further details are given in Appendix \ref{secAppB2} and O'Brien
\etal\ (2022, in preparation). 

\sn {\bf Corrections for Detector Persistence:}\ \ Bright point-like or very
high SB targets (AB\cle 15 mag) in previous images may saturate and create a
positive residual charge that decays exponentially with several time-scales
ranging from minutes to fractions of an hour, and so can persist in subsequent
images with the same instrument in the same {\it or} in a different filter
\citep[\eg][]{Deustua2010, Long2010, Long2012}. A careful analysis of
flat-field errors and persistence in the HUDF data by \citet{Borlaff2019}
removes these effects to SB-levels of $\sim$32.5 AB-\magarc\ in the WFC3/IR
broad-band near-IR filters. We tested for the effects of persistence in the
SKYSURF's WFC3/IR images with an average exposure time of \texp$\simeq$ 500
sec, and concluded that the best sky-SB measuring algorithms of \S\
\ref{sec423} are robust against the rare persistence images left in subsequent
images. For discrete object catalogs (Appendix \ref{secAppC1}), we need to
remove all persistence images as flagged in the \texttt{calwf3} pipeline from
the next few exposures. 

\sn {\bf Corrections for Detector Crosstalk:}\ \ As summarized in, \eg\ the
WFC3 IHB \citep[\eg][]{Deustua2010}, crosstalk is a type of electronic
ghosting that is common in CCD or IR detectors when two or more amplifier
sections are read out by the A/D converters {\it simultaneously}. A bright
source in one amplifier section causes a dim electronic ghosting in other
amplifier section(s) at the corresponding pixels that are read-out at the same
time, in essence, because a spacecraft has no absolute electrical grounding. The
offending signal dumps electrons into the imperfect local ground upon 
read-out, thus reducing the sensed signal by the paired amplifier, hence the
negative sign of the crosstalk signal. This results in a bright point source
(including hot pixels and CRs) or a very high-SB extended target --- as 
read-out by any detector's A/D converter --- generating an area of lower data
numbers in corresponding, mirrored locations of an adjacent detector amplifier
section. Crosstalk happens in both the ACS/WFC, WFC3/UVIS and WFC3/IR
detectors, but not in WFPC2 since its four CCDs are read sequentially. The
crosstalk amplitude is linear with the signal that gives rise to it in the
adjacent amplifier section that is digitized during the same read-out. During a
full-frame, unbinned, four-amplifier readout, the crosstalk between WFC3/UVIS
amplifier section A or C is $\sim$--2$\times$$10^{-4}$ of the source signal,
while for a target in WFC3/UVIS amplifier section B or D, it is
$\sim$--7$\times$$10^{-5}$ of the source signal \citep{Vaiana2010,
Suchkov2012}. For WFC3/IR, crosstalk occurs between amplifiers 1 and 2, or
between amplifiers 3 and 4, and amounts to $\sim$--1$\times$$10^{-6}$ of the
source signal (note the {\it negative} sign of the crosstalk signal in all
cases). For unsaturated sources, crosstalk thus is generally below the
sky-noise, but possibly still noticeable as a dim depression in the sky-SB if
the cause is a large source with high-SB in the adjacent amplifier section.
When it occurs, crosstalk is generally identifiable and correctable to within
0.1\% of the surrounding sky-SB. The most noticeable cases of crosstalk will be
identified during our image flagging procedures in \S\ \ref{sec42} and 
Appendix~\ref{secAppB3}. Further discussion of low-level systematics in the
sky-SB estimates is given in O'Brien \etal\ (2022, in preparation).



\begin{deluxetable*}{| l | lllllll |}
\tablecolumns{6}
\tablewidth{1.0\linewidth}
\tablecaption{Error Estimates$^a$ in Calibration, Zeropoints, Sky-SB 
Measurements, and Thermal Dark Signals
\label{tab:tab5}}
\tablehead{
\colhead{$\vert$\ \ \ \ \ \ \ \ \ Source of Error \ \ \ \ \ \ }                        $\vert$ & 
\colhead{WFPC2     }                                                                           & 
\colhead{ACS/WFC   }                                                                           &
\colhead{WFC3/UVIS }                                                                           &
\multicolumn{3}{c}{--- WFC3/IR ---}                                                            &
\colhead{(\S\S)    }                                                                   $\vert$ \\[-4pt]
\colhead{$\vert$\ \ \ \ \ \ \ \ \ \ \ \ \ \ \ \ }\ \ \ \ \ \ \ \ \ \ \ \ \ \ \ \ \ \ \ $\vert$ &
\colhead{}                                                                                     &
\colhead{}                                                                                     &
\colhead{}                                                                                     &
\colhead{F125W}                                                                                & 
\colhead{F140W}                                                                                & 
\colhead{F160W}                                                                                & 
\colhead{}                                                                             $\vert$ \\[-4pt]
\colhead{$\vert$\ \ \ \ \ \ \ \ \ \ \ \ \ \ (1)}\ \ \ \ \ \ \ \ \ \ \ \ \ \ \ \ \      $\vert$ &
\colhead{(2)}                                                                                  &
\colhead{(3)}                                                                                  &
\colhead{(4)}                                                                                  &
\colhead{(5)}                                                                                  & 
\colhead{(6)}                                                                                  & 
\colhead{(7)}                                                                                  & 
\colhead{(8)}                                                                          $\vert$ 
}
\startdata
Bias/Darkframe subtraction &$\sim$1.0\%       &$\sim$1.5\%     &$\sim$1.5\%     &$\sim$1.0\%     &$\sim$1.0\%     &$\sim$1.0\%     &\ref{sec41}               \\
Dark glow subtraction      &$\sim$2\%         &---             &---             &---             &---             &---             &\ref{sec411}              \\
Postflash subtraction      &---               &$\sim$1\%       &$\sim$1\%       &---             &---             &---             &\ref{sec41}               \\
Global flat-field quality$^b$&$\sim$1--3\%    &0.6--2.2\%      &$\sim$2--3\%    &$\sim$0.5--2\%  &$\sim$0.5--2\%  &$\sim$0.5--2\%  &\ref{sec41}               \\
Numerical accuracy of LES$^c$&\cle 0.2--0.4\% &\cle 0.2--0.4\% &\cle 0.2--0.4\% &\cle 0.2--0.4\% &\cle 0.2--0.4\% &\cle 0.2--0.4\% &\ref{sec423}              \\
Photometric zeropoints$^d$ &$\sim$2\%         &0.5--1\%        &0.5--1\%        &$\sim$1.5\%     &$\sim$1.5\%     &$\sim$1.5\%     &\ref{sec415}              \\
Thermal Dark signal$^e$    &---               &---             &---             &$\sim$0.2\%     &$\sim$0.5\%     &$\sim$2.7\%     &\ref{sec414}, \citetalias{Carleton2022} \\
\hline
Total Error$^f$            &$\sim$4.3\%       &$\sim$3.0\%     &$\sim$3.7\%     &$\sim$2.7\%     &$\sim$2.8\%     &$\sim$3.8\%     &                          \\
Sky-SB low-avg (nW/m$^2$/sr)&---              &---             &                &262--534        &251--513        &240--496        &                          \\
Sky-SB error (nW/m$^2$/sr) &---               &---             &                &\ \ \ 7--14     &\ \ \ 7--14     &\ 15--19        &                          \\
\hline
\enddata
\sn
\tablenotetext{a}{{\it The relative errors} in this table are expressed as a 
percentage of the {\it average} sky-SB values in the HST images, which includes
the typical Zodiacal sky-SB and TD levels. The bottom two rows list both the
lowest and average sky-SB values plus total errors for each in units of
\nWsqmsr\ for the WFC3/IR F125W, F140W, and F160W filters, respectively 
\citepalias[see also Fig.~\ref{fig:fig1} and][]{Carleton2022}.}

\tablenotetext{b}{For WFPC2, the large-scale flat-field errors in the filters
F439W and redwards are \cle 1\%, but the upper bound includes the 1\% error in
the contamination correction and the $\sim$3\% error in the residual CTE
correction. For the less frequently used WFPC2 UV filters, these errors can be
larger.}

\tablenotetext{c}{Numerical accuracy of Lowest Estimated Sky values away from
detected objects (\S\ \ref{sec42}). The LES algorithms also avoid areas of
significant persistence when estimating the sky-SB, which is not included as
an extra term in the error budget.}

\tablenotetext{d}{For WFC3/IR, this includes the $\sim$0.5\% uncertainty in the
applied detector count-rate non-linearity correction (\S\ \ref{sec414}).}

\tablenotetext{e}{The errors in the estimated Thermal Dark signal values for
the F125W, F140W, and F160W filters reflect a plausible range in HST component
temperatures of $\Delta$T$\simeq$$\pm$2 K \citepalias[see Table 2
in][]{Carleton2022}. The TD errors increase from 1.25--1.6 \mum\ due to the
increased black-body contributions as modeled across the WFC3/IR filter set.
For the darkest sky-SB values the additive percentage TD-error is about
2$\times$ larger as quoted here, resulting in the total errors for the darkest
sky-SB values increasing to $\sim$2.8\%, $\sim$2.9\%, $\sim$6.3\% in the F125W,
F140W, and F160W filters, respectively.}

\tablenotetext{f}{The total relative error is the quadratic sum of the
individual relative errors in the calibration, zeropoints, numerical sky-SB
estimates, and dark signals, assuming these are all independent. These errors
are incorporated into our error budget in the Tables in
\citetalias{Carleton2022}. The bottom row shows the errors in the lowest to
average sky-SB values in the previous row, respectively. 
\deleted{\tablenotetext{g}{Sky-SB values can be converted to units of \nWsqmsr\
by multiplying the MJy/sr units by 10$^{-11}$(c/$\lambda_{c}$), where
$\lambda_{c}$ is the filter central wavelength.}}
}
\end{deluxetable*}


\n \subsubsection{WFPC2} \label{sec411} 

\sn Here we summarize the specific considerations for the WFPC2 data used in 
SKYSURF with their error contributions summarized in Table~\ref{tab:tab5}.

\sn {\bf WFPC2 CTE Degradation and Preflash:}\ \ The WFPC2 CTE has gotten
noticeably worse after 8--16 years on-orbit, and so WFPC2 sky-SB measurements
need to be done on pre-flashed images, which subsequently have this pre-flash
level removed. 

\sn {\bf WFPC2-Window Dark Glow:}\ \ The WFPC2 CCD ``Window Glow'' or ``Dark
Glow'' is the largest source of instrumental error for WFPC2, due to low-level
light from the field flattener lenses in front of the CCDs. The window glow is
likely due to irradiation of the MgF$_2$ in the field flattener by energetic
particles (CRs), which may result in both Cerenkov radiation and fluorescence.
There is therefore a correlation between the Dark Glow and the input cosmic ray
(CR) flux with some scatter. Fig. 4.6 of \citet{Biretta2009} shows a {\it
shallow} relation between CR-flux (=input) and Dark Glow (=output) for WFPC2
CCD WF2. The total CR-flux from the CR-only maps produced by SKYSURF can be
used to predict the WFPC2 window glow. The glow is the same for CCDs WF3 and
WF4, substantially higher for CCD PC1, and the lowest for CCD WF2, so we
estimate the sky-SB primarily from the CCD detector WF2, and compare it to those
estimated from WF3 and WF4 as a check.

According to the analytical WFPC2 Dark Current model in the WFPC2 IHB
\citep{Gonzaga2010}~\footnote{\url{
https://www.stsci.edu/hst/instrumentation/legacy/wfpc2}}, at the WFPC2 detector
temperature of T=--88\degree\ C, only about 0.5--1$\times 10^{-4}$ DN/sec of
the measured dark-count rate is due to the usual Dark Current, while about
1--8$\times 10^{-4}$ DN/sec comes from the glowing WFPC2 field flattener. There
is also a very noticeable drop ($\sim$30--50\%) in the dark rate within
$\sim$100 pixels of the edges of each WFPC2 CCD. The lowest ZL sky-SB that we
measure in the WFPC2 filter F606W near the North Ecliptic pole corresponds to
15 DN in 1800 sec \citep{Windhorst1994a, Windhorst1998}. For an average Dark
Glow of 0.77$\pm$0.18 DN in 1800 sec, the error from the Dark Glow subtraction
does not exceed $\sim$1.2\% in V-band at the NEP and is slightly worse in
I-band. The errors in the Dark Glow subtraction are smaller at lower latitudes
and generally do not exceed $\sim$2\%.

\sn {\bf WFPC2 Straylight:}\ \ The orbit-dependent foregrounds such as
Earthshine produce elevated sky-SB levels as discussed and flagged in \S\
\ref{sec43}. In addition, Earthshine propagates through the WFPC2 optical train
in a way that not only elevates the sky-SB on the detector --- as it does for
all HST instruments --- but also produces a recognizable pattern of diagonal
(dark) bands across each detector caused by specifics of the WFPC2 optical
train, in particular the alignment of the OTA and WFPC2 camera pupils
\citep[see \eg\ \S\ 11 of][]{Biretta1995}. These particular straylight
properties occur because the support struts for the repeater mirrors in WFPC2
--- which correct for HST's spherical aberration --- shadow HST's secondary
mirror support struts. For instance, such straylight patterns caused by
Earthshine affected the F300W images taken for the Hubble Deep Field South,
which were mostly taken in HST's Continuous Viewing Zone (CVZ) during orbital
``day-time''. These patterns can be removed as described in, \eg\ \S\ 3.4.2 of
\citet{Casertano2000}. The HST orbital phase monitoring of \S\ \ref{sec43}
flags and ignores such WFPC2 images affected by Earthshine, as their sky-SB
estimates may be affected in a way that is not correctable.

\sn {\bf WFPC2 Decontaminations and Time-Dependent UV-Zeropoints:}\
\citet{Holtzman1995}, \citet{McMaster2002} and \citet{Casertano2000} describe
calibration aspects specific to WFPC2. In orbit from December 1993 till May
2009, the optical train of WFPC2 underwent gradual contamination which affected
its time-dependent sensitivity and zeropoints, especially the WFPC2 UV filters.
Regular decontaminations of the WFPC2 instrument were therefore done, and the
\texttt{calwfc2} pipeline applies post-contamination corrections for the
time-dependent UV-filter zeropoints. Further details can be found in
\citet{McMaster2002} and \S\ 5.2 of the WFPC2 Data Handbook\footnote{\url{
https://www.stsci.edu/hst/instrumentation/legacy/wfpc2}\ \ \ and\\ \url{
https://www.stsci.edu/instruments/wfpc2/Wfpc2\_dhb/wfpc2\_ch53.html\#1920857}}.


\n \subsubsection{ACS/WFC} \label{sec412} 

\sn Here we summarize the specific considerations for the ACS/WFC data used in 
SKYSURF with their error contributions summarized in Table~\ref{tab:tab5}.

\sn {\bf ACS/WFC Dark Current:}\ The ACS/WFC dark-current is $\sim$0.01
e$^-$/pix/s
\citep{Ryon2022}~\footnote{\url{https://hst-docs.stsci.edu/acsihb}}, and has
slowly increased over time due to on-orbit detector degradation, with periodic
drops due to changes in temperature setting in 2006 or the introduction of
postflash in 2015, as shown in Fig. 3 of \citet{Anand2022}. Their Fig. 3 shows
that scatter in the ability to precisely determine the ACS/WFC dark-current
level over the years is $\sim$0.001 e$^-$/pix/s. Their Fig. 2 shows that the 
ability to determine the dark-current level in an individual super-dark frame
is considerably more accurate than this. For the average F606W Zodiacal sky-SB
level of 22.86 AB-\magarc\ \citep[Table 2 of][Fig.~\ref{fig:fig1}
here]{Windhorst2011} and the 0\arcspt 050/pixel scale of the ACS/WFC detector,
$\sim$0.001 e$^-$/pix/s corresponds to a dark-current induced error in the
Zodiacal sky-SB of $\sim$1.5\%. 

\sn {\bf ACS/WFC Flat Fields:}\ \citet{Cohen2020} present ``LP''-flats for
ACS/WFC, which include corrections for both low (``L'') spatial frequency and
pixel-to-pixel (``P'') flat-field variations. From their Figs. 5 and 6, the
errors induced by the ACS/WFC flat-fields are $\sim$0.6--2.2\% of the Zodiacal
sky-SB for medium-length single exposures in our ACS/WFC database in
Table~\ref{tab:tab2}a.

\sn {\bf ACS/WFC Fringing:}\ Multiple reflections between the layers of a CCD
detector can give rise to fringing at longer wavelengths ($\lambda$\cge
750-800 nm), where the amplitude of the fringes is a strong function of the
silicon detector layer thickness and the spectral energy distribution of the
light source, as discussed in the ACS IHB \citep{Ryon2022}. The fringe pattern
is stable and is removed to first order by the flat field for continuum sources
\citep{Ryon2022}. 

\sn {\bf ACS/WFC Red Stellar halos:}\ \ For ACS/WFC, we must correct sky-SB
measurements in the F850LP for effects of the broad red stellar halos in the
aberrated beam that may not be fully captured in the corrected beam (App. 
\ref{secAppB3}). 

\n \subsubsection{WFC3/UVIS} \label{sec413} 

\sn Here we summarize the specific considerations for the WFC3/UVIS data used
in SKYSURF with their error contributions summarized in Table~\ref{tab:tab5}.

\sn {\bf WFC3/UVIS Flat Fields:}\ The WFC3/UVIS global flat-field errors are
$\sim$2--3\% across the detector for most WFC3/UVIS broad-band filters 
\citep[\eg][]{Rajan2010, Mack2015}. 

\sn {\bf WFC3/UVIS Filter Red Leaks and Blue Leaks:}\ \ The WFC3/UVIS filters
were designed to have minimal red leaks for the bluer filters, and very small
blue leaks for the redder filters. A detailed estimate of the WFC3/UVIS
vacuum-UV filter red leaks is given in Fig. 1b and Appendix B.1 of
\citet{Smith2018}. For the WFC3/UVIS optical broadband filters, red leaks are
generally no larger than $\sim$10$^{-5}$--10$^{-4}$ of in-band flux for a flat
spectrum SED. A discussion of the effects from UV filter pinholes on low-SB
measurements is given in Appendix B.2 of \citet{Smith2018}. Any UV filter 
pinhole would imprint a very broad red leak on the image, but because the
WFC3/UVIS filters are placed at a significantly out-of-focus location in the
optical train, pinhole red leak effects are generally dimmer than AB\cge 31
\magarc, or $\sim$1\% of the UV sky-SB. 

\sn {\bf WFC3/UVIS Fringing:}\ \ As in the case of ACS/WFC, fringing may also
affect the sky-SB in the reddest WFC3/UVIS filters, as discussed in the WFC3
IHB \citep{Dressel2021}~\footnote{\url{https://hst-docs.stsci.edu/wfc3ihb}}.

\sn {\bf WFC3/UVIS Internal Reflections:}\ \ Both WFC3/UVIS and IR can have
complex internal reflections when bright stars are in the FOV (see, \eg\ the
Figures in \S\ \ref{sec42}), or produce large artifacts (``dragon's breath'')
when a bright star lands exactly on the edge of the detector masks. Large
artifacts or bright stars are flagged when making object catalogs (Appendix
\ref{secAppB3} \& \ref{secAppC1}), and our code will discard these regions or
images when making sky-SB estimates \citepalias[\S\ \ref{sec42}, \ref{sec43}
and][]{Carleton2022}.

\n \subsubsection{WFC3/IR} \label{sec414} 

\sn Here we summarize the specific considerations for the WFC3/IR data used in 
SKYSURF with their error contributions summarized in Table~\ref{tab:tab5}.
Because SKYSURF's first science results in \citetalias{Carleton2022} come from
the WFC3/IR sky-SB estimates, the known sources of systematic errors that could
affect these estimates are summarized in more detail here. 

\sn {\bf WFC3/IR Blobs and their Correction:}\ WFC3/IR images show several
small (10--15 pixel) blobs that form a stable low-level ($\sim$10--15\% on
average) depression in the foreground \citep{Pirzkal2010} affecting
$\sim$1--2\% of the WFC3/IR pixels. The number of blobs has increased at a rate
of \cle 1 per month to a current total of $\sim$150 blobs
\citep{Olszewski2021}. The WFC3/IR Blobs are believed to be due to ``small
particulate features with reduced QE'' that accumulated on the WFC3 Channel
Select Mechanism \citep[CSM;][]{Bushouse2008}. Specially constructed
``Delta-flat fields'' correct these features significantly, and known blobs
are flagged in the data-quality arrays and ignored in our analysis, so they do
not pose a significant source of error in the SB-estimating algorithms of \S\
\ref{sec42}. 

\sn {\bf WFC3/IR Flat Fields:}\ The latest sky delta-flat fields have been
implemented in the \texttt{calwf3} pipeline. Fig. 2 and 4 of
\citet{Pirzkal2011} show that the flat-field error in WFC3/IR broad-band
filters is generally better than $\sim$0.5--2\% of the average Zodiacal sky-SB,
from the central 800$^2$ pixels of the detector to the edges, respectively
\citep{Mack2021}. To be conservative, we adopt $\sim$2\% in
Table~\ref{tab:tab5} for the WFC3/IR flat-field induced errors, as we cannot
predict {\it where} in the SKYSURF images our algorithms of \S\ \ref{sec42} will
estimate the sky-SB values.

\sn {\bf WFC3/IR Geometry:}\ \ The WFC3/IR detector has 1014$\times$1014
active pixels. To minimize internal reflections, the WFC3/IR detector has a
$\sim$24\degree\ tilt about its x-axis, creating an image elongation of
$\sim$9\%. The WFC3/IR detector therefore covers a rectangular
136\arcs$\times$123\arcs\ FOV with rectangular pixels of 0\arcspt
1341$\times$0\arcspt 1213 on average.

\sn {\bf WFC3/IR Filter Red Leaks and Blue Leaks:}\ \ The WFC3/IR filters were
also designed to have very small red leaks and blue leaks. The blue leaks are
defined in the WFC3 IHB \citep{Dressel2021} as the fraction of erroneous flux
coming from 710--830 nm compared to the expected proper in-band flux. (The
WFC3/IR QE curve is almost flat down to 780 nm but rapidly declines at bluer
wavelengths.\deleted{All WFC3/IR filters and grisms were placed on IR80
substrates, which is a very good blue-light blocker at wavelengths
$\lambda$\cle of 800nm.}) Table 7.4 of the WFC3 IHB shows that for a black-body
with \Teff=5000 K \citepalias[\ie\ representing the reddened Zodiacal spectrum
used in][]{Carleton2022}, the WFC3/IR broad-band filters have a blue leak of
$\sim$2.4$\times$10$^{-7}$--1.7$\times$10$^{-4}$ of the proper in-band flux.
We verified this through numerical integration of the Solar spectrum through
the full F125W filter curve available at
STScI~\footnote{\url{https://www.stsci.edu/hst/instrumentation/reference-data-for-calibration-and-tools/synphot-throughput-tables}}.
This is an important consideration for SKYSURF, as more of the Zodiacal 
sky-SB is generated {\it blueward} of the WFC3/IR filter throughput-curves. 
The worst-case WFC3/IR blue leak is 1.7$\times$10$^{-4}$ of the in-band flux
for the F160W filter \citep{Dressel2021}. This is much smaller than other
systematics that we encounter when measuring absolute sky-SB values in \S\
\ref{sec415}--\ref{sec416}, \ref{sec43} and \citetalias{Carleton2022}.



\vspace*{-0.00cm}
\n\begin{figure*}[!hptb]
\n \cl{
 \includegraphics[width=0.500\txw]{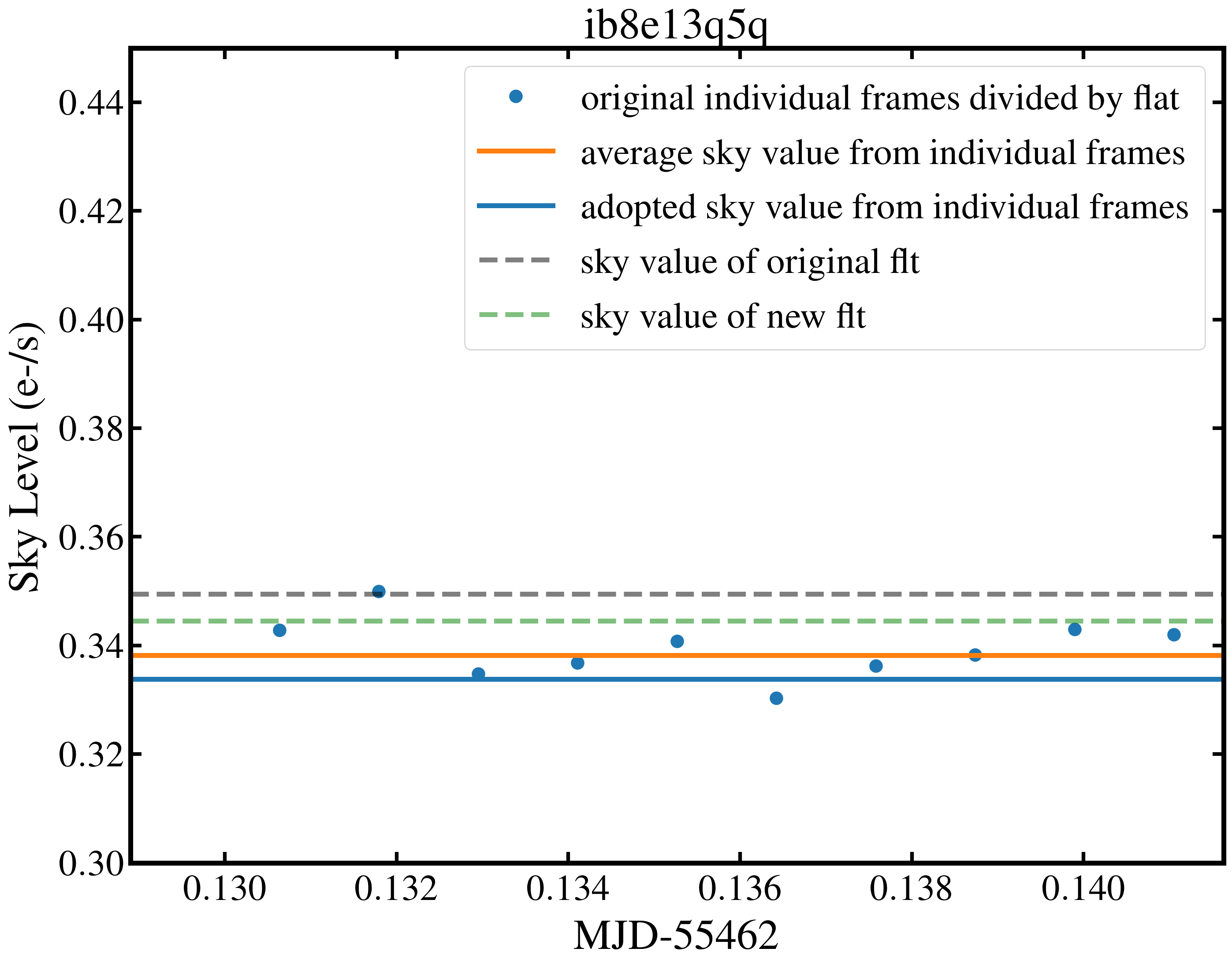}\ 
 \includegraphics[width=0.500\txw]{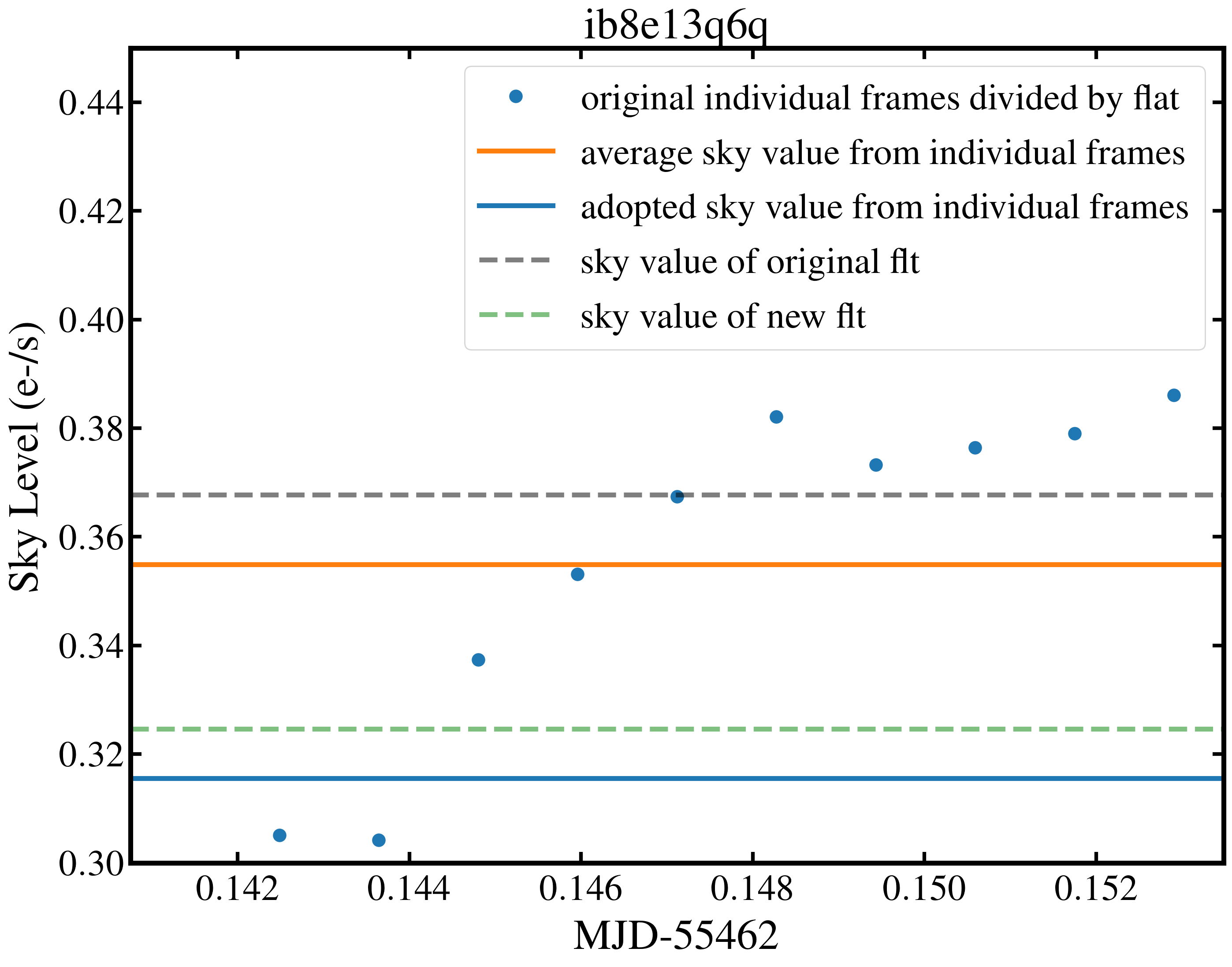}
}

\vspace*{+0.10cm}
\n \caption{
Sky-SB values vs. time in successive half-orbit HST WFC3/IR F125W exposures of
a given high Ecliptic latitude target. The file names on top indicate the
WFC3/IR exposures shown. 
{\bf (a) [Left]:} Sky-SB values vs. time at the start of this orbit, which
typically uses the darker part of the orbit, so the sky-SB values do not change
much over the total exposure length. Time is expressed as Modified Julian Date
(MJD) from the HST FITS headers. The blue data points show the sky-SB values of
individual WFC3/IR ramp reads. The orange line shows the exposure time-weighted
average over all ramps. For WFC3/IR, we adopt the average of the three ramps
with the lowest sky-SB values (blue line), excluding the initial clearing
readouts, which we consider to be the more accurate estimate of the Lowest
Estimated Sky-SB (LES) for that exposure (\S\ \ref{sec422}). {Thus, all
frames in the $\_ima$ file are corrected onto the blue line, and the pipeline is
run on the adjusted image. The sky value of the resulting $\_flt$ image is
shown as the green dashed line, compared with the sky value of the original
$\_flt$ image, shown as the grey dashed line (see \S\ \ref{sec43}).}\ 
{\bf (b) [Right]:} WFC3/IR F125W sky-SB values vs. time at the end of a typical 
HST orbit. The procedures and lines are the same as in Fig.~\ref{fig:fig6}a,
but the differences are now much larger, because in this second half-orbit
WFC3/IR exposure the sky-SB values increase significantly towards the end of
the orbit due to Earthshine, resulting in the downwards correction in the final
adopted sky-SB value (blue line compared to the orange line). The difference
between the two adopted sky-SB levels is consistent to within the error budget
summarized in Table~\ref{tab:tab5}.}
\label{fig:fig6}
\end{figure*}


\sn {\bf WFC3/IR --- Splitting and Analyzing Exposures using Individual
Ramps:}\ \ The WFC3/IR detector read-outs are non-destructive, so all 
individual WFC3/IR exposures consist typically of 8--10 on-the-ramp
sub-exposures, each of which are calibrated to facilitate correction for the
numerous CR-hits and to obtain the desired exposure depth. SKYSURF measures
the sky-SB in each of the 8-10 individual WFC3/IR on-the-ramp sub-exposures,
which enables us to better diagnose the behavior of the sky-SB (\S\
\ref{sec43}) and the Thermal Dark signal \citepalias[][]{Carleton2022} as a
function of orbital phase. An example is shown in
Fig.~\ref{fig:fig6}a--\ref{fig:fig6}b. This process leaves some CRs in the
on-the-ramp sub-exposures, which our robust sky-SB algorithms are designed to
ignore (\S\ \ref{sec43} \& Appendix \ref{secAppB1}). Only the full-ramp 
full-exposure WFC3/IR images that have been CR-filtered are used for SKYSURF's
object counts (Appendix \ref{secAppC}). 

\sn {\bf WFC3/IR Count-Rate Non-Linearity Correction:}\ \ \citet{Riess2010}
and \citet{Riess2019} discuss the changing WFC3 ZP due to its detector
Count-Rate Non-Linearity (CRNL) over a very large dynamic range in flux. As a
result, the WFC3/IR ZPs are slightly different for the bright calibration stars
compared to faint galaxies, whose average SB is close to or below that of the
Zodiacal sky-SB. In all its filters, the WFC3/IR zeropoints are subject to CRNL
of about 0.049$\pm$0.005 mag (or 4.5$\pm$0.5\%) over the {\it full range} of 
16 AB-mag (6.4 dex) in flux per pixel, or $\sim$+0.0077$\pm$0.0008 mag (or
$\sim$+0.71$\pm$0.07\%) over each 2.5 mag (1 dex) interval of flux per pixel. 
This detector CRNL is a charge trapping problem, where up to 4.5\% of the faint
object and faint SB-flux are lost in detector traps, and is the opposite of
persistence, which is mapped and removed as described in \S\ \ref{sec41}.
WFC3/IR standard stars are measured in the total flux range of 12\cle AB\cle 21
mag, and of course cover an area about the size of the WFC3/IR PSF, which is
\cge 0.0163 arcsec$^2$. The sky-SB levels in the WFC3/IR broad-band filters are
of order 22.6--22.3 AB-\magarc\ \citep[Table 2 of][]{Windhorst2011} 
(``\citetalias{Windhorst2011}''), or $\sim$26.9 mag/PSF area. Hence, the SKYSURF
sky-SB is measured at levels $\sim$6--15 mag dimmer than the pixel-signal of
the standard stars used to determine the WFC3 zeropoints. We verified that the
standard WFC3 pipeline \texttt{calwf3} corrects all pixel SB-levels accordingly
for detector CRNL.\deleted{\citet{Riess2021} show that the WFC3/IR CRNL
corrections enable a measurement of the local Hubble constant with a $\sim$1\
\kmsMpc accuracy using Cepheids and SNe over a wide flux range.} In SKYSURF,
typical sky-SB levels are corrected {\it upwards} by
$\sim$[(6--15)/16]$\times$4.5\% $\simeq$+(1.7--4.2)\% to match the standard
star flux scale, while objects with higher SB per pixel are corrected by
correspondingly less. Hence, SKYSURF does {\it not} need to make further
corrections for WFC3/IR detector count-rate non-linearity, but we do account
for its $\sim$0.5\% uncertainty in our error budget below. 

\sn {\bf WFC3 Dark Current Monitors:}\ \ The WFC3/IR Dark Current (DC) has
been monitored over the years \citep[\eg][]{Sunnquist2017a, Sunnquist2017b,
Sunnquist2019}. In particular, these studies show some trends in the measured
WFC3/IR DC levels and structure with the four different IR detector quadrants.
Also, these studies found trends of the measured WFC3/IR DC levels with
telescope temperature and with HST's of orbital phase (Sun Altitude; \S
\ref{sec43}\ \& Appendix \ref{secAppA22}) during the Dark Current frame
observations. In O'Brien \etal\ (2022, in preparation), we study the WFC3/IR
quadrant variations in more detail, but find them, in general, to be \cle
0.5\%, \ie\ smaller than most other systematic uncertainties in measuring the
sky-SB values (\S\ \ref{sec42}), and hence not requiring additional
quadrant-dependent DC corrections. \citet{Sunnquist2017b} suggest that the rms
variations in dark current level are $\sim$0.0069 e$^-$/s, or $\sim$1\% of the
typical sky-SB in Fig.~\ref{fig:fig1}, which is incorporated into the error
budget of Table~\ref{tab:tab5}. 

The fact that the WFC3/IR Dark Current levels appear to be a function of Sun
Altitude in \citet{Sunnquist2017b} prompted us to take a closer look at how much
of the electronic Dark Current level is temperature-dependent --- \eg\ due to
temperature variations in the detector A/D converters, as
\citet{Sunnquist2017b} suggest --- and how much additional Thermal Dark {\it
signal} may be caused by HST's temperature variations across its orbit and as a
function of Sun Altitude. These aspects are addressed in \S\ \ref{sec43} and 
\citetalias{Carleton2022} as needed for the current SKYSURF results, and in
more detail by Carleton \etal\ (2022b, in preparation) and O'Brien \etal\ (2022, in
preparation), which will include a correction for temperature-dependent dark
current variations using the \citet{Sunnquist2017b} prescription. 

\sn {\bf WFC3/IR Thermal Dark Signal:}\ \ For the WFC3/IR channel, we need to
distinguish between the WFC3/IR Dark Current and its Thermal Dark signal. The
former is mostly an electronic {\it current} in the WFC3/IR detector that
depends on the temperature of the detector and its read-out electronics, while
the latter is caused by Planck black-body {\it photons} generated in HST OTA,
the WFC3 housing, and the WFC3/IR detector enclosures. We need to have the best
possible estimates of both to analyze the sky-SB values measured in WFC3/IR
images. We will throughout refer to the WFC3/IR Thermal Dark {\it signal} as
``TD'' or ''TD {\it signal}''. The WFC3/IR TD signal was predicted in Table
7.11 of the WFC3 IHB \citep{Dressel2012, Dressel2016, Dressel2021}, where it is
referred to as a ``thermal dark current''. 

The IR detector Dark Current is rather well measured out to the WFC3 shutter,
which is a WFC3/IR filter slot filled with aluminum. The WFC3/IR TD signal is
more uncertain and depends somewhat on the exact thermal history of each HST
orbit. Predictions of the Thermal Dark signal can be made with the python
$synphot$
package~\footnote{\url{https://www.stsci.edu/hst/instrumentation/reference-data-for-calibration-and-tools/synphot-throughput-tables}}.
Tables with thermal foreground levels are incorporated into the $synphot$
tool, based on the observed temperatures of the HST optical and instrument
components discussed in Appendix A of \citetalias{Carleton2022}. In summary, our
calculations of the WFC3/IR TD signal assume temperatures of the HST primary
and secondary mirror of T$\simeq$287 K, temperatures for the WFC3 pick-off
mirror and the corrective optics mirrors inside WFC3 and the IR channel of
T$\simeq$273 K, and temperatures for the two outer enclosures and the one inner
cold enclosure holding the WFC3/IR detector of T$\simeq$223--173 K,
respectively. 

To ensure that our sky-SB measurements are properly corrected for TD signal, 
\citetalias{Carleton2022} will conduct sky-SB estimates measurements in the
three WFC3/IR broad-band filters F125W, F140W, and F160W, and correct all three
for the TD signal predicted for the known ambient temperatures of the HST
components. Our $synphot$ analysis reveals that the predicted TD signal is
strongly wavelength dependent. For the F125W and F140W images, the vast
majority of the TD signal comes from the WFC3/IR cold enclosure, and not from
the two telescope mirrors plus the five optical-path mirrors inside WFC3, due
to their much smaller solid angle as seen from the IR detector and the fact
that their Planck SB is still minimal at the long-wavelength cutoff of these
two bluer filters. In the F160W images, the TD contribution from the 7 mirrors
in front of the WFC3/IR cold enclosure increases significantly. In
\citetalias{Carleton2022}, we will therefore consider a range of plausible TD
values given the HST component temperatures discussed in their Appendix A.
Depending on the temperature of these HST components, the predicted TD signal
may vary by up to 30\%. In the F160W filter, this corresponds to $\sim$2.7\% of
the average sky-SB, which is folded into Table~\ref{tab:tab5}. 

\n \subsubsection{SKYSURF Zeropoints and Zeropoint Monitors} \label{sec415} 

\sn For SKYSURF, we need to know HST's zeropoints (ZPs) as accurately as
possible from 1994--2020. Photometric stability over the years is different
from photometric calibration or ZP errors. We therefore consider below in order:
1) photometric ZP definition in the context of PSF wings; 2) the actual 
photometric zeropoints; and 3) stability and drifts of photometric ZPs over the
years. This is followed by a discussion of ZP specifics for each HST instrument
used in SKYSURF. 

\sn {\bf 1) Zeropoint Definition and ZP Use:}\ For each of the HST instruments
used in SKYSURF, the Data Handbooks and ISRs provide instrumental zeropoints
for an object with a total flux of 1.000 \eminpixsec. These ZPs are usually
offered for ``compact objects'' within radii $r$\cle 0\arcspt 4, {\it and} for
an ``infinite radius aperture''. Because of the Encircled Energy (EE) curve of
the spherically aberration-corrected PSF, the EE-values for point sources
decrease from 91\% at $r$\cle 0\arcspt 4 in the F606W filter to $\sim$84\% in
the F125W, F140W and F160W filters due to HST's broader PSFs at the longer
wavelengths. Since the HST EE-curves approach 100\% enclosed flux for $r$\cle
6\arcspt 0, the ``infinite radius apertures'' integrate the total fluxes out
to $r$\cle 6\arcspt 0. For details, we refer to the ISRs by
\citet[\eg][]{Ryan2016}, \citet{Deustua2016}, and \citet{Medina2022}. Most
faint galaxies that SKYSURF detects to AB\cle 26.5 mag in single $\sim$500 sec
SKYSURF exposures have half-light radii \re$\sim$0\arcspt 2--0\arcspt 3 (\S\
\ref{sec47} \& Appendix \ref{secAppC}), so their total flux is not fully
captured by the ``compact object'' aperture ZPs at radii $r$$\le$0\arcspt 4.
Hence, to obtain more accurate total magnitudes for slightly resolved faint
galaxies, as well as accurate sky-SB levels in the empty image areas (\S\
\ref{sec42}), we will always use the ZPs for ``infinite radius apertures'' in
SKYSURF. For the current SKYSURF study, the ``infinite radius apertures'' ZPs
provide the best absolute sky-SB values. This assumes that for a given
sky-pixel and {\it constant} sky-SB, the HST PSF will leak as much flux into
neighboring pixels as the neighboring pixels will leak into that pixel. 

\sn {\bf 2) Photometric Zeropoints:}\ The instrument ZPs are derived from
white dwarf standard star observations, which for most filters can be done to
$\sim$1\% accuracy \citep[\eg][]{Bohlin2020}. The ACS/WFC and WFC3/UVIS and IR
instrument zeropoints have been stable to $\sim$1--2\% \citep{Kalirai2010b,
Ryon2022, Dressel2016, Dressel2021} with the caveats discussed below, allowing
their use as {\it absolute photometers}. When comparing results from very
similar broad-band filters in different instruments, small corrections for
differences in central wavelength may need to be applied (see, \eg\
Fig.~\ref{fig:fig2}a here; \citetalias{Windhorst2011},
\citetalias{Driver2016b}; and the ICRAR zeropoint transformation
tool \citep{Koushan2021}~\footnote{\url{http://transformcalc.icrar.org}}). This is relevant when
doing the objects counts over a wide dynamic range in similar filters from
different instruments. Details on the filter zeropoints and aperture
corrections needed for faint object fluxes and faint object counts are provided
by Tompkins \etal\ (2022, in preparation). 

\sn {\bf 3) ZP Stability and Drifts over the Years:}\ We need to monitor and
carefully account for panchromatic WFPC2, ACS/WFC, and WFC3/UVIS and IR
zeropoint changes vs. time, some of which may be due to modest accumulation of
contaminants onto the HST optics and in each camera over 11--18 years. This is
especially true for the HST WFC3/UVIS ultraviolet filters F225W, F275W, and
F336W. We discuss the relevant ZP details for each of the cameras below. 

\mn {\bf WFPC2:}\ \ WFPC2's Archival data spans over 16 years (1994--2009)
with photometric zeropoints stable to within $\sim$2--3\% \citep{Holtzman1995},
and $\sim$1--2\% in F439W and redwards. The WFPC2 photometric zeropoints are
taken from Table 10 of \citet{Holtzman1995} using their equation 9. Section 5.7
and Table 5.10 of the WFPC2 Data Handbook\footnote{\url{
https://www.stsci.edu/instruments/wfpc2/Wfpc2\_dhb/wfpc2\_ch53.html\#1920857}}
summarize the end-of-mission uncertainties in the WFPC2 calibration aspects
and time-dependent zeropoints, which are summarized in Table~\ref{tab:tab5}.
The WFCP2 {\it ATODGAIN} was read from the FITS header, and the appropriate
gain ratio from \citet{Holtzman1995} was applied to correct the zeropoints for
the different CCDs to AB-magnitudes. 

\mn {\bf ACS/WFC:}\ \ The STScI ACS group has monitored the ACS zeropoints
since its 2002 Shuttle launch on Servicing Mission 3 (SM3B). True ACS zeropoint
changes are smaller than $\sim$1--2\%, with drifts of order $\sim$0.1--0.2\%
per year. SKYSURF uses the ACS zeropoints
website~\footnote{\url{http://acszeropoints.stsci.edu}} which provides this
up-to-date information as a function of observing date. We use these
time-dependent functions to determine the best zeropoints for the observation
dates for all ACS/WFC images. From the scatter in the time-dependent
zero-points in Fig. 2 of \citet{Bohlin2020}, we will conservatively take this
ZP-trending to be no better than $\sim$1\% in accuracy, which is incorporated in
the error budget of Table~\ref{tab:tab5}. 

\mn {\bf WFC3/UVIS:}\ \ STScI has also monitored the WFC3/UVIS zeropoint
changes since its 2009 Shuttle launch on SM4 \citep[\eg][]{Ryan2016,
Deustua2016, Bajaj2019, Calamida2021}. Long-term WFC3 zeropoint changes can
amount to $\sim$3\% \citep{Ryan2016} but are in part due to definition changes
in the \texttt{calwf3} pipeline. True WFC3 zeropoint changes are smaller than
this, with drifts of order $\sim$0.1--0.2\% per year \citep{Calamida2021}.
SKYSURF uses the WFC3/UVIS zeropoints websites~\footnote{\footnotesize{
\url{https://www.stsci.edu/contents/news/wfc3-stans/wfc3-stan-issue-33-october-15},
and\\
\url{https://www.stsci.edu/hst/instrumentation/wfc3/data-analysis/photometric-calibration/uvis-photometric-calibration}}}, 
which provide this up-to-date information. In particular, \citet{Calamida2021}
summarize well-sampled time-dependent zeropoint changes for most WFC3/UVIS
filters from 2009--2019. For the WFC3/UVIS UV-filters with central wavelengths
below 300 nm, the ZPs show an increase in sensitivity of $\sim$+0.5\%/yr for the
first two years, followed by a slowly declining sensitivity over time in
subsequent years. All filters redwards of F300X have seen a slow, nearly
linear and somewhat wavelength-dependent decline in ZP-sensitivity over time,
with well-determined slopes that typically range between --0.1\%/yr to
--0.2\%/yr. E.g., for the most used WFC3/UVIS F606W and F814W filters
(Table~\ref{tab:tab2}b), this steady decline in ZP amounts to about --1.8\% and
--1.0\% over 11 years, respectively. The linear fits of the
\citet{Calamida2021} zeropoint drifts have rms errors of $\sim$0.5\%, and
provide the best available data on the time-dependent WFC3/UVIS zeropoints as
of 2021. To implement these slowly time-dependent ZPs as a function of
wavelength and time, we therefore downloaded the WFC3/UVIS images again in
early 2022 with \texttt{calwf3} pipeline version 3.6.2 on AWS, which updated
all FITS headers with the proper time-dependent ZPs. For details of this
process, we refer to O'Brien \etal\ (2022, in preparation). The resulting
WFC3/UVIS ZP errors as a function of time are estimated have a $\sim$1\%
accuracy, as listed in Table~\ref{tab:tab5}. 

\mn {\bf WFC3/IR:}\ \ The WFC3/IR photometric zeropoints have roughly remained
constant to within $\sim$1.5\% (rms) in the 10 years after its SM4 launch
without a noticeable time-dependent decline \citep[Fig. 7 of][]{Bajaj2019},
unlike those of WFC3/UVIS. Some of the $\sim$1.5\% scatter is due to residual
persistence, which with proper dithering and drizzling can be further reduced
\citep{Bajaj2019}. We will therefore take the error in the WFC3/IR broad-band
ZPs to be $\sim$1.5\% in Table~\ref{tab:tab5}. The WFC3/IR ZPs used in the
F125W, F140W, and F160W filters are 26.232, 26.450, 25.936 AB-mag,
respectively, for an object with 1.000 \eminpixsec.~\footnote{\footnotesize{
\url{https://www.stsci.edu/hst/instrumentation/wfc3/data-analysis/photometric-calibration},\\
\url{https://www.stsci.edu/hst/instrumentation/wfc3/data-analysis/photometric-calibration/ir-photometric-calibration},
see also \url{http://svo2.cab.inta-csic.es/svo/theory/fps3/index.php}}}


\n \subsubsection{Sky-SB Error Budget Thus Far} \label{sec416} 

\mn Table~\ref{tab:tab5} summarizes the error budget for WFPC2, ACS/WFC,
WFC3/UVIS and WFC3/IR resulting from the considerations in \S\
\ref{sec411}--\ref{sec415} \& \ref{sec42}. For the WFC3/IR filters F125W and
F140W, the total relative error is $\sim$2.7--2.8\% of the average sky-SB
level, while for F160W it is $\sim$3.8\%, which is indicated by the brown
horizontal short-dotted line in Fig.~\ref{fig:fig1}. The dominant sky-SB
errors in Table~\ref{tab:tab5} are {\it multiplicative} in nature, \ie\ the
global flat-field and photometric zeropoint errors. {Because of the way the
flat-fields and photometric zeropoints are created, these two main relative
error sources are independent, so that adding them in quadrature is justified.
O'Brien et al. (2022, in preparation) present a comparison between the ACS,
WFC3/UVIS and WFPC2 sky-SB measurements over 26 years (\eg\ $\sim$49,000 images
in the F606W filter alone), enabling them to confirm our relative errors in
Table~\ref{tab:tab5} and --- together with the WFC3/IR data --- compare
SKYSURF's sky-SB values to panchromatic Zodiacal models.} 

The predicted WFC3/IR Thermal Dark signal error is modest in the F125W and
F140W filters, but increases significantly for the F160W filter
\citepalias{Carleton2022}, and therefore becomes a dominant {\it additive}
error in the F160W sky-SB estimates. For this reason, the bottom two rows of
Table~\ref{tab:tab5} list {\it both the lowest and the average} sky-SB values
plus the total errors for each in units of \nWsqmsr\ for the WFC3/IR F125W,
F140W, and F160W filters, respectively. For the {\it darkest} sky-SB values the
percentage TD error is about 2$\times$ larger than the errors quoted for the
{\it average} sky-SB values, increasing the total errors for the darkest sky-SB
values to $\sim$2.8\%, $\sim$2.9\%, $\sim$6.3\% in the F125W, F140W, and F160W
filters, respectively. In \citetalias{Carleton2022}, we will consider {\it both
the darkest and average sky-SB values} in these filters, and will therefore
propagate the errors of Table~\ref{tab:tab5} accordingly into our results.
\citetalias{Carleton2022} includes a method to avoid images with significant
straylight for sky-SB estimates, which will have its own error considerations. 

\n \subsection{Optimal Object-Free Sky-SB Estimates in All Individual SKYSURF 
Exposures} \label{sec42} 

\sn A critical part of SKYSURF is measuring the sky-SB in between the discrete
objects as accurately as possible. For this, we introduce in the next 
subsections different algorithms to estimate the sky-SB as well as possible in
between discrete objects, which are primarily faint galaxies, Galactic stars,
and CRs. To test these algorithms, we check them against simulated images with 
{\it known input} sky-SB values, as well as realistic distributions of galaxies,
stars, and CRs. Here, we summarize how those simulations were done, how sky-SB
measurements were made with different algorithms, and how these were verified.
Details of these WFC3/IR image simulations and all nine SKYSURF methods used to
estimate the sky-SB are given in O'Brien \etal\ (2022, in preparation). 

\n \subsubsection{Simulation of Images with Known Sky-SB and Realistic Cosmic
Ray, Star and Galaxy Counts} \label{sec421} 

\sn Following the observed panchromatic galaxy and star counts of
\citet{Windhorst2011}, we generated 784 simulated WFC3/IR F125W images with
{\it known} input sky-SB values, and realistic surface densities and magnitude
distributions of stars and galaxies. The 784 simulated images cover the range
of expected sky-SB values given the exposure time distribution in WFC3/IR F125W
of Fig.~\ref{fig:fig5}. For about half the images we include the expected
increase in sky-SB and its resulting sky-gradient in typical HST exposures
towards the Earth's limb at the end of each HST orbit, as discussed in \S\
\ref{sec43}. 

The {\it total} AB-magnitudes of both the imposed galaxy and star counts use
the same WFC3/IR F125W AB-mag zeropoint of \S\ \ref{sec415} for an object with
1.000 \eminpixsec. The simulated stellar and galaxy profiles are broadened
with the WFC3 PSF, while the galaxy profiles also have the ellipticity
distributions superimposed as observed for faint galaxies \citep{Odewahn1997}.
Finally, the images add a {\it known} sky-SB and corresponding rms sky-noise
level. A variety of Zodiacal sky-SB levels are used with a range of 
sky-SB$\simeq$0.628--3.14 \eminpixsec\ for the range of exposure



\vspace*{-0.00cm}
\n\begin{figure*}[!hptb]
\n\cl{
\includegraphics[width=1.050\txw]{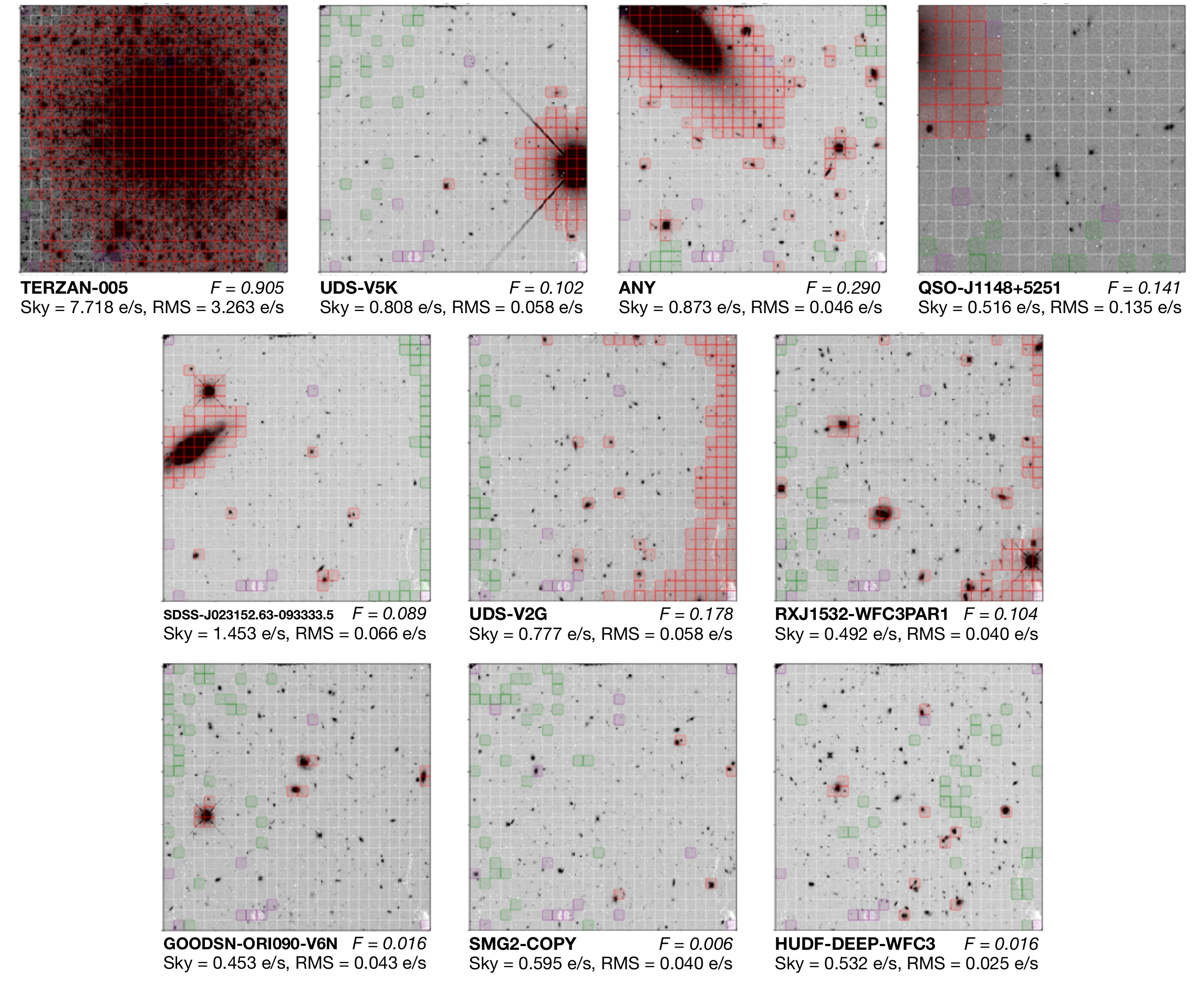}
}


\vspace*{+0.30cm}
\n \caption{
Examples of the robust HST sky-SB estimator (\S\ \ref{sec422}--\ref{sec423})
applied to a sample of WFC3/IR F125W images. {Purple boxes indicate cells
not used in the sky-SB determination because they contain bad pixels, and red 
boxes indicate cells not used because the object finder identified these cells 
as affected by a bright object or its outskirts. The remainder boxes are used
to determine the sky-SB in each image, while green boxes represent the lowest
5\% values of the boxes that were used to estimate the Lowest Estimated Sky
values in each image, which is defined in \S\ \ref{sec422}. The HST Target Name
is given in each case, as well as the fraction $F$ of total boxes flagged as
red, plus the sky-SB and its rms values in the non-flagged areas.} The top row
show images that are substantially covered and affected by large objects so
that at best only the opposite image sides or image corners can be used to
estimate the LES value in that image, which still may have higher sky-SB than
the true ZL+DGL+EBL in that direction of the sky. The middle row shows examples
of images with measurable low-level gradients. The bottom row of images are
relatively empty fields, where SKYSURF's LES value is less biased by large
discrete objects and may be closer to the ZL+DGL+EBL level, as discussed in \S\
\ref{sec4}. This robust automated sky-SB estimation algorithm for WFC3/IR is
also adapted for ACS/WFC, WFC3/UVIS and WFPC2 images, and is essential to make
SKYSURF's sky-SB estimates across the sky.}
\label{fig:fig7}
\end{figure*}


\n times of \texp$\simeq$50--1300 sec. Following the instrument parameters in
Table 7.11 of the WFC3 IHB \citep{Dressel2021}, the rms noise includes the
sky-noise of the imposed Zodiacal sky-level, the WFC3/IR read-noise, and Dark
Current noise. Including all of these yields a typical simulated WFC3 F125W
signal of $\sim$1.1 \eminpixsec, so that an average 500 sec WFC3/IR exposure
has a total imposed sky-level of $\sim$556 \emin\ with an rms noise of
$\sim$23--24 \emin. 

In total, 784 WFC3 images were simulated by two independent teams, one
following the above analytical galaxy image description that is based on 
well-defined statistical properties of real WFC3 images
\citep[\eg][]{Windhorst2011}, while the other team used {\it real} WFC3 galaxy
parameters as observed in CANDELS and 3DHST \citep{vanderWel2012, Skelton2014}.
The latter method automatically produces the observed size distribution of \S\
\ref{sec47} and the observed $b/a$ distribution of \citet{Odewahn1997}, so
these parameters did not need to be imposed by the second team. These images
look realistic, with the exception that real HST images show galaxies that are
more irregular than the exponential disk or generalized Sersic light profiles.
We validated our simulations by running the star-counts and galaxy counts on
these images following the procedures described in \S\ \ref{sec47} and Appendix
\ref{secAppC}, and concluded that the required surface densities, count slopes,
and galaxy size and $b/a$ distributions were imposed correctly. This set of
simulated WFC3/IR images with {\it known} input sky-SB levels was subsequently
analyzed by {\it different} independent teams that assessed which algorithms
most closely recover the {\it known} input sky-SB levels. This provides the
proof of concept that SKYSURF can measure the sky-SB levels in between the
discrete objects accurately, independent of their actual cause.


\n \subsubsection{SKYSURF Algorithms to Obtain the Lowest Estimated Sky-SB in
Each Exposure} \label{sec422} 

\sn SKYSURF's specific goal is to measure the Lowest Estimated Sky-SB in each
image that is not due to a statistical variation nor image defects, but is
closest to the true sky-SB that SKYSURF intends to constrain. Throughout, we
will refer to this Lowest Estimated Sky-SB level as ``LES'' or the ``LES
method''. In object-free areas of SKYSURF images taken during the darkest part
of each HST orbit, the LES level will be as close as HST can measure to the
true ZL+DGL+EBL level in that direction and time of the year. This is because
all contaminating signals outside the HST instruments (\eg\ Earthshine,
scattered Sun and Moon light) add as a positive signal --- only some of the
instrument calibration errors in \S\ \ref{sec41} could add a dim negative
signal to the images. Hence, SKYSURF needs to find this LES level in each
filter in each direction of the sky, accounting also for the Earth's Limb,
Moon, and Sun Angle at that time, and flagging images that may have higher
straylight values, as discussed in \S\ \ref{sec43}. 

Several different teams developed independent sky-SB estimation methods to
recover the sky-SB levels imposed in the simulated images by the two separate
teams in \S\ \ref{sec421}. These sky-SB estimation teams had {\it no} knowledge
of the actual sky-SB levels imposed by the two independent simulation teams.
The only aspects they were made aware of are that the simulated images had a
plausible range of sky-SB levels and exposure times (Fig.~\ref{fig:fig5}), 
surface densities of Galactic stars, galaxies with size- and ellipticity
($b/a$) distributions, as well as CR distributions. This process can be easily
expanded to include ACS/WFC, WFC3/UVIS, or WFPC2 filters and pixel scales, to
check on the wavelength dependence of the reliability of our algorithms that
estimate the sky-SB using the LES method. 

A total of nine methods were created to estimate LES levels in each image, each
taking a different approach in identifying the best algorithm to use on real
SKYSURF images (Fig.~\ref{fig:fig7}). These methods vary in their approach:
some calculate a clipped mean, while others incorporate the \ProFound\
\citep{Robotham2017} package which was designed to robustly create an
interpolated sky-SB map from an image. Several of the methods in 
Fig.~\ref{fig:fig7} attempt to estimate the sky-SB by taking into account that
the sky-SB might vary across a single image, \ie\ indicating the presence of a
sky-SB gradient (\S\ \ref{sec43}--\ref{sec44}). 

In several of the methods, SKYSURF accomplishes this by dividing each
1014$\times$1014 pixel WFC3/IR image into a 26$\times$26 grid of square
sub-regions or boxes that contain $39\times39$ pixels each. Every box thus
contains 1521 pixels, which is sufficient to make an estimate of its own sky-SB
value accurate to $\sim$2.6\% (\ie\ 1/39) of the sky-rms value in that box if it
contained no objects or image defects, that is, if the sky in that box were
truly featureless and flat. Next, we implemented a procedure that excluded
boxes based on their rms value or the number of pixels flagged in the Data
Quality extension of the image, which identifies bright or faint objects or
known defects in that box. The green boxes in Fig.~\ref{fig:fig7} represent the
lowest 5\% values of the {\it total} grid of boxes that was used to estimate
the Lowest Estimated Sky values in each image, \ie\ those without detectable 
objects. We emphasize that our LES method {\it does not look at the lowest 5\%
of all pixel values} in each WFC3/IR image, an estimate that would certainly be
biased to be well below the



\vspace*{-0.00cm}
\hspace*{-1.00cm}
\n\begin{figure*}[!hptb]
\n\cl{
 \includegraphics[width=1.00\txw,angle=-0]{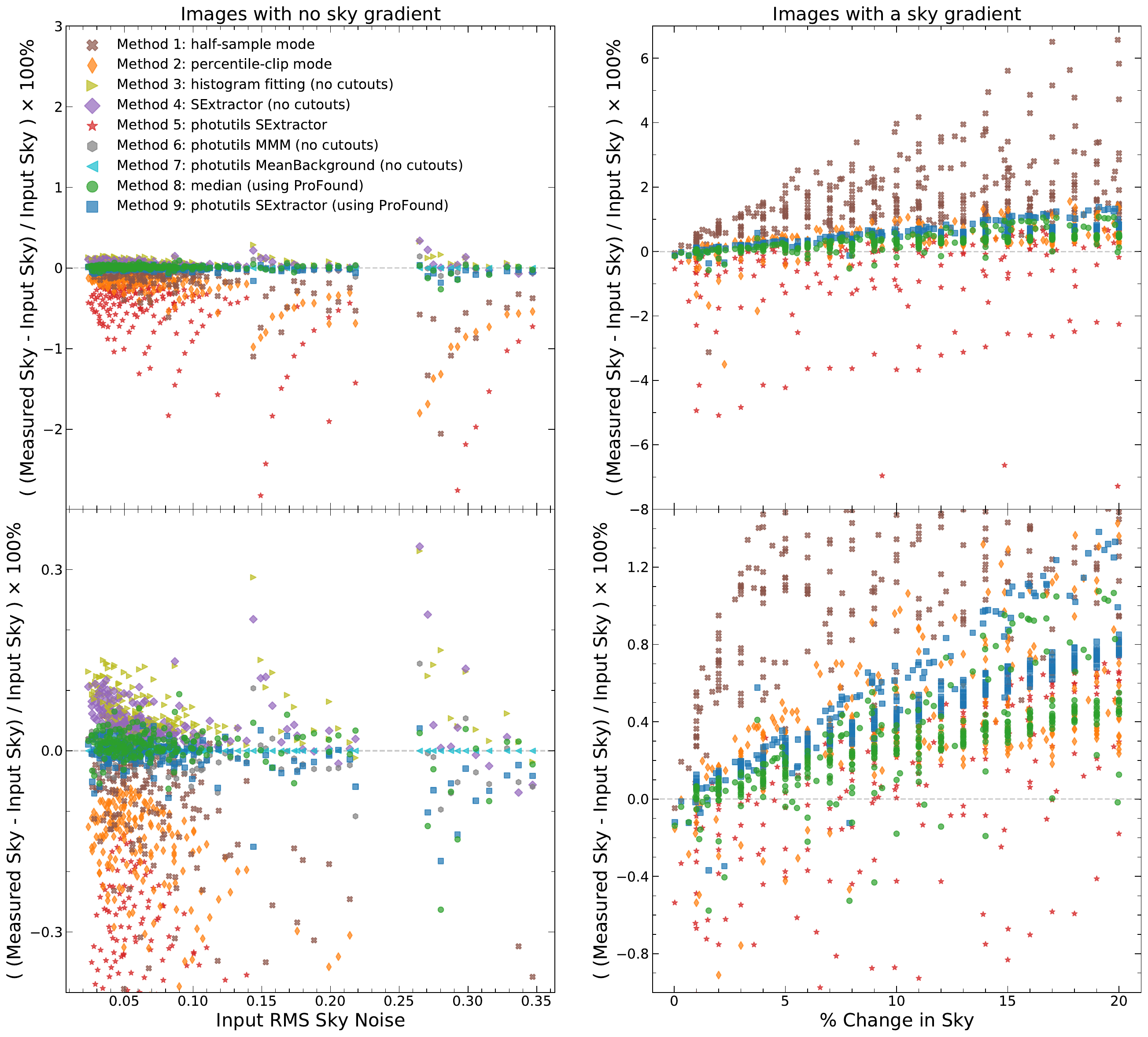}
 }

\vspace*{+0.20cm}
\n \caption{
{\bf [Left panels; Bottom panel (c) is an enlargement of top (a)]:} Systematic
comparison of our nine independent sky-SB measurement methods against the known
inputs from 784 simulated WFC3/IR F125W images with known sky-SB levels and
realistic CR levels, star- and galaxy counts superimposed. Methods 8 and 9
that are based on \ProFound\ (green and blue-filled squares) generally provide
measured sky-SB levels that differ \cle 0.1--0.2\% from the known simulated
input values, as long as the WFC3/IR sky-noise is \cle 0.24 \eminpixsec, \ie\ 
exposures with \texp$\ge$200 sec in the broad-band filters. 
{\bf [Right panels; bottom panel (d) is an enlargement of top (b)]:} As in
Fig.~\ref{fig:fig8}ac, but simulating images with up to 20\% sky-SB gradients
corner-to-corner or edge-to-edge. These simulated image gradients are in excess
of any gradients expected in the Zodiacal Light on arcmin scales (\S\ 
\ref{sec423}). For images with gradients that are 5--10\% of the lowest sky-SB
value, methods 8 (which uses medians based on \ProFound) and 9 (Photutils using
\SExtractor\ and \ProFound) provide measured sky-SB levels that are
generally within 0.4\% from the simulated input values when the gradients are
\cle 10\%, and within 0.2\% from the input values when the gradients are \cle
5\% corner-to-corner or edge-to-edge across the simulated image.}
\label{fig:fig8}
\end{figure*}


\n true image mode or peak. {\it Instead, the LES method uses the lowest 5\% of
all boxes} that were selected to not contain barely detectable faint objects,
bright object outskirts, and/or image defects. We show in \S\ \ref{sec423} that
using {\it this 5\% threshold of all boxes with the lowest sky-SB values} most
accurately recovers the known input sky-SB values from the simulations in \S\
\ref{sec421}.

In \citetalias{Carleton2022}, we use the best of these algorithms to estimate
the LES values for all 34,400 WFC3/IR images in the F125W, F140W, and F160W
images in Table~\ref{tab:tab3}. To check if the distribution of the green boxes
in Fig.~\ref{fig:fig7} showed any noticeable correlation with the GDC
structure across the detector, we collapsed the green boxes from all 34,400
WFC3/IR images into one 26$\times$26 density grid. The WFC3/IR detector is on
HST's optical axis, and due to its $\sim$24\degree\ tilt about its x-axis, its
pixels cover each a somewhat rectangular area of the sky with a $\pm$4.5\% GDC
variation that stretches mostly along the y-axis \citep[see \eg\ Fig. B.3
of][]{Dressel2021}. The average locations of the green boxes to estimate LES
sky-SB values across 34,400 exposures showed {\it no} correlation with the GDC
structure across the detector. Nor did they avoid the ``fiducial point'' in the
WFC3/IR center where the GO user is recommended to place their small targets,
suggesting that the average GO observer uses WFC3/IR more as a survey
instrument than for individual (large) object imaging. In summary, the green
boxes in Fig.~\ref{fig:fig7} sample the LES values rather randomly across the
detector when considering all 34,400 WFC3/IR exposures. This then justifies the
sky-SB estimation in the undrizzled flat-fielded $\_flt$/$\_flc$ images (\S\
\ref{sec41}). We will henceforth assume that the LES method estimates the
lowest sky-SB in each exposure well enough to proceed.


\n \subsubsection{Algorithm- and Scene-Dependent Reliability of Recovered
Input Sky-SB} \label{sec423} 

\sn Here, we summarize the main algorithmic results and discuss the best
algorithms to estimate the Lowest Estimated Sky-SB in the absence and presence
of sky-SB gradients. By implementing the filtered gridding method of \S\
\ref{sec422}, we are also able to indirectly detect gradients across an image.
We flag and remove sub-regions that are brighter than the mean sky-SB level +
the mean sky-SB rms level, then remove images where more than 30\% of
sub-regions are flagged as non-sky regions. A comparison of the results from
these nine methods is shown in Fig.~\ref{fig:fig8}. 

In the {\it absence} of sky-gradients, Fig.~\ref{fig:fig8}ac show that method 3
(based on histogram fitting; yellow triangles), and methods 8 and 9
\citep[based on \ProFound-catalogs;][green and blue squares]{Robotham2017}
generally provide measured sky-SB levels to well within 0.2\% from the known
input values. In methods 8 and 9, the simulated sky-SB values are best recovered
when the {\it lowest 5\% of the sky boxes' SB-values are used as an estimator}.
Method 4 (based on \SExtractor-catalogs; purple diamonds) performs worse at
sky-levels with lower S/N-ratio. For methods 8 and 9, the
$\vert$observed--simulated$\vert$ difference is as small as $\sim$0.1\%, as
long as the WFC3/IR sky-noise is \cle 0.24 \eminpixsec\ (\ie\ \texp$\ge$200 sec
for its broad-band filters).

In retrospect, the resulting $\sim$0.1\% accuracy of estimating the known 
input sky-SB values with the best algorithms makes sense. In \S\ \ref{sec422}
we estimated that in the best possible case of a clean flat sky over an entire
image, each of the 676 boxes of 1521 pixels could estimate its sky-value to an
accuracy of 2.6\% of the sky-SB in that box. For a perfectly flat sky these
methods may thus recover the input sky-SB to no better than 
2.6\%/$\sqrt{676}$$\simeq$0.1\% of the input sky-value. This is indeed about
the best accuracy at which methods 8 and 9 recover the known input sky-SB
values in Fig.~\ref{fig:fig8}ac. 

Fig.~\ref{fig:fig8}bd show that in the {\it presence} of simulated sky-SB
gradients that amount to 10--20\% of the lowest sky-SB value edge-to-edge,
method 8 provides measured sky-SB levels generally within $\sim$0.4\% from the
simulated input values when the gradients are \cle 10\%, and within $\sim$0.2\%
when the imposed sky-SB gradients are \cle 5\% edge-to-edge or corner-to-corner.
These simulated image gradients are in excess of any gradients expected in the
ZL on arcmin scales. Significant gradients with amplitudes \cge 10\% can be
caused in real HST images by: a) straylight (\S\ \ref{sec43}); b) large
artifacts; or c) large targeted objects with real astrophysical gradients
(Appendix \ref{secAppB3}). SKYSURF flags images with strong gradients or large
targets or artifacts, estimates their LES values (at lower accuracy) to help
diagnose these causes of straylight and straylight gradients, and where needed
discards them for sky-SB analysis or object counts. 

In conclusion, the histogram-fitting method 3 provides LES sky-SB values with
the simplest algorithm available that is accurate to within $\sim$0.2\% in
SKYSURF images {\it without gradients} (Fig.~\ref{fig:fig8}c), but it is not
proven in the presence of strong gradients. The percentile-clip method 2 works
better in the presence of gradients, but has larger errors (\cle 0.3\%) in
images without gradients. Hence, in the first science analysis of
\citetalias{Carleton2022} we will use method 3 and discard images with higher
sky-SB levels and significant gradients. Methods 4, 8 and 9 are considerably
more CPU intensive, since they require complete \SExtractor- or
\ProFound-catalogs object catalogs to be made in advance. Overall, method 8 
can provide the most accurate LES sky-SB values to within $\sim$0.1\% in
SKYSURF images {\it without gradients}, and to within $\sim$0.2--0.4\% in
images {\it with gradients that are less than 5--10\%} edge-to-edge or
corner-to-corner, respectively. This method will be applied to all individual
SKYSURF images in O'Brien \etal\ (2022, in preparation), which requires the
entire SKYSURF database to be processed with \ProFound.


\n \subsection{Orbital Sky-SB Dependence and Possible Sources of Straylight}
\label{sec43} 

\sn We define our usable sample of SKYSURF images as a function of orbital
parameters that may affect the measured image sky-SB values: Limb Angle (LA),
Moon Angle (MA), Sun Angle (SA), Sun Altitude \alphaSun\ above the Earth's
Limb, the Illuminated Earth Fraction (IEF), HST's Roll Angle (RA), and HST's
position in the Continuous Viewing Zone (CVZ). The MA and SA are defined as
the angular distance between the HST pointing-direction and the center of the
Moon or Sun during the exposure, respectively. Sun Altitude \alphaSun\ is
defined as the Sun--Earth--HST angle, which determines the amount of Sunlight
scattered off the bright Earth that can reach the HST aperture. SKYSURF does
ephemeris and visual checks to identify images that got too close to: (a) the
Earth's Limb, including those taken at the bottom of the CVZ, (b) the Sun;
and/or (c) the Moon. Any one of these can corrupt the sky-SB measurements, and
in some cases also the SKYSURF source catalogs. Some, but not all of this
information is available in the HST Engineering telemetry data, \ie\ the
``\jit''-files that come with the raw FITS images. Where needed, we compute the
parameters LA, MA, SA, and \alphaSun\ for each SKYSURF exposure. Appendix
\ref{secAppA1} gives details on how the HST orbital parameters are calculated
at any given time since its April 24, 1990 launch or its subsequent Shuttle
Servicing Mission reboosts. In our current analysis, only sky-SB measurements
from images with sufficient blank sky are used for the final SKYSURF analysis
(\S\ \ref{sec42}). 

A full ray-tracing of the HST straylight from first principles is beyond the
scope of the SKYSURF project, and so we start with an {\it empirical} approach
to trace the sky-SB measurements as a function of the orbital parameters LA,
MA, SA, and \alphaSun, and define the ranges in these parameters where the
SB-measurements are generally not significantly enhanced above the minimum in
each direction, as described in \citet{Caddy2021}. For the 249,861
ACS+WFPC2+WFC3 images, we calculate these angles and identify the angle ranges
that may have resulted in enhanced straylight levels. Such exposures may still
be usable for panchromatic object counts, in which case they are used in our
modified drizzled pipeline (\S\ \ref{sec46}). 

Some examples are shown in Fig.~\ref{fig:fig6}a--\ref{fig:fig6}b. Here, 
Fig.~\ref{fig:fig6}a shows the WFC3/IR F125W sky-SB values of the first 
half-orbit exposure at the start of an orbit, which typically uses the darker
orbital part, so that the sky-SB values do not change much over the total
exposure length. The blue data points show the sky-SB values of {\it individual
WFC3/IR ramp read-outs} as estimated with the best algorithms in \S\
\ref{sec423}. The orange line in Fig.~\ref{fig:fig6}a shows the exposure
time-weighted average over all ramps. The default \texttt{calwf3} calibration
procedure of this exposure uses all these values, which results in the grey
dashed line as the sky-SB value of the final $\_flt$ file, which is biased by
the ramps with the highest sky-SB. When we run this through \texttt{calwf3}, we
find the green dashed line as the sky value of the final $\_flt$ file, which
still does not reflect the lowest possible sky-SB value for that exposure well.
For WFC3/IR, we therefore adopt the average of the {\it three} ramps with the
lowest sky-SB values (blue line), excluding the initial clearing readouts. We
consider the blue line to be the more representative value of the Lowest
Estimated Sky-SB for that exposure. 

Fig.~\ref{fig:fig6}b shows the WFC3 F125W sky-SB at the end of a typical HST 
orbit, which in this case is the second half-orbit exposure observed directly
after the one shown in Fig.~\ref{fig:fig6}a. The procedures and lines are the
same as in Fig.~\ref{fig:fig6}a, but the differences are now much larger
because in this second half-orbit WFC3/IR exposure the sky-SB values increase
significantly towards the end of the orbit due to Earthshine. The resulting
blue line indicates the final adopted sky-SB value compared to the 
\texttt{calwf3} average (orange line). Over the course of this orbit, the
dispersion of our reprocessed $\_flt$ sky-SB values is smaller than the
dispersion in the original $\_flt$ sky-SB values, so our procedure results in
more consistent estimates of the Lowest Estimated Sky-SB than the default
\texttt{calwf3} pipeline. 

\begin{figure*}[!hptb]
\n \cl{
 \includegraphics[width=1.000\txw]{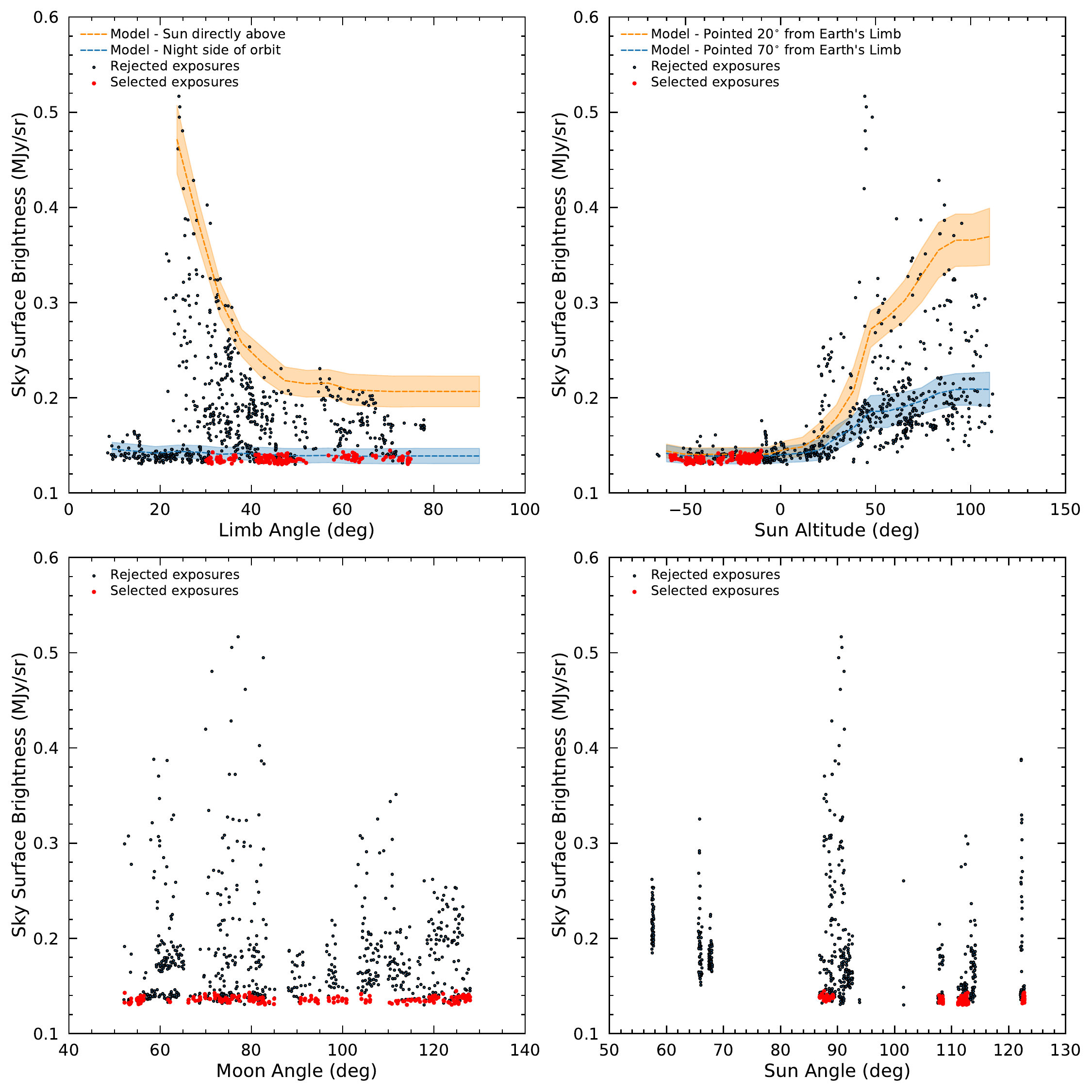}
}

\vspace*{+0.00cm}
\n \caption{
Combined boundaries in orbital parameters that yield the darkest sky-SB values
(red points) as defined using the GOODS North ACS F850LP dataset of \S\
\ref{sec43} \& Appendix \ref{secAppA2}. 
{\bf (a) [Top Left]:}\ Total sky-SB vs. Earth's Limb Angle (LA). The orange
model shows the elevated sky-SB due to Earthshine for Sun Altitude
\alphaSun$\simeq$90\degree\ on the orbital day side with Sun Angle fixed at
SA$\sim$90\degree. A strong exponential increase is seen for LA\cle
40--45\degree, but even fields with LA\cle 70--80\degree\ on the day side can
have an elevated sky-SB. The blue line plus uncertainty wedge indicates the
\citet{Caddy2022} model on the night side for \alphaSun$\simeq$--10\degree, and
shows a marked decrease in sky-SB. 
{\bf (b) [Top Right]:}\ Sky-SB vs. Sun Altitude \alphaSun. The orange and blue
models show that lower LA's result in steeper relationships for \alphaSun\cge
10--20\degree. For \alphaSun\cle --10\degree, the sky-SB is lowest and remains
approximately constant. 
{\bf (c) [Bottom Left]:}\ Sky-SB vs. Moon Angle. For MA\cge 50\degree, the Moon
Angle is not a major driver of straylight. 
{\bf (d) [Bottom Right]:}\ Sky-SB vs. Sun Angle. The large scatter for each Sun
Angle cluster is primarily due to the variation in Sun Altitude and Limb Angle
for each exposure. The minimum estimated sky-SB closest to the true ZL+DGL+EBL
level is reached for observations that combine: LA\cge 30\degree, \alphaSun\cle
--10\degree, SA\cge 80\degree, {\it and} MA\cge 50\degree\ (red points).}
\label{fig:fig9}
\end{figure*}

The GOODS North data of \citet{Giavalisco2004} were used as an initial case
study to investigate the sky-SB in HST images as a function of the orbital
parameters, and reduce the impact on sky-SB measurements of: (a) Earthshine due
to the proximity of the Earth's Limb; (b) the Sun Altitude above the Earth's
Limb to minimize Sunlight scattered off the bright Earth; (c) the Moon; {\it
and} (d) the Sun during each exposure to minimize their straylight. The data
and model predictions are shown together with their uncertainty wedges in
Fig.~\ref{fig:fig9}. Details are given in \citet{Caddy2021}, \citet{Caddy2022},
and Appendix \ref{secAppA2}. The ACS F850LP-filter data in GOODS-North were
chosen for this study, because: (1) it is the closest filter in central
wavelength to the WFC3/IR filters used to present our first SKYSURF results;
(2) the high Ecliptic Latitude (\bEcl$\simeq$57\degree) of GOODS-North reduces
the impact of true ZL variations on the analysis; and (3) its very large sample
size. These data consist of sky-SB measurements from 1018 ACS F850LP images in
the GOODS North field, which is in HST's CVZ and covers a range of Earth Limb
Angles, Sun Altitudes, Sun Angles, and Moon Angles recorded in the HST
telemetry data. Only exposure times in the range of 240--1800 seconds were used
for this study. 

In Fig.~\ref{fig:fig9}a--\ref{fig:fig9}d, the measured sky-SB values are
colored by orbital parameters that meet the SKYSURF criteria for avoiding
straylight contamination. Black points are exposures that do {\it not} meet our 
criteria and red points are those that do. In summary, the {\it combined
boundaries in orbital parameters that yield the darkest sky-SB values} as
indicated by the red points are: (a) Earth's Limb Angle LA\cge 30--40\degree
to avoid Earthshine; (b) Sun Altitude above the Earth



\vspace*{-0.30cm}
\hspace*{-0.00cm}
\n


\n \alphaSun\cle --10\degree\ (\ie\ orbital night side) to minimize Sunlight
scattered off the bright Earth; (c) Moon Angle MA\cge 50 \degree; {\it and} (d)
Sun Angle SA\cge 80\degree\ to avoid straylight into the HST optics,
respectively. 

Imposing these constraints as detailed below, the standard deviation of the
GOODS-N sky-SB values is reduced from 0.055 MJy/sr to 0.002 MJy/sr with an
overall decrease in mean sky-SB of 0.138--0.179 MJy/sr, respectively. These
selection criteria can thus minimize the impact of Earthshine, Sun Altitude, 
Sun Angle and Moon Angle on the HST's sky-SB measurements. These limits are
more extreme than those recommended by the HST Instrument Handbooks, and we
impose them on all our panchromatic sky-SB estimates for ACS/WFC, WFC/UVIS and
WFC/IR that sub-select HST's orbital phases with the lowest straylight 
(O'Brien \etal\ 2022, in preparation). Further details on straylight resulting
from the Earth's Limb, the Sun Altitude above the bright Earth and the Sun
Angle with respect to HST's pointing direction, as well as the Moon angle are
given in \S\ \ref{secAppA2}.


\n \subsection{Definition and Strategy for Treatment of Sky-SB Gradients} 
\label{sec44} 

\sn We need to estimate and preserve the Lowest Estimated Sky-SB (LES) level in
all SKYSURF images for subsequent sky-SB analysis. This is currently {\it not}
the default option in AstroDrizzle~\footnote{
\url{https://hst-docs.stsci.edu/drizzpac/chapter-5-drizzlepac-software-package/5-2-astrodrizzle-the-new-drizzle-workhorse}}\ 
\citep{Avila2015}. Hence, we run AstroDrizzle on the entire SKYSURF database
with the {\it LES sky-SB preserved} in each sequence of observations in a given
filter that was taken in the same visit. HST exposures can have low-level
gradients if they were pushed too close to the Earth's limb, or if they were
taken at too small Moon Angles or Sun Angles. An example is given in
Fig.~\ref{fig:fig6}b. AstroDrizzle normally removes a gradient surface before
drizzling, using the \texttt{MDRIZSKY} keyword that preserves the sky, so the
subtracted sky-level or sky surface can be reinstated. We use \ProFound\
\citep{Robotham2017} to map low-level gradients with spline surface fits to the
sky-SB in between all discrete objects detected in the SKYSURF images,
preserving the spline surfaces in separate FITS files for later diagnostics. In
this process, we subtract the gradient from the \ProFound\ surface, but {\it
not} the lowest reliable DC-level of the sky-SB, as that is one of the key
parameters SKYSURF is measuring. Details are given by Tompkins \etal\ (2022, in
preparation). 

Images at low Galactic latitude (\bII\cle 30\degree), or images of large
bright-object targets may have real astrophysical gradients in their sky-SB due
to substantial DGL, such as Galactic star-formation regions and the outskirts
of nearby galaxies. In that case, any gradient must not be removed (\eg\ the
top right two panels in Fig.~\ref{fig:fig7}). This is monitored and flagged by
SKYSURF (blue column in Fig.~\ref{fig:fig3}), with feedback to the SKYSURF 
SB-measurements and drizzling database and input from the LA, MA, SA monitoring
in \S\ \ref{sec43}, to help identify which images at $\vert$$b_{II}$$\vert$\cle
30\degree\ may be dominated by DGL or by large bright-object targets, and not
by orbital straylight.


\n \subsection{Definition of SKYSURF Drizzle Families and Drizzle Footprints}
\label{sec45} 

\sn SKYSURF drizzles {\it all} images to the same pixel scale of 0\arcspt
060/pixel, including the single exposures, to provide proper photometry for
both sky-SB measurements and discrete object catalogs (\S\ \ref{sec41}). This
includes the latest practices in AstroDrizzle for the identification and
masking of CRs and Charge Transfer Efficiency (CTE) trails as a function of
image-stack depth. Appendix \ref{secAppB1}--\ref{secAppB2} describe the
application of CR-removal and Charge Transfer Inefficiency trail removal over
{\it wider} apertures in the \texttt{calwf3} pipeline, which is applied to
SKYSURF's drizzling process. 

To drizzle the SKYSURF database, we generated instrument+filter-specific
on-sky footprints for all for 220,657 images with \texp$\ge$200 sec (out of a
total of 249,861 SKYSURF images), starting with their APT files. This allowed
us to organize the SKYSURF database in preparation for the drizzling process,
and to assess its total area (\S\ \ref{sec32}), which we need for accurate
galaxy counts averaged over as many independent HST fields as possible (\S\
\ref{sec252}). 

In total, there are 11,965 APT footprints to be drizzled in SKYSURF using a
linking distance of \cle 6 arcminutes. Each group contains all images in all
available filters, whose centers are within 6 arcminutes of their nearest
neighbor. The most frequently observed group has 10,761 exposure files (\ie\
the HUDF), while the least frequently observed groups have only one exposure in
one filter and are mostly SNAPshots. The images in each group in the same
filter {\it and} instrument are then drizzled into separate mosaics. With an
average of $\sim$4--5 exposures per HST orbit (Fig.~\ref{fig:fig5} \&
Table~\ref{tab:tab4}), the SKYSURF database amounts to $\sim$55,000 HST orbits,
or the equivalent of \cge 15 years of HST observing time assuming an average of
$\sim$3600 science orbits per year.\deleted{(Each calendar year has about 5500
HST orbits, but a fraction is not used for science due to calibration or
engineering requirements, South Atlantic Anomaly (SAA) transitions, HST safings
or downtime, etc.).} We use all 11,965 APT footprints together with SKYSURF's
initial SB-measurements to assess which images are useful for the final set of
reliable sky-SB measurements, and for the final object \SExtractor\ and
\ProFound\ catalogs and all-sky object counts (Appendix \ref{secAppC}). Using
the same matching algorithm, the 11,965 APT footprints contain the 4,858
independent fields of \S\ \ref{sec32} that are more than 1\degree\ away from
the nearest APT footprint. 

\n \subsection{Drizzling all SKYSURF Images in AWS with Lowest Estimated 
Sky-SB Preserved per Image Family} \label{sec46} 

\sn The last unconventional step critical for SKYSURF is that when drizzling the
images, we preserve the sky-level into all images that AstroDrizzle normally
removes. This sky-level as defined in \S\ \ref{sec42} is carefully monitored by
SKYSURF as a function of orbital phase for all multiple exposures in a given
filter visit per target (\S\ \ref{sec43}). 

In a first drizzle run, the \ProFound\ sky-surface maps of \S\ \ref{sec42}
are subtracted from each and its LES level is added back in. In the second
drizzle run, all images in a given filter from the first run have their own
LES value subtracted, followed by adding to each image the lowest LES value
from that visit that went into the mosaic. Hence, each mosaic is drizzled
while reinstating the lowest sky-SB value from that visit (\S\
\ref{sec42}--\ref{sec43}). In the case of WFC3/IR --- where we have more
detector reads per exposure in a given visit --- the average over the lowest
three sky-SB values from that exposure gets reinstated to represent its sky-SB.
The justification of this procedure was given in Fig.~\ref{fig:fig6},
Fig.~\ref{fig:fig9}, and \S\ \ref{sec43}. In all cases, the subtracted
\ProFound\ sky surface maps and the reinstated constant sky-SB levels are
preserved as FITS files or in the FITS header, respectively, so future users
can apply them differently for other purposes. Mosaic weight-maps are also
created in this process, which preserve the information on the signal-to-noise 
ratio in each mosaic pixel that follows from the sky-SB levels in the 
original images that were drizzled.

We use Amazon Web Services~\footnote{\url{https://aws.amazon.com/}} for mass
cloud computations without transferring all data files whenever possible,
since the entire HST Archival database resides in AWS and is being kept
up-to-date there by STScI. We use AWS to run our modified ACS and WFC3
pipelines on the HST Archival data until the end-product is as reliable as it
can be. Most of the code is Python, which works well in AWS. AWS can run code
or pipelines in Python, R, or C$^{++}$. A few SKYSURF tasks originally
developed under a FORTRAN or IRAF environment do not run on AWS and are run on
the SKYSURF servers at our home institutions. 

On AWS we can reprocess the HST ACS/WFC and WFC3/UVIS and IR Archive within a
few months at modest costs for each iteration. We do this for the entire SKYSURF
dataset. Running \ProFound\ on all images in the entire SKYSURF database 
takes one full month on AWS. Central storage space of 40 TB houses the
different levels of datasets for processing and distribution at ASU, plus 
another 40 TB to store SKYSURF's expanded data products. In total, 80 TB of
disk space at ASU and ICRAR hosts the final SKYSURF database and products.
These will be made available via MAST. Details on SKYSURF's drizzle products as
they become available and their applications are given by Carter \etal\ (2022,
in preparation).



\vspace{-0.00cm} 
\hspace{-0.50cm}
\n\begin{figure*}[!hptb]
\n\cl{
 \includegraphics[width=1.00\txw,angle=-00]{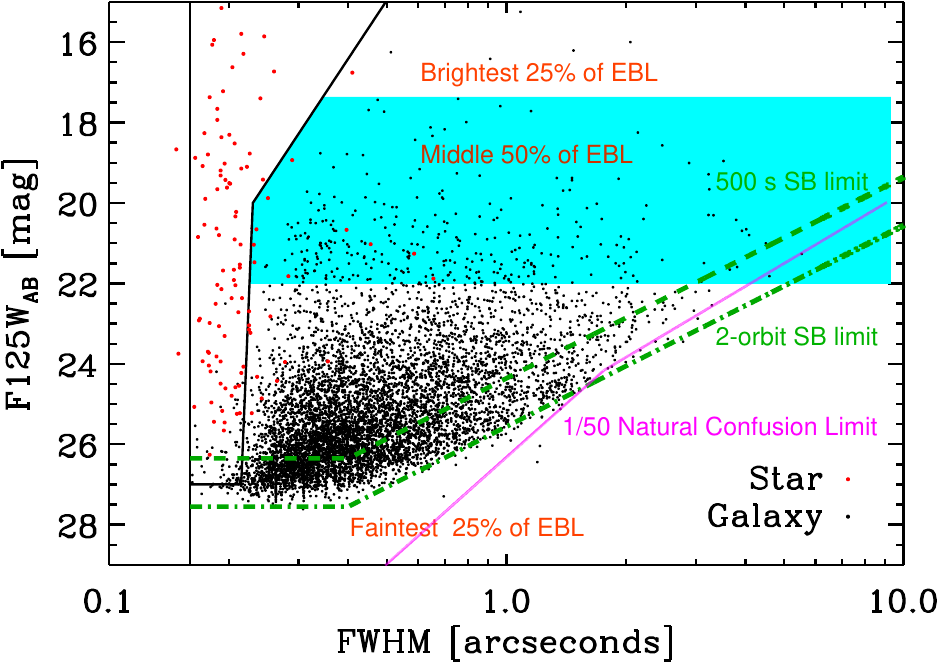}
}

\vspace*{-0.00cm}
\n \caption{
The star-galaxy separation procedure shown here for WFC3 F125W images 
\citep[adapted from][]{Windhorst2011} (``\citetalias{Windhorst2011}''). Plotted
are total AB-magnitude vs. \SExtractor\ image diameter $FWHM$. Objects with
image diameter less than the PSF-FWHM ($\simeq$ 0\arcspt 14; 
\citetalias{Windhorst2011}) are image defects and are discarded. Stars
(thin vertical filament of red dots) and galaxies (black dots) are separated
using cuts in FWHM and magnitude, similar to the black slanted line above from
\citetalias{Windhorst2011}. This is effective at separating stars and galaxies in WFC3/IR data, which
were separated using FWHM cuts in various filters. At AB$\simeq$18 mag, the star
counts and galaxy counts reach similar surface densities at intermediate to
high Galactic latitudes (Fig.~\ref{fig:fig11}), and star-galaxy separation at
these fluxes is straightforward at HST resolution. The star-galaxy separation
becomes less reliable for fluxes fainter than AB$\simeq$26 mag. The red arrows
indicate the F125W flux ranges where the brightest 25\%, the middle 50\%, and
the faintest 25\% of the discrete EBL level, respectively, are produced 
(Fig.~\ref{fig:fig2}d). Green dashed lines indicate the SB-limits for an average
single SKYSURF F125W exposure of \texp=500 sec (Fig.~\ref{fig:fig5}) and for
the 2-orbit F125W images of \citetalias{Windhorst2011}, whose data is plotted here. The pink line
indicates the natural confusion limit derived from the integrated J-band counts
in Fig.~\ref{fig:fig2}c (represented here as a broken power law), as discussed
in \S\ \ref{sec47}. Both these limits are relevant for estimating if a
significant fraction of low-SB galaxies may have been missed at faint fluxes to
explain any significant diffuse EBL \citepalias{Carleton2022}.}
\label{fig:fig10}
\end{figure*}




\vspace{-0.00cm} 
\hspace{-0.50cm}
\n\begin{figure*}[!hptb]
\n\cl{ 
 \includegraphics[width=1.000\txw,angle=-00]{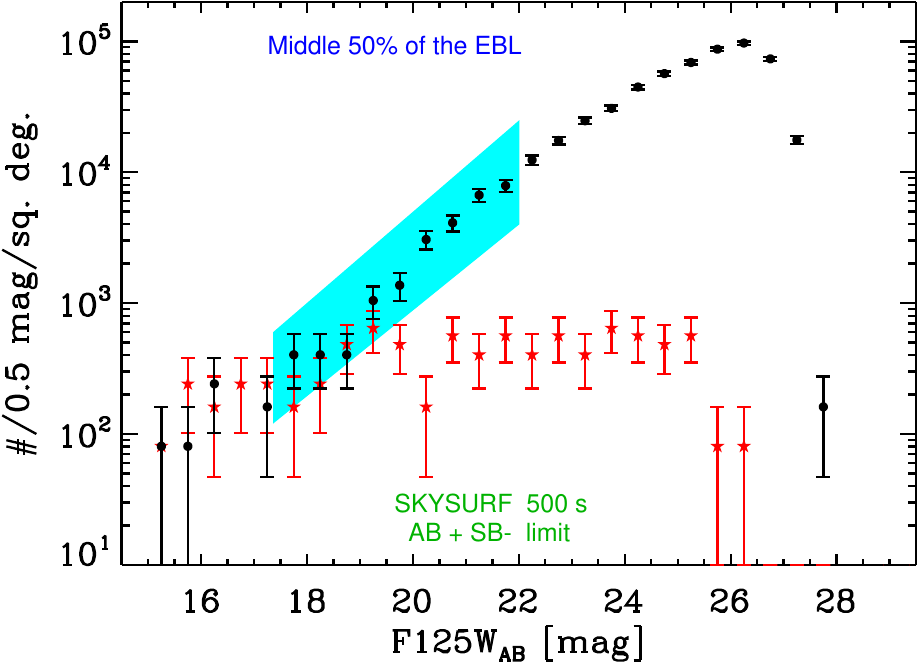}
}

\vspace*{-0.00cm}
\n \caption{
Differential star counts {\bf (red asterisks)} and galaxy counts {\bf 
(black-filled circles)} from the WFC3 ERS F125W images adapted from
\citet{Windhorst2011}, with the star-galaxy separation optimized from
Fig.~\ref{fig:fig10}. At AB$\simeq$18 mag, the star counts and galaxy counts
reach similar surface densities at intermediate to high Galactic latitudes, but
star-galaxy separation at these fluxes is straightforward at HST resolution 
(Fig.~\ref{fig:fig10}). The blue box indicates the 17.4\cle \JAB\cle 22 mag
range where the {\it middle 50\%} of the discrete EBL is produced in the J-band
(Fig.~\ref{fig:fig2}d). The vertical green line indicates the total flux level
of AB$\simeq$22 mag. Brighter than this limit, even shallow SKYSURF images with
typical exposures times \texp$\sim$500 sec are substantially complete for all
galaxies with FWHM-sizes\cle 3\arcspt 0 (Fig.~\ref{fig:fig10}).}
\label{fig:fig11}
\end{figure*}


\n \subsection{Star-Galaxy Separation, Catalog Reliability and Completeness} 
\label{sec47} 

\sn Our second main SKYSURF goal is to have highly reliable panchromatic HST
object catalogs across the sky (Fig.~\ref{fig:fig3}--\ref{fig:fig4}). SKYSURF
therefore needs to define accurate, conservative completeness limits for each
object catalog. We need to determine at what magnitude bin the counts in each
filter and field start to turn over from the expected power-law slope (see
\citealp{Windhorst2011}, \citealp{Driver2016b}; Fig.~\ref{fig:fig2} \&
\ref{fig:fig10}--\ref{fig:fig11} here). As a preliminary test, we consider
images taken as part of WFC3/IR ERS program here \citepalias{Windhorst2011}.
Because of SKYSURF's large dynamic range in flux, we can afford to cut off each
individual filter catalog in each HST field at rather high significance levels.
From the deeper fields, we can then determine at which flux levels each
shallower catalog becomes incomplete when we do the star and galaxy counts in
each field, typically at the \cge 5--6$\sigma$ level for point sources. When
all catalogs are combined, SKYSURF can provide accurate object counts for
16\cle AB\cle 28 mag. Details are given in Tompkins \etal\ (2022, in
preparation), Carleton \etal\ (2022b, in preparation), and Goisman \etal\ (2022,
in preparation), including simulations to quantify incompleteness as a function
of total flux, size, and SB. 

{\bf Star-Galaxy Separation Method:}\ The black slanted line in 
Fig.~\ref{fig:fig10} illustrates SKYSURF's star-galaxy separation method of
\citetalias{Windhorst2011}, where objects were classified as stars in at least
three available filters from the 10-band filter set. The average SKYSURF field
will have on average only $\sim$two filters available for star-galaxy separation
(Table~\ref{tab:tab4}). Hence, SKYSURF's star-galaxy separation process will
be modified to work on fewer filters as the image drizzling, catalog production
and object counting proceeds over all SKYSURF fields. The blue wedge in 
Fig.~\ref{fig:fig10} indicates the flux range 17.4\cle AB\cle 22 mag in F125W
where galaxies contribute the central 50\% of the EBL integral. The red arrows
indicate the F125W flux ranges where the brightest 25\%, the middle 50\%, and
the faintest 25\% of the discrete EBL are produced, as determined from 
Fig.~\ref{fig:fig2}d (\S\ \ref{sec23}). 

{\bf Catalog Reliability:}\ For 17.4\cle AB\cle 22 mag, a total of 37 objects
were classified as stars (red dots) to the left of the black slanted--vertical
line in Fig.~\ref{fig:fig10}. To the right of the black slanted line, there are
328 objects (black dots) classified as galaxies for 17.4\cle AB\cle 22 mag, and
10 objects that were classified as stars in other filters, but are
misclassified in the galaxy region in the F125W filter, although some of these
could still be stars. Using such a star-galaxy separation method over this
relatively bright magnitude range at HST resolution thus yields a fraction of
objects misclassified as galaxies of $\sim$10/328$\simeq$3.0\%. The ratio of
classified stars-to-galaxies in this magnitude range is about
$\sim$37/328$\simeq$11\%. Because the star-to-galaxy ratio remains small at
most Galactic latitudes with $\vert$$b_{II}$$\vert$\cge 30\degree, SKYSURF's
{\it galaxy} samples will thus in general be more reliable
($R$$\simeq$1--(10/328)$\simeq$97\%) than its {\it stellar} samples (reliability
$R$$\simeq$1--(37/47)$\simeq$78\%). 

{\bf Catalog Completeness:}\ The two green dashed lines in Fig.~\ref{fig:fig10}
show the SB-limits for a typical single SKYSURF exposure of \texp=500 sec and
for a 2-orbit exposure stack, which are $\sim$25.6 and $\sim$26.8 \magarc,
respectively. We use the distribution of points in Fig.~\ref{fig:fig10} to
estimate the catalog completeness. In the average SKYSURF exposure time of
$\sim$2 orbits per filter, we expect the SKYSURF completeness limit for compact
objects to be AB\cle 26--27 mag depending on the filter used, and \cle 28 mag
for the deeper HST fields. For many of the shallowest SKYSURF exposures, the
completeness limit is AB\cle 25.5--26 mag
(Tables~\ref{tab:tab1}--\ref{tab:tab3} and Fig.~\ref{fig:fig11}). 

The pink line in Fig.~\ref{fig:fig10} indicates the ``natural confusion'' limit
derived from the {\it integrated} J-band counts in Fig.~\ref{fig:fig2}c
(represented here as a broken power law). For galaxy FWHM-sizes larger than the
natural confusion limit at any given total flux, there is a \cge 2\%
probability of objects overlapping {\it due to their own finite object sizes},
rather than due to the much smaller instrument PSF FWHM. Object finding
algorithms may start to have issues deblending a noticeable fraction of
(lower-SB) objects at sizes larger than this limit, if such objects existed
\citep{Windhorst2008, Windhorst2021}. Indeed, very few objects are detected to
the right of {\it either} the SB-limit and the natural confusion limit in
Fig.~\ref{fig:fig10}. The most conservative of the two limits thus seems to be
bounding the detected galaxy samples at any \re-value. (For deeper HST images,
the natural confusion limit remains fixed, but the SB-limit improves with image
depth, so the natural confusion limit thus becomes as important as the
SB-limit). Together, these two limits are relevant for estimating whether a
significant fraction of low-SB galaxies may have been missed at faint fluxes to
explain a significant excess of diffuse EBL
\citepalias[Fig.~\ref{fig:fig1} and][]{Carleton2022}. Further discussion is
given by \citet{Windhorst2021} and Kramer \etal\ (2022, in preparation).

We use the two-orbit F125W data in Fig.~\ref{fig:fig10} (lower green dashed
lines) to estimate the incompleteness due to SB-selection in an average 500 sec
SKYSURF exposure (top green dashed lines) and natural confusion (pink line). 
This allows us to estimate the amount of discrete, integrated, and extrapolated
EBL not yet subtracted from SKYSURF's sky-SB measurements in
\citetalias{Carleton2022}. For this calculation, we assume that the fraction of
faint low-SB objects missing in typical 500 sec F125W exposures due to
SB-selection {\it is mostly detected} in the deeper two-orbit data of 
\citetalias{Windhorst2011} shown in Fig.~\ref{fig:fig10}. This estimate thus
does not yet include any population of faint, very low-SB objects that might be
missing from the deepest HST images altogether to the right of these lines in
Fig.~\ref{fig:fig10}, which we address in \citetalias{Carleton2022}. To assess
the amount of possibly missing discrete, integrated, and extrapolated EBL in
Fig.~\ref{fig:fig2}bd due to SB-selection, we estimate the 500 sec catalog
incompleteness in each AB-magnitude slice from the fraction of objects {\it
known} to exist in deeper HST images between the two green SB-boundaries in
Fig.~\ref{fig:fig10}. This incompleteness amounts to $\sim$1\% at
AB$\simeq$22.0 mag and increases to $\sim$26\% at AB$\simeq$26.0 mag. For the
eight 0.5-mag slices between 22\cle \JAB\cle 26 mag, the incompleteness
correction due to SB-selection is represented by: 

\begin{equation}
Incompleteness\ Correction\ =\ 1.0\ +\ [1.00\ +\ 6.184\ (J_{AB} - 22.0\ 
mag)]/100\%, 
\label{eq:eq3}
\end{equation}

\n where the quantity between square brackets is the best fit to the estimated
percentage of {\it known} missing objects as a function of total AB-mag. For
AB\cge 26.5 mag, sample incompleteness is not considered here, because the
average 500 sec SKYSURF exposure does not reach fainter than this limit. We
note that Eq.~\ref{eq:eq3} would imply a SB-incompleteness of $\sim$45\% at
AB\cle 29 mag. The SB-incompleteness at AB\cle 29 mag is possibly that large,
given the much deeper HUDF data and SB-limits plotted in Fig. 2 of
\citet{Windhorst2008} and \citet{Windhorst2021}. Much deeper JWST images will
be needed to address the SB-incompleteness for objects with AB\cge 26--30 mag.
Further considerations of SB-selection against ultra-diffuse low-SB galaxies in
deep surveys are discussed in, \eg\ the SMUDGES survey of \citet{Zaritsky2019},
the Subaru Hyper Suprime-Cam survey of \citet{Greco2018}, and future SKYSURF 
studies. For our average shallow SKYSURF exposures and catalogs, we adopt
Eq.~\ref{eq:eq3} as the {\it minimum} SB-incompleteness correction that needs
to be applied for 22.0\cle \JAB\cle 26.5 mag to the discrete integrated EBL for
objects that are known to exist in deeper HST images, but that are missing in
the average 500 sec SKYSURF exposures. This is correction is applied in
\citetalias{Carleton2022} to fully estimate our iEBL+eEBL values.



\vspace{-0.00cm} 
\hspace{-0.50cm}
\n\begin{figure*}[!hptb]
\n\cl{ 
 \includegraphics[width=1.000\txw,angle=-00]{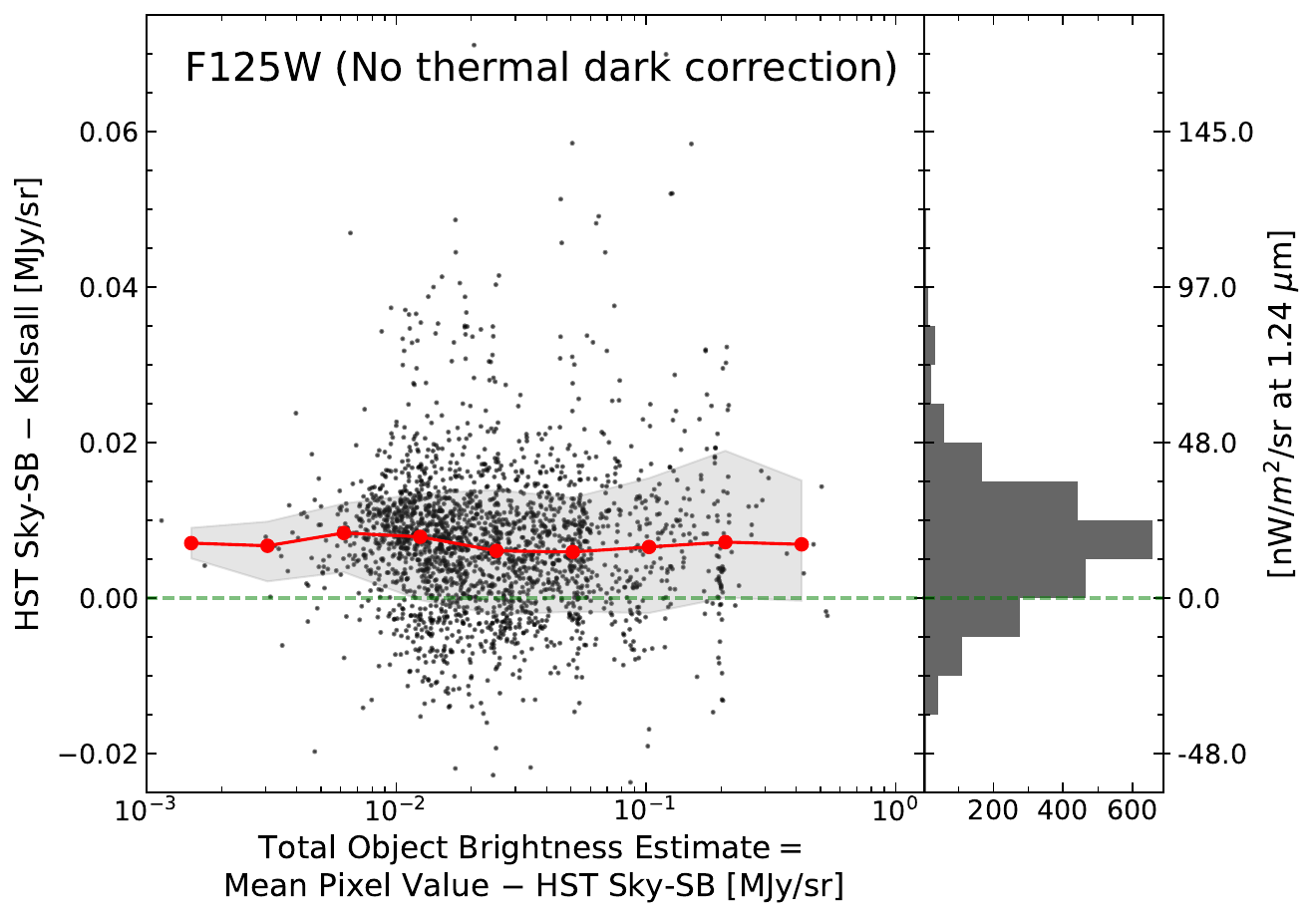}
}

\vspace*{-0.00cm}
\n \caption{
Difference in observed HST WFC3/IR F125W sky-SB and the \citet{Kelsall1998}
model prediction versus total {\it residual object brightness} after
subtracting from each image the best sky-SB estimate of \S\ \ref{sec42}. We
only include images taken with a Sun Altitude less than 0\degree\ and where no
more than 30\% of the sub-regions are masked in Fig.~\ref{fig:fig7}. To
estimate the total {\it residual} object brightness, we calculate the mean
pixel value of every image and then subtract the measured LES sky-SB level of
that image. The average 1.25 \mum\ HST--Kelsall sky-SB difference amounts to
$\sim$0.0071 MJy/sr or $\sim$17 \nWsqmsr. We find no significant trend between
the HST--Kelsall sky-SB in each image and its total sky-subtracted object
brightness. This confirms the validity of our LES method to measure sky-SB
values that are largely free of residual object flux.}
\label{fig:fig12}
\end{figure*}


\n \subsection{Proof-of-Concept: HST--Kelsall 1.25 \mum\ sky-SB Difference vs.
net Object SB in each Image} \label{sec48} 

\sn It is possible that stray light from discrete objects may contaminate our
sky-SB measurements even with the best algorithms of \S\ \ref{sec42}. This
could be due to residual instrumental effects not recognized in \S\
\ref{sec41}, or due to straylight from or the extended light-profiles of 
bright objects. While the HST cameras are in general well baffled against
straylight, there are known locations just inside or outside the detector FOV 
where light within the telescope can scatter back onto or within the detector,
thus raising the measured sky-SB levels between the discrete objects (\S\
\ref{sec413}).

To provide proof of concept of our methods in this paper, we therefore
investigated whether the (sky-subtracted) total object brightness is 
correlated with levels of diffuse light that may be present or left in each
image. We consider the total flux coming from objects in each image to be the
mean pixel-value of the $\_flt$-image minus the LES sky-SB level as measured 
in \S\ \ref{sec42}. For this test, we exclude images that may be contaminated
by Earthshine or Sun Light scattered off the Earth by only considering Sun
Altitudes \alphaSun\cle 0\degree\ (\S\ \ref{sec43}). We also exclude images
that are significantly saturated with objects, where more than 30\% of the
sub-regions are masked out in Fig.~\ref{fig:fig7}. 

We perform this test on the WFC3/IR F125W images for two reasons: 1) we use
WFC3/IR's ability to flag and remove most Cosmic Rays during each on-the-ramp
exposure, so that each image mean-pixel value is not skewed by bright CRs, but
instead traces the sky-SB plus the total object flux in the image; 2) we can
directly compare the LES sky-SB in each WFC3/IR F125W image to the
\citet{Kelsall1998} COBE/DIRBE J-band Zodiacal model prediction made in the
{\it same direction and time of the year} (\ie\ at the same RA, Decl., t). We
assume here that the HST--Kelsall sky-SB differences are a measure of diffuse
light that may be present in the F125W images. Details of the applied 
\citet{Kelsall1998} model and the analysis of the HST-Kelsall sky-SB
differences are given in \citetalias{Carleton2022}. 

In essence, we are testing here if the darkest WFC3/IR F125W images --- where
we can measure LES values close to the Zodiacal sky-SB --- are largely free of
residual object flux that affects the sky-SB. The results are shown in
Fig.~\ref{fig:fig12}, where the F125W HST--Kelsall sky-SB differences are {\it
nearly constant at $\sim$0.0071 MJy/sr or $\sim$17 \nWsqmsr\ over a factor of
100 in total object flux} along the horizontal axis. Hence,
Fig.~\ref{fig:fig12} shows no significant trend between the HST--Kelsall sky-SB
difference in each image and its total sky-subtracted object brightness. This
confirms the validity of our LES method to estimate sky-SB values that are
largely free of residual object flux. We will return to this residual
HST--Kelsall sky-SB difference in all three WFC3/IR filters in
\citetalias{Carleton2022}. 

In summary, it is unlikely that a residual diffuse light level is affected
significantly by instrumental effects where light from discrete objects
scatters back into image pixels that are part of the foreground sky-SB. This
suggests that discrete objects (i.e. extended stellar or galaxy
light-profiles) can be ruled out as a major source of diffuse light.
\deleted{In conclusion, residual diffuse light that HST sees in excess of the
\citet{Kelsall1998} model does not significantly depend on the total
sky-subtracted object brightness that the image contained to begin with.}


\n \section{Discussion} \label{sec5}

\mn In this section we briefly summarize how SKYSURF's methods and error 
budget of \S\ \ref{sec4} propagate into the constraints on diffuse light that 
HST may see in excess of Zodiacal models. We confine ourselves to the WFC3/IR
F125W filter used as an example throughout this paper. We refer to 
\citet[][]{Carleton2022} [or ``\citetalias{Carleton2022}''] for a detailed
discussion of any diffuse light signal HST may see --- or limits thereto --- at
1.25--1.6 \mum\ in excess of the \citet{Kelsall1998} and \citet{Wright1998}
Zodiacal models. 

The F125W filter has 6810 full on-the-ramp exposures (Table~\ref{tab:tab3}), 
of which 2337 survive the sub-selection in \S\ \ref{sec48}. 
Fig.~\ref{fig:fig12} suggests an average HST--Kelsall sky-SB difference of
+0.0071$\pm$0.0138 (rms) MJy/sr, or $\sim$17$\pm$33 (rms) \nWsqmsr. These are
the values {\it before} subtraction of the WFC3/IR Thermal Dark signal and the
DGL for each exposure. \citetalias{Carleton2022} discusses in detail the
corrections that needs to be made to the total HST WFC3/IR diffuse sky-signal,
which include subtracting the best fit TD signal from a variety of HST component
temperatures, and the DGL level as estimated for each image location in the
sky.\deleted{In summary, the TD signal in the WFC3/IR F125W filter is
$\sim$0.0012 MJy/sr, while the DGL correction is \cge 0.0009 MJy/sr for most
pointings at high Galactic latitude.} 

The discussion of the HST calibration, Zeropoint, sky-SB measurements and TD
signal errors in \S\ \ref{sec41}--\ref{sec42} and Table~\ref{tab:tab5} suggest
a $\sim$2.7\% total error in our ability to estimate the F125W sky-SB. 
Subtraction of the DGL has its own errors, as does the comparison to any of
the Zodiacal Light models. For instance, the error in the \citet{Kelsall1998}
ZL model prediction is 15 \nWsqmsr\ at 1.25 \mum\ and 6 \nWsqmsr\ at 2.2 \mum.
This amounts to $\sim$3\% of their average {\it predicted} Zodiacal sky-SB. In
\citetalias{Carleton2022}, we thus need to add these ZL model errors in
quadrature to the total sky-SB estimation error of Table~\ref{tab:tab5} --
appropriately interpolated for the F140W and F160W filters --- before
constraints can be placed on diffuse light that HST may see in excess of the
Zodiacal Light models. We now have all the tools in place to do so. 

Last, we return to the most recent diffuse light limits that have been placed
in the visual--near-IR wavelength range, as summarized in \S\ \ref{sec2} and 
Fig.~\ref{fig:fig1}. {The CIBER spectra of \citet{Matsuura2017} suggest a
diffuse light component in excess of the {\it discrete} EBL of 25.6$\pm$12.9 
\nWsqmsr\ at 1.1 \mum\ and 29.7$^{+15.9}_{-9.9}$ \nWsqmsr\ at 1.6 \mum\ compared to
the \citet{Kelsall1998} model prediction as seen from 1 AU. \citet{Sano2020}
reanalyze the weekly COBE/DIRBE data and suggest the existence of a total
(isotropic) diffuse light component of $\sim$45$\pm$10 \nWsqmsr\ in excess of
the \citet{Kelsall1998} model at 1.25 \mum, which amounts to $\sim$34$\pm$10
\nWsqmsr\ after subtracting the iEBL+eEBL component of 11.2$\pm$0.9 \nWsqmsr\
\citep{Koushan2021, Carleton2022}. We plot these constraints as the purple symbols
in Fig.~\ref{fig:fig1}. \citet{Korngut2022} use spectra from 3 recent CIBER
rocket flights to estimate the Equivalent Width (EW) of the Ca triplet around
8542 \AA, extrapolate their result to the \citet{Kelsall1998} model prediction
at 1.25 \mum, and find a diffuse light component of 46$\pm$19 \nWsqmsr\ at 1.25
\mum. This estimate has a larger error, but because it is based on the
Ca-triplet, they suggest a residual (spheroidal) Zodiacal component that is not
present in the \citet{Kelsall1998} model.} \citet{Lauer2021, Lauer2022} present
0.608 \mum\ object counts and sky-SB measurements from New Horizons images
taken at 43--51 AU (blue points with error bars in Fig.~\ref{fig:fig1}), where
their single image obtained at 51 AU from the Sun suggests a much dimmer
diffuse light signal of $\sim$8.1$\pm$1.9 \nWsqmsr\ at 0.608 \mum. They discuss
a number of explanations for such a signal. 

\citetalias{Carleton2022} will discuss the HST diffuse light constraints at
1.25--1.6 \mum\ from SKYSURF in the context of this recent work. In O'Brien
\etal\ (2022, in preparation) we will present the sky-SB in the broad-band
ACS/WFC, WFC3/UVIS, and WFC3/IR filters of
Table~\ref{tab:tab1}--\ref{tab:tab3}, and compare these results to various
Zodiacal Light models at 0.2--1.7 \mum. Table~\ref{tab:tab1}--\ref{tab:tab2}
shows that the full SKYSURF data set will have $\sim$49,000 images in the F606W
filter alone that use all three HST cameras ACS/WFC, WFPC2, and WFC3/UVIS,
allowing us to check for camera-dependent systematics. SKYSURF's goal is then
to use all these results to better constrain the Zodiacal Light models and how
much diffuse light may exist at 0.2--1.7 \mum\ as seen from Low Earth Orbit.


\n \section{Summary and Conclusions} \label{sec6} 

\mn In this paper, we present an overview and describe the rationale, methods,
and first results from the Hubble Space Telescope Archival project
``SKYSURF''. The following are our main highlights and results: 

\sn \bul (1) SKYSURF uses HST's unique capability as an {\it absolute
photometer} on timescales of decades to measure the 0.2--1.7 \mum\ all-sky
surface brightness from 249,861 WFPC2, ACS, and WFC3 exposures. Of these,
220,657 exposures have \texp$\ge$200 sec and cover 16,822 HST FOVs that are
being drizzled in on average $\sim$8 exposures per filter and $\sim$1.8
filters per target. 

\sn \bul (2) Among these, $\sim$1400 constitute independent HST fields spread
across the sky that are suitable for galaxy counts that average over Cosmic
Variance, and cover an area of \cle 10 deg$^2$. For object detection and
catalogs, our zeropoint errors are \cle 1.5\%, while for sky-SB estimates our
total errors are \cle 3--4\% of the Zodiacal sky-SB. 

\sn \bul (3) SKYSURF's panchromatic Legacy dataset is designed to constrain the
diffuse UV-near-IR sky-SB components: Zodiacal Light (inner Solar System),
Kuiper Belt Objects (outer Solar System), Diffuse Galactic Light, integrated
and extrapolated discrete Extragalactic Background Light (iEBL/eEBL), {\it
and} the diffuse Extragalactic Background Light (dEBL). 

\sn \bul (4) Many of SKYSURF's goals require a non-standard re-processing of
these HST images that we summarize, which include, \eg\ monitoring the
instrument zeropoints over 11--18 years, identifying and removing image
gradients where needed, using wider object apertures to remove the outskirts of
stars and galaxy images, as well as trails from CRs and Charge Transfer 
Inefficiency effects, and restoring into each drizzled image footprint the
lowest estimated sky-level from each visit in a given image/filter combination.

\sn \bul (5) We present simulations of HST WFC3/IR images with known
sky-values and sky-gradients, and realistic CR distributions and star plus
galaxy counts. We test nine different algorithms that measure the ``Lowest
Estimated Sky'' (LES) in each image away from the discrete objects and
identified the best algorithm that recovers the inserted LES values to within
0.2\% when there are no image gradients, and within 0.2--0.4\% in the presence
of 5--10\% image gradients. Images with larger gradients are flagged for
SKYSURF sky-SB analysis. Sky-SB values and image gradients are checked against
Earth Limb Angle, Sun Altitude, Sun Angle, and Moon Angle, to help identify the
cleanest and darkest subset of the SKYSURF images. 

\sn \bul (6) Our WFC32/IR sky-SB estimates in the F125W, F140W, and F160W
filters have errors of $\sim$2.7--3.8\% of the average sky-SB, respectively. 
We compare our WFC3/IR F125W sky-SB estimates to the \citet{Kelsall1998}
J-band Zodiacal model prediction for the same direction and time of the year. We
find no significant trend between the 1.25 \mum\ HST--Kelsall {\it sky-SB
difference} for each image and its total sky-subtracted object brightness. This
confirms the validity of our LES method to estimate sky-SB values that are
largely free of residual object flux. 

\sn \bul (7) Last, we will release to MAST our SKYSURF's panchromatic legacy
products with the publication of the respective sequel papers: optimized
object catalogs, absolute calibrated panchromatic sky-SB data, and models of
the ZL, DGL, and EBL, with pointers and documentation at the SKYSURF website
at ASU (\url{http://skysurf.asu.edu}). Through this first SKYSURF paper, we
invite community feedback, so we may further improve our methods as the entire
SKYSURF database gets reprocessed. 

\bn \acknowledgements We dedicate this paper to Dr. Jan Hendrik Oort, who during
his long and very productive life predicted the existence of the Oort Cloud, 
which he sometimes doubted could be observed (Dr. M. J. A. Oort, private
communication). 

{We thank the referee for thoughtful suggestions that helped improve the
presentations of this paper.} We thank Annalisa Calamida, Susana Deustua,
Sylvia Baggett, Jay Gallagher, Phil Korngut, Tod Lauer, John Mather, Peter
McCullough, Marc Oort, Marc Postman, Michael Tompkins, and Meenakshi Wadhwa for
helpful suggestions. We thank Dr. Larry Petro for his help in the early stages
of this project. We thank Javier Calunga for his help in setting up our SKYSURF
servers, and Mark Stevens for his continuous help in maintaining the SKYSURF
server website. We thank HST Archive staff at STScI for getting a very large
amount of HST Archival data to us without any major hiccups, and for their
expert advice on HST component temperatures. All of the data presented in this
paper were obtained from the Mikulski Archive for Space Telescopes (MAST). This
project is based on observations made with the NASA/ESA Hubble Space Telescope
and obtained from the Hubble Legacy Archive, which is a collaboration between
the Space Telescope Science Institute (STScI/NASA), the Space Telescope European
Coordinating Facility (ST-ECF/ESA), and the Canadian Astronomy Data Centre
(CADC/NRC/CSA). Some image simulations were based on observations taken by the
3D-HST Treasury Program (GO 12177 and 12328) with the NASA/ESA HST, which is
operated by the Association of Universities for Research in Astronomy, Inc.,
under NASA contract NAS5-26555. 

We thank Ms. Desiree Crawl, Prof. Thomas Sharp, and the NASA Space Grant
Consortium in Arizona for consistent support of our many undergraduate SKYSURF
researchers at ASU during the pandemic. We acknowledge support for HST
programs AR-09955 and AR-15810 provided by NASA through grants from the Space
Telescope Science Institute, which is operated by the Association of
Universities for Research in Astronomy, Incorporated, under NASA contract
NAS5-26555. 

We also acknowledge the indigenous peoples of Arizona, including the Akimel
O'odham (Pima) and Pee Posh (Maricopa) Indian Communities, whose care and
keeping of the land has enabled us to be at ASU's Tempe campus in the Salt
River Valley, where this work was conducted.

\sn \software{
Astropy: \url{http://www.astropy.org} \citep{Robitaille2013, Astropy2018};\ 
IDL Astronomy Library: \url{https://idlastro.gsfc.nasa.gov} \citep{Landsman1993};\ 
Photutils: \url{https://photutils.readthedocs.io/en/stable/} \citep{Bradley20};\
\ProFound: \url{https://github.com/asgr/ProFound} \citep{Robotham2017};\ 
\ProFit: \url{https://github.com/ICRAR/ProFit} \citep{Robotham2018};\ 
\SExtractor: SourceExtractor:
\url{https://www.astromatic.net/software/sextractor/} or
\url{https://sextractor.readthedocs.io/en/latest/} \citep{Bertin1996}. 
}

\facilities{
Hubble Space Telescope Mikulski Archive \url{https://archive.stsci.edu};\ 
Hubble Legacy Archive (HLA) \url{https://hla.stsci.edu};\ Hubble Legacy
Catalog (HLC) \url{https://archive.stsci.edu/hst/hsc/}
}


\bibliographystyle{aasjournal}


\bibliography{references_paperI}{}



\appendix

\n \section{HST Orbital Parameters and Orbital Straylight}\ 
\label{secAppA} 

\sn In Appendix \ref{secAppA1}--\ref{secAppA2}, we give a summary of the
calculation of HST's orbital parameters, as well as of straylight from the
Earth, Sun and Moon across HST's orbital phases. In Appendix A of
\citetalias{Carleton2022} we also give summary of HST's known thermal behavior
as documented in its telemetry data. 

\n \subsection{Calculation of HST Orbital Parameters} \label{secAppA1} 

\sn The Earth's Limb Angle (LA) is recorded in the \jit-files every few seconds
during an exposure. SKYSURF retrieves this information for all ACS/WFC3, WFPC2,
and WFC3/UVIS and IR images. The run of LA's across a WFC3/IR exposure has to
be recomputed using the post-observing ephemerides of HST. We developed code
to generalize this, and obtain a good match to these \jit-values compared to
ephemerides. The LA-predicting 
software~\footnote{\url{https://www.qsl.net/kd2bd/predict.html} (IDL)\ \ and\
\ \url{https://github.com/npirzkal/EarthPlot} (Python)} calculates the exact
{\it past} HST orbital parameters (in 6-dimensional space $\vec{s}$ + velocity
$\vec{v}$ coordinates: $x$, $y$, $z$; $v_x$, $v_y$, $v_z$) in the geocentric
system for any time in the past, which can be retrieved from the NORAD 
database~\footnote{\url{http://www.celestrak.com/NORAD/archives/request.php}}, 
where HST is satellite number 20580. The NORAD satellite archive returns a file
that needs to be split into individual HST orbital files, done with IDL
routine \texttt{split\_hstorbits.pro}. The results are saved in a
\texttt{.predict} folder, and the LA is calculated with \texttt{limbangle.pro},
which updates the \texttt{predict} files for each orbit. Together with the past
HST ephemerides, these calculations also include the Illuminated Earth Fraction
(IEF) as a parameter that can affect the amount straylight in the individual
SKYSURF images (\S\ \ref{sec43}). Because of Earth-limb straylight possibly
reflecting differently off the HST OTA cover at different off-nominal roll
angles (RAs), this code can also help trace the sky-SB as a function of
off-nominal RA and IEF as needed \citep{Caddy2022}. 

It should be noted that COBE's orbit is at an average Low Earth Orbit (LEO)
altitude of 885 km, while HST is at an average orbital altitude of $\sim$539 
km. The HST orbital altitude was somewhat higher early in the mission, and it
has slightly decayed during the 1991, 2002, and 2013 Solar maxima, but some
altitude was recovered due to reboosts from several of its five Shuttle
Servicing Missions. Given the very low atmospheric emission at 539--885 km
altitude in directions well away from the Earth (other than Geocoronal lines
such as \Lya at 1216 \AA\ and He II at 1.083 \mum), we will assume that the
{\it orbital altitude difference} between the HST and COBE is {\it not} the
main source of any differences in 1.25--1.6 \mum\ sky-SB measurements between
the two missions discussed in \citetalias{Carleton2022}. This is supported by
MSISE-90 Upper Atmospheric models of the
Earth~\footnote{\url{http://www.braeunig.us/space/atmos.htm}}, which lists that
the mean atmospheric pressure is 2.27$\times$10$^{-7}$ Pa at 540 km, while it
is 1.04$\times$10$^{-8}$ Pa at 885 km.

\n \subsection{HST Orbital Straylight}\ \label{secAppA2} 

\sn This section gives further details on straylight due to proximity to the
Earth's Limb, the Sun Altitude above the bright Earth and the Sun Angle with
respect to HST's pointing direction, as well as due to the Moon angle.

\n \subsubsection{Earth's Limb Angle} \label{secAppA21} 

\sn Eliminating levels of scattered Earthshine contamination in exposures used
in deriving sky-SB levels is paramount to the core science goals of SKYSURF.
All orbital parameters that may impact the Earthshine intensity in exposures
must be considered. SKYSURF thus needs to identify all exposures impacted by
Earthshine or scattered light, and flag those. To ensure that Earthshine is
limited to a minimum, one would need to select only those fields that were
taken in orbital dark time, which meet the \texttt{SHADOW} or \texttt{LOW-SKY}
Special Requirements. STScI developed these options for this purpose early in
the HST mission, as can be seen in the current version of APT. However, not
every HST observer who asked for orbital \texttt{SHADOW} or \texttt{LOW-SKY}
time in their APT file did, in fact, receive it, nor did every observer who
received orbital \texttt{SHADOW} or \texttt{LOW-SKY} time ask for it. 

We flag all individual images that came too close to the Earth's limb (LA\cle
40--60\degree), irrespective of whether the observer asked for orbital dark
time or not. For WFC3/IR, we do this also using individual time-samples {\it
within} an exposure as in Fig.~\ref{fig:fig9}a--\ref{fig:fig9}b. Since the
orbital dark time does not generally exceed $\sim$1800 sec per HST orbit, the
simplest approach is to sub-select only those exposures in the range 200--1800
sec. Exposures longer than 2100 sec often run into elevated sky due to
proximity to the Earth's limb. An example is given by the green 0.23 and 1.6
\mum-points in Fig.~\ref{fig:fig1}, which are for these reasons higher than the
expected Zodiacal SB-values that are normalized to the Solar spectrum, as
discussed in \citet{Windhorst2018}. 

Fig.~\ref{fig:fig9}a shows the sky-SB as a function of Earth's Limb Angle. 
The orange line and model uncertainty wedge of \citet{Caddy2022} show the
elevated sky-SB due to Earthshine for Sun Altitude 
\alphaSun$\simeq$90\degree\ on the orbital day side with the SA fixed at
$\sim$90\degree. The model cuts off at LA$\sim$20\degree\ due to HST scheduling
constraints. A strong exponential increase is seen for fields taken on the day
side of the orbit with LA\cle 40--45\degree, but even fields with LA\cle
70--80\degree\ on the day side have an elevated sky-SB. The blue line plus
uncertainty error wedge indicates the \citet{Caddy2022} model on the night side
for \alphaSun$\simeq$--10\degree, and shows a marked decrease in sky-SB. The
model extends to LA$\simeq$10\degree\ due to scheduling constraints on the night
side. Deviation from an otherwise flat relation is seen for fields with LA\cle
20\degree\ on the night side. Exposures that have 10\cle LA\cle 20\degree\ and
also \alphaSun\cle 0\degree\ have reduced sky-SB, but with somewhat increased
scatter due to the shallow Limb Angle. 

In summary, Earthshine increases strongly as a function of Limb Angle. This is
also visible in FIGS WFC3/IR grism data \citep[\eg][]{Pirzkal2014, Brammer2015,
Brammer2016}. The 1.083 \mum\ Geocoronal He II line can be quite bright and
contaminate WFC3 near-IR images in the F105W and F110W filters and the G102
grism \citep{Brammer2014, Pirzkal2020}, so F105W and F110W images with
increased sky-SB levels are also flagged for this reason.
\citetalias{Carleton2022} shows that the HST WFC3/IR F125W and COBE/DIRBE
J-band filters exclude the 1.083 \mum\ Geocoronal He II line, if it were
present at low Limb Angles. The WFC3/IR filters F125W, F140W, and F160W used
for our first SKYSURF results in \citet[][]{Carleton2022} are thus free of the
brightest known Geocoronal emission lines. 

\n \subsubsection{Sun Altitude and Sun Angle} \label{secAppA22} 

\sn In addition to the Earth's Limb Angle, the Sun Altitude above the Earth's
Limb, the Sun Angle, and the Moon Angle [both measured from HST's pointing
direction] over the duration of an exposure must be identified and constrained
to minimize Sunlight scattered off the bright Earth, or off the IPD cloud at 
angles too close to the Sun. The results of this analysis are shown in
Fig.~\ref{fig:fig9}c--\ref{fig:fig9}d. 

\sn {\bf Sun Altitude (\alphaSun)}:\ \ The Sun Altitude limit is found to be
the most influential parameter in reducing Earthshine contamination, followed
by the Limb Angle limit. For SKYSURF data the Sun Altitude limit is chosen to
be all exposures with \alphaSun\cle --10\degree. Earthshine directly below the
location of HST can be assessed independently from surface flux measurements at
0.3--5 \mum\ wavelengths by the CERES (Clouds and the Earth's Radiant Energy
System) satellite over the duration of {\it the same} Low-Earth orbit
\citep{Wielicki1996}. This enables us to monitor Earthshine levels {\it in real
time} on the day side of an HST orbit underneath the telescope, where the
intensity of Earthshine cannot be accurately predicted due to the complex
nature of the weather systems below \citep{Caddy2021}. This work suggests that
higher Earthshine levels are found to potentially contaminate HST fields with
Sun Altitudes values \alphaSun\cge --10\degree\ or more on the day side of an
orbit. Fig.~\ref{fig:fig9}b shows the sky-SB as a function of Sun Altitude
\alphaSun. The orange and blue lines plus model uncertainty wedges of 
\citet{Caddy2022} show that lower LA's result in steeper relationships for Sun
Altitudes \alphaSun\cge 10--20\degree\ due to the increased Sunlight scattered
off the bright Earth. In this CVZ dataset --- and in many others --- the Sun
Altitude can be \cge 90\degree\ and in GOODS-N up to 110\degree\
(Fig.~\ref{fig:fig9}b), or $\sim$20\degree\ towards HST's aperture as seen from
HST's ``local Zenith''. Such angles almost certainly increase Solar straylight
off the bright Earth, depending on HST's roll angle and how its aperture door
faces the illuminated Earth. For \alphaSun\cle --10\degree, the sky-SB is
lowest and remains approximately constant. As a result, no day-side exposures
are used in the final SKYSURF sky-SB analysis. This constraint considerably
reduces the amount of usable data in the archive but is necessary to prevent
significant Earthshine contamination. 

\sn {\bf Sun Angle}:\ \ Fig.~\ref{fig:fig9}d shows that the minimum HST sky-SB
levels increase for Sun Angles\cle 80\degree, including on the night side of an
orbit, in part due to an increase in ZL intensity closer to the Sun. Standard
HST scheduling constraints of SA\cge 50\degree\ are observed. HST observations
are often scheduled at or near SA$\simeq$90\degree\ or ``nominal roll'' angle
to ensure optimal illumination of the solar array. Sun Angles are therefore
clustered corresponding to the date of the exposures. The large scatter in the
sky-SB levels around each Sun Angle is primarily due to the variation in Sun
Altitude and Limb Angle for each exposure. Due to the degeneracy between
elevated ZL and scattered Sunlight inside the HST telescope tube, it is not
possible to identify the dominant cause of elevated sky-SB for Sun Angles
SA\cle 80\degree. As a result, we will avoid Sun Angles SA\cle 80\degree\ for
our final SKYSURF sample. 

\n \subsubsection{Moon Angle} \label{secAppA23} 

\sn Fig.~\ref{fig:fig9}c shows the sky-SB as a function of Moon Angle. For
the Moon, the HST avoidance angle is typically MA\cge 50\degree. At MA\cge 
50\degree, the Moon Angle is not the major driver of increased sky-SB. 

\n \section{Specific SKYSURF Requirements for Image Drizzling and Image
Filtering} \label{secAppB} 

\sn In this section, we discuss the application of wider-apertures for the
removal of trails from cosmic rays (CRs) and Charge Transfer Inefficiency
effects during drizzling. We also give details on SKYSURF's methods to
identify and remove HST images with large artifacts, and HST images with large
targets that overfill the FOV. In all cases, fields that are discarded during
the flagging process in Appendix \ref{secAppB3} must have a known reason for
their omission to avoid biasing the remaining sample with galaxy counts in
Appendix \ref{secAppC} that are too low or too high. Further details are given
in Tompkins \etal\ (2022, in preparation).

\n \subsection{Application of Wider-Aperture Cosmic Ray Removal during
Drizzling} \label{secAppB1} 

\sn SKYSURF has several unique aspects to optimize its CR rejection
procedures: (a) do wider-aperture CR rejection to get accurate sky-SB
measurements; (b) drizzle images that mostly overlap; and (c) properly drizzle
all targets with one ACS, WFPC2, or WFC3 exposure per filter. These latter
steps are needed to increase the statistics for SKYSURF sky-SB measurements.
The HLA does not process all HST images, and does not drizzle HST targets with
one exposure per filter, nor drizzles targets with partly overlapping images.
SKYSURF needs to carry out both steps to increase the statistics for SKYSURF
sky-SB measurements and apply the GDC to each output image to enable us to do
faint object photometry across each FOV. 

Non-Standard CR rejection is required for SKYSURF. For reliable discrete object
counts, we need to remove CRs reliably during drizzling. \citet{Miles2018} and
\citet{Miles2020} recovered more than 1.4 billion CRs that have hit ACS, WFC3 or
STIS images, which took a few days on AWS to process. The CR-flux varies by
10--20\% across the Solar Cycle, in a way that is anti-correlated with the
Solar Maximum activity \citep{Miles2018}. SKYSURF uses the fact that CRs are
usually sharper than the FWHM of the image PSF to remove them from images with
only 1--2 exposures \citep[\eg][]{vanDokkum2001}. For most targets with \cge
3--4 exposures per filter per target (Table~\ref{tab:tab4}), the CR filtering
will work more reliably by comparing the sky-flux in each aligned pixel
\citep[\eg][]{Windhorst1994a}. 

The CR rejection is customized in this processing, in particular ensuring that
faint regions around bright CRs are accurately rejected and accounted for, so
as not to impact the final surface brightness measurements. To make sure this 
is done correctly, we run the pipeline on both the $\_flt$ and the
CTE-corrected $\_flt$ files ($\_flc$), with two different CR rejection masks. We
then use the $\_flt$ images for the sky-histogram, and $\_flc$ images to get the
best possible total fluxes (Appendix \ref{secAppB2}). This uses wider
apertures to optimize the CR-masking, and is part of our modified SKYSURF
pipeline. AstroDrizzle performs this process as follows: a) all images,
starting with single exposures per filter, get drizzled on a grid with a common
--- Gaia DR2 based \citep{Gaia21} --- World Coordinate System (WCS); and b) run
\SExtractor\ and \ProFound\ on all images, including the single exposures per
target, where typically 5--10\% of the pixels are affected by CRs. Most
algorithms in \S\ \ref{sec42} can nonetheless measure their sky-SB accurately,
and \ProFound\ \citep{Robotham2017} has been tested to verify it can do so
also. All multiple orbital --- or multi-orbit --- exposures per filter on a
given target are then drizzled into a final mosaic for each filter on that
target (\S\ \ref{sec32} \& \ref{sec45}). The lowest sky-SB measurement is
preserved for each visit during this drizzling process (\S\ \ref{sec46}), with
sky gradients mapped and removed (unless they are part of the astrophysical
scene). Both the removed sky-gradients and the restored sky-levels are
preserved as separate FITS files or in the FITS header, respectively, for later
diagnostics or other uses. 

SKYSURF is processing all HST images with exposure times \texp$\ge$200 sec 
(220,657 WFPC2, ACS, and WFC3 exposures in 16,822 HST FOVs
(Table~\ref{tab:tab4}) through these special-purpose, non-standard DrizzlePac
steps on AWS, including the use of custom-measured sky values, and preserving
the full information about any variation in sky between different exposures. 

\n \subsection{Application of Wider-Aperture Removal of Charge Transfer 
Inefficiency Trails} \label{secAppB2} 

\sn To get the most reliable sky-SB measurements, SKYSURF also makes
corrections for Charge Transfer Efficiency (CTE) degradation. We assess and
correct biases due to the increase of CTE effects over time, \ie\ trails
behind bright point sources, including CRs. CTE trails that bleed into the
sky-SB foreground are not removed in standard processing. The standard 
pipeline therefore applies the CTE-correction code of \citet{Anderson2010} to
all images before drizzling. If not corrected, CTE trails may affect the sky
pixel histogram, and may also accentuate read-noise artifacts. During the
90--100 sec ACS and WFC3 readout times, CRs are also accumulated during
readout, which are harder to deal with. We therefore use the non-CTE-corrected
$\_flt$ files to remove those CRs. SKYSURF then grows the standard CR masks a
little larger than normal to remove faint CTE trails in the final drizzled
images. For a subset of the data per detector, we test our SB-measurements {\it
before and after} removal of CTE trails \citep{Anderson2010} as a check. We find
no significant difference in the sky-SB estimates. 

We use the $\_flc$ images to get the sky-histograms and the best possible
total object fluxes. With \SExtractor\ or \ProFound, we then define all objects
on the $\_flc$ images, and next grow the CR-, CTE-, and object-masks on the
$\_flt$ images to ensure that all of the CTE-trailed CR-flux and object-flux
is included in the CR-, CTE-, and object-rejection apertures. Since SKYSURF's
dataset has a large dynamic range in flux, we can afford to do wider-aperture
removal of CRs: false negatives are acceptable (\eg\ some faint point-like
objects mistakenly removed as CRs in some images), but false positives are not
(\eg\ faint CRs left in the images). The optimal parameter settings for blind
global CR-masking for SKYSURF thus need to be done over more conservative,
wider-apertures than for default CR-removal. 

The CTE-correction code of \citet{Anderson2010} has been improved over the
years \citep[\eg][]{Anderson2021}, which is now available under
\texttt{calwfc3} pipeline version 3.6.0 and higher. Our WFC3/UVIS images were
retrieved in spring 2020 using the latest available \texttt{calwfc3} version 
3.5.0 at that stage, which implemented the 2016 version of the CTE correction.
The difference between the 2016 and 2021 CTE corrections are primarily a
somewhat noisier sky-image and perhaps somewhat over corrected faint object
fluxes using the 2016 CTE correction, and a somewhat smoother sky with somewhat
under-corrected faint object fluxes using the 2021 CTE
correction~\footnote{\footnotesize{see \eg\ Fig. 1 and 2 of 
\url{https://www.stsci.edu/hst/instrumentation/wfc3/performance/cte}}}. To
quantify this for SKYSURF, we ran all WFC3/UVIS F850LP images with both
\texttt{calwfc3} pipelines (versions 3.5.0 and 3.6.2 in early 2022) on AWS. We
found indeed that the rms sky-noise in object-free areas was $\sim$2.4\% lower,
while the average object-free sky-SB was 0.8\% higher with the 2021 pipeline.
Details are given by O'Brien \etal\ (2022, in preparation). The 0.8\%
uncertainty in sky-SB induced by the different CTE-correction methods is
acceptable for SKYSURF and folded into the error budget of Table 5 for
WFC3/UVIS. The 2.4\% higher average sky-rms values resulting from
\texttt{calwfc3} version 3.5.0 are not of immediate concern, as they are easily
offset by SKYSURF's statistics from over 40,000 WFC3/UVIS images. Both the
WFC3/UVIS 2016 and 2021 CTE corrections are preserved until it can be
determined which one provides more reliable total fluxes and counts of faint
objects across the sky, as that is the second main goal of SKYSURF. Details 
on this process are given by Tompkins \etal\ (2022, in preparation). 

\n \subsection{Identifying and Removal of HST Images with Large Artifacts and
Large Targets}\label{secAppB3} 

\sn To get reliable panchromatic object catalogs and sky-SB measurements, 
SKYSURF needs to reliably identify and remove all major image artifacts. These
can be due to, \eg\ scattered (star) light, satellite trails, ``dragon's
breath'', edge-glow, straylight from (brighter) stars just outside the FOV, and
image persistence or crosstalk in the ACS, WFPC2, or WFC3 images (\S\
\ref{sec41}). Some examples are shown in, \eg\ \citet{Gosmeyer2017} and
\citet{Dauphin2022}. STScI has spent considerable effort to catalog these image
artifacts in WFC3, and the ACS group has done the same mostly in F606W and
F814W, used as a starting point and guidance to do the flagging. Anomalies have
been recorded since ACS and WFC3's launch in 2002 and 2009, respectively, in
the QL database of ``Visually Noted Anomalies'' (VNA's). The VNA database is
available for all non-proprietary data from the WFC3 and ACS archives in MAST.

\deleted{SKYSURF also needs to accurately remove bright-star halos. One of the
most difficult tasks for the HLA was masking the areas affected by bright stars.
\citet{Porterfield2016} present a scattered-light anomaly analysis of \cge
8500 ACS images. Our best algorithms of \S\ \ref{sec42} identify and remove
stellar halos, \eg\ in ACS z-band, where red stellar halos can be large 
depending on the star's color. SKYSURF uses the Gaia DR2 catalog \citep{Gaia21}
to identify and remove the halos of stars \cle 19 mag. We do not need to remove
all stars with AB$\simeq$17--19 mag from the images, since they are generally
not saturated, so they are useful for star counts across the sky from 17\cle
AB\cle 26.5 mag. \ProFound\ identifies all (likely) saturated stars with AB\cle
16 mag, and traces their outer halos, which are removed using large-enough
buffer zones with \ProFit\ to mask them before doing sky-SB measurements 
(Appendix \ref{secAppC12}).}

The ability of SKYSURF to accurately carry out the Zodiacal foreground
measurements hinges on a robust, largely automated rejection algorithm of all
images with known large and bright foreground objects. These objects are often,
but not always, obvious from the title of the observing program or the Object
Name in the FITS header. SKYSURF uses information in the HST FITS headers and
Engineering telemetry (\jit\ files), through catalogs on NED and SIMBAD, and/or
by visual inspection of WFPC2, ACS, and WFC3 images to flag: (a) {\it large}
targets that HST pointed at; (b) very large random objects that overfill the
ACS or WFC3 FOV; or (c) known or unknown large image defects. Examples are,
\eg\ large Galactic objects or star-forming regions, as well as nearby NGC and
RC3 galaxies with a scale-length $r_e$ large enough to overfill the FOV. Either
one of these can corrupt the SKYSURF source catalogs, and often also the SKYSURF
sky-SB measurements, as illustrated in the top row of Fig.~\ref{fig:fig7}. 
These large targets get flagged by SKYSURF collaborators in the same stage
where the large artifacts and visual anomalies are found and flagged 
(\eg\ Fig.~\ref{fig:fig3} \& Fig.~\ref{fig:fig7}). 

\deleted{Blank sky survey fields are generally fully usable for sky-SB
measurements and object counts (\eg\ bottom two rows of Fig.~\ref{fig:fig7}).
Once images are properly CR cleaned and drizzled, SKYSURF makes a version of
the images that masks out objects/areas of the original HST target intent for
the purpose of foreground star- and background galaxy counts. If this is not
possible, such images are discarded for object counts. After excluding all
large targeted objects and their obvious neighbors, the background objects in
those images well away from the targeted foreground objects may be usable for
unbiased object counts. When in doubt, SKYSURF discards an entire image from the
Galactic star counts and background galaxy counts.}

After removing the fields with large artifacts and/or large HST targets, we
expect that $\sim$30\% of the HST FOVs that are \cge 1\degree\ apart on the
sky, or $\sim$1400 independent HST fields, will be usable for independent
galaxy counts in a given filter that will significantly average over Cosmic
Variance (\S\ \ref{sec252}, \ref{sec32} \& \ref{sec45}). Further details are
given in Tompkins \etal\ (2022, in preparation).

\n \section{SKYSURF Panchromatic Object Catalogs and Star-Galaxy Separation} 
\label{secAppC} 

\sn In this section, we summarize SKYSURF's methods to generate object
catalogs and perform star-galaxy separation, using both the \SExtractor- and
the \ProFound-packages, as well as SKYSURF's method to correct for Galactic
extinction for both galaxies and stars. 

\n \subsection{SKYSURF Object Catalogs} \label{secAppC1} 

\sn Reliable object catalogs are essential for SKYSURF's goals. Hence, SKYSURF
uses two independent object finder algorithms on the drizzled ACS/WFC and
WFC3/UVIS and IR images as illustrated in Fig.~\ref{fig:fig3}. For the SKYSURF
source finding process, we use traditional object finder \SExtractor\
\citep{Bertin1996} and the more recent object-finding code \ProFound\
\citep{Robotham2017} and profile-fitting code \ProFit\ \citep{Robotham2018}.
Both packages were designed to deblend close objects and find the optimal
object total fluxes. SKYSURF compares the \ProFound\ source catalogs to those
from \SExtractor\ to quantify which algorithm yields the most robust object
total fluxes and sky-SB measurements in between the discrete objects (\eg\
Fig.~\ref{fig:fig3} \& Fig.~\ref{fig:fig7}--\ref{fig:fig8}). 

\n \subsubsection{Object Catalogs with \SExtractor} \label{secAppC11} 

\sn For SKYSURF source finding, we first use the traditional \SExtractor\
algorithm \citep{Bertin1996} that the astronomical community has been using for 
over 25 years, but on our newly produced SKYSURF mosaics with CR-removal over
wider apertures, while the lowest estimated sky-SB from each drizzle family is
left in the drizzled images. Each catalog is manually inspected for quality
control, as described in Appendix \ref{secAppB3}. Object catalogs are then
passed on to the star-galaxy separation module described in \S\ \ref{sec47} \&
Appendix \ref{secAppC2}. Questionable catalogs are flagged for either
refinement of \SExtractor\ parameters or removal from the survey, \ie\ in the
case of overcrowded fields. Once the best possible catalogs are obtained with
\SExtractor\ and visually verified, star-galaxy separation is performed by
examining plots of source brightness vs. spatial extent (\ie\ magnitude versus
full-width at half-max (FWHM; see Fig.~\ref{fig:fig10}). 

\n \subsubsection{Object Catalogs with \ProFound} \label{secAppC12} 

\sn The \ProFound\ \citep{Robotham2017} source finding and \ProFit\
\citep{Robotham2018} object profile-fitting codes are used for the source
finding and source definition aspects of SKYSURF. These packages were designed
to get the most reliable image sky-surface, object light profiles and {\it
total magnitudes} for panchromatic object counts. \ProFound\ is optimized to
trace the shape of the object and then dilate around this shape to some
convergence limit. 

\n \subsection{SKYSURF Star-Galaxy Separation and Galactic Extinction
Corrections} \label{secAppC2} 

\sn SKYSURF requires accurate star-galaxy separation in each field, which must
be highly reliable to get the best possible {\it galaxy} counts over $\sim$1400
independent HST fields covering 16\cle AB\cle 28 mag. SKYSURF therefore needs to
do accurate star-galaxy separation on all images in each filter in an optimized
way that is both automatic and highly reliable. \ProFound\ can do star counts
reliably in ground-based images to AB\cle 20 mag as confirmed by Gaia
\citep[see \eg][]{Koushan2021}. At HST resolution, reliable star counts can be
done to AB\cle 25--26 mag (Fig.~\ref{fig:fig10}--\ref{fig:fig11} here;
\citealp{Windhorst2011}). At total fluxes fainter than AB$\sim$25 mag, the
galaxy counts clearly dominate the star counts by a factor of \cge 100
(Fig.~\ref{fig:fig11}) at intermediate to high Galactic latitudes, so we do not
need to aim for very accurate star-galaxy separation for the purpose of galaxy
counts at AB$\simeq$25--26 mag.

The panchromatic magnitudes of all SKYSURF galaxies --- and hence their galaxy
counts --- are corrected for Galactic extinction using \citet{Schlafly2011} as
a function of (\lII, $b^{II}$). The panchromatic galaxy counts will likely only
be reliable and be done for Galactic latitudes $\vert$$b_{II}$$\vert$\cge
30\degree, where the extinction corrections are generally accurate and
relatively small. No extinction correction has been applied to the star counts
in Fig.~\ref{fig:fig11}. As the SKYSURF catalogs proceed, bright stars (AB\cle
19 mag) are verified with the Gaia DR2 catalog \citep{Gaia21}. Given their Gaia
parallax distance from the Sun, their Galactic extinction correction thus 
depends on where Gaia places the star compared to the 3D Galactic dust maps of 
\citet{Green2019}, which are based on a joint analysis that combines
Pan-STARRS, 2MASS, and Gaia data. Stars fainter than those measured with Gaia
(AB\cge 20 mag) are likely much closer than the brighter Gaia stars, with
typical disk scale heights of a few 100 pc \citep[\eg][]{Pirzkal2009, Ryan2011,
Ryan2017}, and may not need a full extinction correction, as they are likely in
front of most of the Galactic dust. Further details on SKYSURF's star-galaxy
separation methods, object catalogs, and Galactic extinction corrections made
are given by Carleton \etal\ (2022b, in preparation) and Tompkins \etal\ (2022,
in preparation).


\ve 

\vspace*{-1.00cm}
\n \section{Acronyms Used in SKYSURF} \label{secAppD}
{
\mn 
}{
\begin{tabular}{ll}
\hline
\hline
\n Acronym   &Explanation                              \\[-2pt]
\hline
AB-mag       &--2.5 log (Object-Flux / Zeropoint-Flux) \\  
ACS          &Advanced Camera for Surveys \\  
AGN          &Active Galactic Nucleus \\  
APT          &Astronomers Proposal Tool \\
ASU          &Arizona State University \\  
AWS          &Amazon Web Services \\
CCD          &Charged Coupled Device \\  
CDM          &Cold Dark Matter \\  
CERES        &Clouds and the Earth's Radiant Energy System \\
CIB          &Cosmic Infrared Background \\
COB          &Cosmic Optical Background \\
COBE         &Cosmic Background Explorer \\
COS          &HST's Cosmic Origins Spectrograph \\
CR           &Cosmic Ray \\
CTE          &Charge Transfer Efficiency \\ 
CV           &Cosmic Variance \\
CVZ          &Continuous Viewing Zone \\
DC           &(Electronic) Dark Current \\
DGL          &Diffuse Galactic Light \\
DIRBE        &Diffuse Infra-Red Background Experiment \\ 
EBL          &Extragalactic Background Light \\ 
dEBL         &diffuse Extragalactic Background Light \\ 
eEBL         &extrapolated Extragalactic Background Light \\ 
iEBL         &integrated Extragalactic Background Light \\ 
ERS          &(HST WFC3) Early Release Science program \\
FOC          &HST's Faint Object Camera \\  
FOS          &HST's Faint Object Spectrograph \\  
FOV          &Field of View \\  
FWHM         &Full-Width Half-Maximum \\  
GDC          &Geometrical Distortion Corrections \\
GOODS        &Great Orbiting Observatories Deep Survey \\  
\HAB         &H-band (1.6 \mum) AB-mag \\  
HDF          &Hubble Deep Field \\  
HLA          &Hubble Legacy Archive \\
HLC          &Hubble Legacy Catalog \\
HST          &Hubble Space Telescope \\  
HUDF         &Hubble UltraDeep Field \\  
HWHM         &Half Width Half Maximum (=0.5$\times$FWHM) \\  
ICL          &Intra-Cluster Light \\
IEF          &Illuminated Earth Fraction \\ 
IGL          &Intra-Group Light \\
IPD          &InterPlanetary Dust \\
\hline
\end{tabular}
} \label{tab:tab8}

\ve 

\vspace*{-0.00cm}
{
\cl { D.\ \ ACRONYMS USED IN SKYSURF (continued) }
\mn 
}{
\begin{tabular}{ll}
\hline
\hline
\n Acronym  &Explanation                               \\[-2pt]
\hline
IRAF         &Image Reduction and Analysis Facility \\  
ISM          &Interstellar Medium \\
\JAB         &J-band (1.25 \mum) AB-mag \\  
Jy           &Jansky or Flux Density unit (=10$^{-26}$\WsqmHz) \\
KBOs         &Kuiper Belt Objects \\
LA           &Earth's Limb Angle \\
LEO          &Low Earth Orbit \\
LES          &Lowest Estimated Sky-SB \\ 
LFS          &Lowest Fitted Sky-SB \\
MA           &Moon Angle \\
MAST         &Mikulski Archive for Space Telescopes \\
NED          &NASA Extragalactic Database \\ 
NEP          &North Ecliptic Pole \\  
NICMOS       &Near-Infrared Camera and Multi-Object Spectrograph \\  
OCC          &Oort Cloud Comets \\
OTA          &Optical Telescope Assembly \\
PAM          &Pixel Area Map \\
PSF          &Point Spread Function \\  
QSOs         &Quasi Stellar Objects \\
RA           &HST Roll Angle \\
R.A.         &Right Ascension \\
RC3          &Third Reference Catalog of Bright Galaxies \\  
SAA          &South Atlantic Anomaly \\
SA           &Sun Angle \\
SB           &Surface Brightness \\  
SDSS         &Sloan Digital Sky Survey \\  
SED          &Spectral Energy Distribution \\ 
SEP          &South Ecliptic Pole \\
SFR          &Star-Formation Rate \\ 
SF           &Star-Forming \\
SM           &Servicing Mission \\
STIS         &Space Telescope Imaging Spectrograph \\
STScI        &Space Telescope Science Institute \\  
TD           &Thermal Dark {\it signal} \\
TNOs         &Trans-Neptunian Objects \\
UVIS         &WFC3 UV--Visual channel \\  
UV           &Ultraviolet ($\sim$0.1--0.3 \mum) \\  
WFC3         &HST's Wide Field Camera 3 \\  
WFPC2        &HST's Wide Field Planetary Camera 2 \\  
WF/PC        &HST's Wide Field/Planetary Camera \\
ZL           &Zodiacal Light \\
\hline
\end{tabular}
} \label{tab:tab9}

\ve 


\end{document}